\documentclass[preprint]{aastex61}

\newcommand\aastex{AAS\TeX}

\accepted{March 9, 2023}
\shorttitle{\aastex\ ULIRG 1--2 kpc nuclear dense molecular gas}
\shortauthors{Imanishi et al.}
\begin{document}

\title{Dense Molecular Gas Properties of the Central Kpc of Nearby
Ultraluminous Infrared Galaxies Constrained by ALMA Three 
Transition-line Observations}

\correspondingauthor{Masatoshi Imanishi, Shunsuke Baba}
\email{masa.imanishi@nao.ac.jp, shunsuke.baba.astro@gmail.com}

\author[0000-0001-6186-8792]{Masatoshi Imanishi}
\affil{National Astronomical Observatory of Japan, National Institutes 
of Natural Sciences (NINS), 2-21-1 Osawa, Mitaka, Tokyo 181-8588, Japan}
\affil{Department of Astronomy, School of Science, Graduate
University for Advanced Studies (SOKENDAI), Mitaka, Tokyo 181-8588, Japan}
\affil{Toyo University, 5-28-20, Hakusan, Bunkyo-ku, Tokyo 112-8606, Japan}

\author[0000-0002-9850-6290]{Shunsuke Baba}
\affil{Kagoshima University, Graduate School of Science and Engineering, 
Kagoshima 890-0065, Japan}
\affil{National Astronomical Observatory of Japan, National Institutes 
of Natural Sciences (NINS), 2-21-1 Osawa, Mitaka, Tokyo 181-8588, Japan}

\author[0000-0002-6939-0372]{Kouichiro Nakanishi}
\affil{National Astronomical Observatory of Japan, National Institutes 
of Natural Sciences (NINS), 2-21-1 Osawa, Mitaka, Tokyo 181-8588, Japan}
\affil{Department of Astronomy, School of Science, Graduate
University for Advanced Studies (SOKENDAI), Mitaka, Tokyo 181-8588, Japan}

\author[0000-0001-9452-0813]{Takuma Izumi}
\affil{National Astronomical Observatory of Japan, National Institutes 
of Natural Sciences (NINS), 2-21-1 Osawa, Mitaka, Tokyo 181-8588, Japan}
\affil{Department of Astronomy, School of Science, Graduate
University for Advanced Studies (SOKENDAI), Mitaka, Tokyo 181-8588, Japan}
\affil{Department of Physics, Graduate School of Science, Tokyo
Metropolitan University, 1-1 Minami-Osawa, Hachioji-shi, Tokyo
192-0397, Japan} 




\begin{abstract}

We report the results of ALMA 1--2 kpc-resolution, three 
rotational transition line (J=2--1, J=3--2, and J=4--3) observations of
multiple dense molecular gas tracers (HCN, HCO$^{+}$, and HNC) for
ten nearby (ultra)luminous infrared galaxies ([U]LIRGs).
Following the matching of beam sizes to 1--2 kpc for each (U)LIRG,
the high-J to low-J transition-line flux ratios of each molecule and
the emission line flux ratios of different molecules at each J
transition are derived. 
We conduct RADEX non-LTE model calculations and find that, under a wide
range of gas density and kinetic temperature, the observed
HCN-to-HCO$^{+}$ flux ratios in the overall (U)LIRGs are naturally
reproduced with enhanced HCN abundance compared to HCO$^{+}$. 
Thereafter, molecular gas properties are constrained
primarily through the use of HCN and HCO$^{+}$ data and the adoption
of fiducial values for the HCO$^{+}$ column density and
HCN-to-HCO$^{+}$ abundance ratio.  
We quantitatively confirm the following: 
(1) Molecular gas at the (U)LIRGs' nuclei is dense 
($\gtrsim$10$^{3-4}$ cm$^{-3}$) and warm ($\gtrsim$100 K). 
(2) Molecular gas density and temperature in nine ULIRGs'
nuclei are significantly higher than that of one LIRG's nucleus. 
(3) Molecular gas in starburst-dominated sources tends to be 
less dense and cooler than ULIRGs with luminous AGN signatures.
For six selected sources, we also apply a Bayesian approach by
freeing all parameters and support the above main results.
Our ALMA 1--2 kpc resolution, multiple transition-line data of
multiple molecules are a very powerful tool for scrutinizing
the properties of molecular gas concentrated around luminous
energy sources in nearby (U)LIRGs' nuclei.

\end{abstract}

%\keywords{galaxies: active --- galaxies: nuclei --- quasars: general ---
%galaxies: Seyfert --- galaxies: starburst --- submillimeter: galaxies}



\section{Introduction} \label{sec:intro}

Galaxies whose infrared (8--1000 $\mu$m) luminosities are 
L$_{\rm IR}$ $=$ 10$^{12-13}$L$_{\odot}$ and 
L$_{\rm IR}$ $=$ 10$^{11-12}$L$_{\odot}$ are referred to as 
ultraluminous infrared galaxies (ULIRGs) and luminous infrared
galaxies (LIRGs), respectively \citep{sam96}.
Most (U)LIRGs exhibit strong infrared dust thermal radiation which is much
brighter than the UV-optical emission, and are found in major mergers
of gas-rich galaxies in the nearby universe at $z <$ 0.3
\citep[e.g.,][]{sam96}.  
Starburst activity and/or luminous active galactic nucleus (AGN)
activity powered by a mass-accreting supermassive black hole (SMBH) at
nuclear regions hidden behind the curtain of obscuring gas and dust,  
are considered the primary energy sources for the observed large
infrared luminosities of nearby (U)LIRGs. 
However, estimating starburst and AGN contribution to the
large luminosities is challenging because of very large extinction 
in nearby (U)LIRGs' nuclei. 

Numerical simulations have predicted that molecular gas is 
transferred to nuclear regions through merger-induced processes 
\citep[e.g.,][]{jun09,mor19} and a large amount of dense
($\gtrsim$10$^{3-4}$ cm$^{-3}$) molecular gas is observed in nearby merging 
(U)LIRGs' nuclei \citep[e.g.,][]{gao04,jun09,ima19,ued21,ima22}.
Such dense molecular gas can (1) form stars and become an
important fuel toward a central SMBH (i.e., enhancing AGN activity)
and (2) be affected by the hidden energy sources of (U)LIRGs, both
radiatively and mechanically. 
Scrutinizing the physical and chemical properties of dense molecular
gas at the (U)LIRGs' nuclei is vital to better understand the nature
of these regions.
Rotational J-transition emission lines of dense molecular gas tracers 
are found in the (sub)millimeter wavelength range at 0.3--3.5 mm,
where extinction effects are very small.
(Sub)millimeter molecular rotational J-transition lines can thus be a 
very useful tool for the scrutinization.

Spectral line energy distribution (SLED) of CO at (sub)millimeter 
and far-infrared (30--300 $\mu$m), based on large-beam-sized 
observations with the Herschel Space Observatory and/or ground-based 
single-dish telescopes ($\gtrsim$5$''$ or $\gtrsim$5 kpc at $z
\gtrsim$ 0.05), has often been used to identify the presence of
luminous AGN activity, in addition to starburst activity, in nearby
(U)LIRGs. 
This is because high-J (J $\gtrsim$ 4--5) CO emission can be stronger in 
an AGN than in a starburst when normalized to low-J (J = 1--2) CO
luminosity \citep[e.g.,][]{hai12,per14,mas15,ros15,lu17}. 
This can be attributed to the fact that dense and warm ($\gtrsim$100
K) molecular gas in the vicinity of a luminous AGN can excite CO to
high-J levels more efficiently than in normal star-forming regions
\citep[e.g.,][]{mei07,spa08}.  
SLED of molecules with higher dipole moments (e.g., HCN and HCO$^{+}$)
has also been used to probe the physical conditions of dense molecular
gas in nearby (U)LIRGs. However, this has mostly been conducted with 
large-beam-sized ($\gtrsim$5$''$) observations using single dish
(sub)millimeter telescopes \citep[e.g.,][]{kri08,gre09,pap14,isr22}. 
Dense molecular gas properties of energetically dominant compact 
($\lesssim$1--2 kpc) nuclear regions of nearby ULIRGs
\citep[e.g.,][]{soi00,dia10,ima11,per21} may not be best probed using
these low-angular-resolution data, because emission from spatially
extended ($\gtrsim$a few kpc) host galaxies can be a severe contaminant. 
Probing only nuclear dense molecular gas through
higher-spatial-resolution ($\lesssim$1--2 kpc) observations is highly
desirable.  

ALMA can facilitate high-angular-resolution ($\lesssim$1$''$) and 
high-sensitivity observations of HCN, HCO$^{+}$ and HNC (bright dense
molecular gas tracers; \citet{ler17}) at multiple J transitions. 
Results of $\lesssim$1$''$-resolution HCN, HCO$^{+}$, and HNC
observations of nearby (U)LIRGs' nuclei at J=1--0, J=2--1, J=3--2, and
J=4--3 have been reported 
\citep[e.g.,][]{ima13a,ima13b,ima14,sco15,aal15a,aal15b,ima16b,mar16,man17,sli17,ima18,sai18,ima19,ima21,sak21,ima22}. 
However, no study has conducted detailed investigations on the
physical conditions of nuclear dense molecular gas based on the
combination of multiple J-transition line data and non-LTE modeling, 
except for the nearest ULIRG Arp 220 at $z =$ 0.018
\citep[e.g.,][]{tun15,sli17,man17}. 
Now that three J-transition line data of HCN, HCO$^{+}$, and HNC
are in hand for multiple nearby (U)LIRGs, the time is ripe to do such
investigations. 

In this paper, we combine available J=2--1, J=3--2, and J=4--3 data 
%---
\footnote{
Redshifted J=1--0 lines for HCN, HCO$^{+}$, and HNC were not observable
using ALMA before 2022, in case a source is located at $z \gtrsim$ 0.06.}
%---
of HCN, HCO$^{+}$, and HNC previously obtained from our ALMA observations 
\citep{ima13a,ima13b,ima14,ima16a,ima16b,ima18,ima21,ima22}.
After matching the beam sizes of all the J-transition line data of all
molecules to the same value for each object (1--2 kpc), we perform the
aforementioned investigations.  
Throughout this study, we adopt the cosmological parameters H$_{0}$
$=$ 71 km s$^{-1}$ Mpc$^{-1}$, $\Omega_{\rm M}$ = 0.27, and
$\Omega_{\rm \Lambda}$ = 0.73. 
In addition, molecular line flux ratios are calculated in Jy km s$^{-1}$. 
Density and temperature mean H$_{2}$ volume number
density (n$_{\rm H_2}$) and kinetic temperature (T$_{\rm kin}$), respectively,
unless otherwise stated.

\section{Targets} 

The targets of this study are nine ULIRGs and one LIRG, previously
observed at J=2--1, J=3--2, and J=4--3 of HCN, HCO$^{+}$, and HNC, with 
$\lesssim$1--2 kpc resolution, through our ALMA programs  
\citep{ima13a,ima13b,ima14,ima16a,ima16b,ima18,ima21,ima22}. 
Table \ref{tab:objects} summarizes the basic properties of the (U)LIRGs.
We selected nearby (U)LIRGs with different levels of AGN's
energetic contributions to bolometric luminosity, estimated using 
optical, infrared, hard X-ray ($>$10 keV), and (sub)millimeter
spectroscopic energy diagnostic methods. 
Two sources (NGC 1614 and IRAS 13509$+$0442) and a secondary 
fainter nucleus of IRAS 12112$+$0305 (SW) are regarded as
starburst-dominated, without discernible luminous AGN signatures, whereas
the remaining ULIRGs' nuclei are diagnosed as containing luminous AGNs.
More detailed explanations of the individual (U)LIRGs can be
found in \citet{ima16b}. 
Although the observed ten (U)LIRGs are neither statistically complete
nor unbiased, their molecular line studies will provide important
information of (1) the general properties of nuclear dense molecular gas
in nearby (U)LIRGs at $\lesssim$1--2 kpc resolution and 
(2) possible difference of molecular gas properties between (U)LIRGs'
nuclei with and without luminous AGN signatures.  
 
%%%%%%%%%% Table 1 (objects) %%%%%%%%%
\begin{deluxetable*}{lcccrrrrcccc}[bht!]
\tabletypesize{\scriptsize}
%\rotate
\tablecaption{Basic Properties of Nine Ultraluminous and One Luminous 
Infrared Galaxies \label{tab:objects}} 
\tablewidth{0pt}
\tablehead{
\colhead{Object} & \colhead{Redshift} & 
\colhead{d$_{\rm L}$} & \colhead{Scale} & 
\colhead{f$_{\rm 12}$} & 
\colhead{f$_{\rm 25}$} & 
\colhead{f$_{\rm 60}$} & 
\colhead{f$_{\rm 100}$} & 
\colhead{log L$_{\rm IR}$} & 
\colhead{Optical} & 
\colhead{AGN}  &
\colhead{IR/submm/X} 
\\
\colhead{} & \colhead{} & \colhead{[Mpc]} & \colhead{[kpc/$''$]}  
& \colhead{[Jy]}
& \colhead{[Jy]} & \colhead{[Jy]} & \colhead{[Jy]}  &
\colhead{[L$_{\odot}$]} & \colhead{Class} & \colhead{IR [\%]} &
\colhead{AGN} \\   
\colhead{(1)} & \colhead{(2)} & \colhead{(3)} & \colhead{(4)} & 
\colhead{(5)} & \colhead{(6)} & \colhead{(7)} & \colhead{(8)} & 
\colhead{(9)} & \colhead{(10)} & \colhead{(11)} & \colhead{(12)} 
}
\startdata
NGC 1614\tablenotemark{A} & 0.0160 & 68 & 0.32 & 1.38 & 7.50 & 32.12 & 34.32 
& 11.7 & HII$^{a,b}$ (Cp$^{c}$) & $<$5 & \\ 
IRAS 06035$-$7102 & 0.0795 & 356 & 1.5 & 0.12 & 0.57 & 5.13
& 5.65 & 12.2 & LI$^{d}$ & 22$^{+3}_{-10}$ & Y$^{h,i,j,k}$ \\
IRAS 08572$+$3915 & 0.0580 & 256 & 1.1 & 0.32 & 1.70 & 7.43  & 4.59  & 12.1 &
LI$^{e}$(Sy2$^{c}$) & 86$^{+2}_{-3}$ & Y$^{l,m,n,o,p,q,r}$ \\   
IRAS 12112$+$0305 & 0.0730 & 326 & 1.4 & 0.12 & 0.51 & 8.50 & 9.98 & 12.3 & 
LI$^{e}$ (Sy2$^{c}$) & $<$0.7 & Y$^{s,t}$ (NE nucleus) \\     
IRAS 12127$-$1412 & 0.1332 & 620 & 2.3 & $<$0.13 & 0.24 &
1.54 & 1.13 & 12.2 & LI$^{e}$ (HII$^{c}$) & 88$\pm$3 & Y$^{h,n,p,r}$ \\
IRAS 13509$+$0442 & 0.1364 & 636 & 2.4 & 0.10 & $<$0.23 &
1.56 & 2.53 & 12.3 & HII$^{e}$ (Cp$^{c}$)  & $<$0.03 & \\
IRAS 15250$+$3609 & 0.0552 & 243 & 1.1 & 0.16 & 1.31 & 7.10
& 5.93 & 12.0 & LI$^{a}$ (Cp$^{c}$) & 51$^{+4}_{-5}$ & Y$^{h,i,r,u}$ \\
Superantennae\tablenotemark{B} & 0.0617 & 273 & 1.2 & 0.22 & 1.24 & 5.48 &
5.79 & 12.1 & Sy2$^{b,d,f,g}$ & 24$\pm$4 & Y$^{v,w,x,y}$ \\    
IRAS 20551$-$4250 & 0.0430 & 188 & 0.84 & 0.28 & 1.91 & 12.78 & 9.95 & 12.0 
& LI or HII$^{d}$ (Cp$^{c}$) & 26$\pm$3 & Y$^{h,z,aa}$ \\
IRAS 22491$-$1808 & 0.0776 & 347 & 1.5 & 0.05 & 0.55 & 5.44
& 4.45 & 12.2 & HII$^{c,e}$ & $<$0.07 & Y$^{t}$ \\  
\enddata

\tablenotetext{A}{Also known as IRAS 04315$-$0840. This is a LIRG.}

\tablenotetext{B}{Also known as IRAS 19254$-$7245.}

\tablecomments{
Col.(1): Object name. 
Col.(2): Redshift adopted from ALMA dense molecular line data 
\citep{ima16b}. 
Col.(3): Luminosity distance in Mpc. 
Col.(4): Physical scale in kpc arcsec$^{-1}$. 
Col.(5)--(8): f$_{12}$, f$_{25}$, f$_{60}$, and f$_{100}$ are 
{\it IRAS} fluxes at 12 $\mu$m, 25 $\mu$m, 60 $\mu$m, and 100 $\mu$m,
obtained from \citet{kim98}, \citet{san03}, or the IRAS
Faint Source Catalog (FSC).
Col.(9): Decimal logarithm of infrared (8$-$1000 $\mu$m) luminosity
in units of solar luminosity (L$_{\odot}$), calculated using
$L_{\rm IR} = 2.1 \times 10^{39} \times$ D(Mpc)$^{2}$
$\times$ (13.48 $\times$ $f_{12}$ + 5.16 $\times$ $f_{25}$ +
$2.58 \times f_{60} + f_{100}$) ergs s$^{-1}$ \citep{sam96}.
Only NGC 1614 is a LIRG.
Col.(10) Optical spectroscopic classification. 
``Sy2'', ``LI'', ``HII'', and ``Cp'' mean Seyfert 2, LINER, HII-region, 
and starburst$+$AGN composite, respectively.
$^{a}$: \citet{vei95}.
$^{b}$: \citet{kew01}.
$^{c}$: \citet{yua10}.
$^{d}$: \citet{duc97}.
$^{e}$: \citet{vei99}.
$^{f}$: \citet{mir91}.
$^{g}$: \citet{col91}.
Col.(11): Infrared spectroscopically estimated bolometric contribution 
of AGNs in \% by \citet{nar10} for all ULIRGs and
by \citet{per15} for the LIRG NGC 1614.
Col.(12): Presence (``Y'') of signatures of luminous obscured AGNs, 
including optically elusive buried ones, in the infrared 3--40 $\mu$m
and/or hard X-ray ($>$10 keV) and/or (sub)millimeter spectra, 
in other representative references:
For IRAS 12112$+$0305 NE and IRAS 22491$-$1808, certain AGN signatures are
found only in the (sub)millimeter range, where the extinction effects are
significantly smaller than those in the infrared and X-rays \citep{hil83}.
$^{h}$: \citet{ima10}.
$^{i}$: \citet{spo02}.
$^{j}$: \citet{dar07}.
$^{k}$: \citet{far09}.
$^{l}$: \citet{dud97}.
$^{m}$: \citet{imd00}.
$^{n}$: \citet{ima06}.
$^{o}$: \citet{spo06}.
$^{p}$: \citet{ima07}.
$^{q}$: \citet{arm07}.
$^{r}$: \citet{vei09}.
$^{s}$: \citet{ima16b}.
$^{t}$: \citet{ima18}.
$^{u}$: \citet{sti13}.
$^{v}$: \citet{ris03}.
$^{w}$: \citet{ima08}.
$^{x}$: \citet{bra09}.
$^{y}$: \citet{ima21}.
$^{z}$: \citet{ris06}.
$^{aa}$: \citet{san08}
}
\end{deluxetable*}
%%%%%%%%%%%%%%%%%%%%%%%%%%%%%%%%%%%

%Further details of these ten galaxies are described by \citep{ima21}.

\section{Observations and Data Analysis} 

The details of the majority of HCN, HCO$^{+}$, and HNC line
observations have been reported in previous publications 
\citep{ima13a,ima13b,ima14,ima16a,ima16b,ima18,ima21,ima22}, 
and thus are not repeated here.
However, the J=4--3 data for HCN, HCO$^{+}$, and HNC for IRAS
06035$-$7102, and HNC J=4--3 data for IRAS 08572$+$3915 were not shown
in the previous publications.
The log of these observations in ALMA Cycle 5 is summarized in
Appendix A.  

Because different J-transition line data were obtained at different
times, with different beam sizes (different array configurations),
first, the beam sizes of the multiple J-transition molecular line data
were matched with the same value for each object using the CASA
task ``imsmooth'' \citep{CASA}. 
We adopt 1 kpc if all data were obtained with beam sizes smaller than
1 kpc. 
This is because (1) the large infrared luminosity of nearby ULIRGs is 
usually dominated by compact ($\lesssim$1--2 kpc) nuclear regions 
\citep[e.g.,][]{soi00,dia10,ima11,per21} and 
(2) we aim to provide the clearest view of the nuclear dense
molecular gas properties of nearby (U)LIRGs by minimizing possible
contamination from spatially extended ($\gtrsim$a few kpc)
star-forming regions in the host galaxies. 
For certain fractions of ULIRGs, certain data exhibit beam sizes
larger than 1 kpc.  
For these cases, the smallest possible beam size (1.5, 1.6, or 2 kpc)
is adopted. 
Table \ref{tab:beam} summarizes the beam sizes adopted for individual
objects.  
For IRAS 08572$+$3915, the beam size of J=4--3 data taken in Cycle 0
is significantly larger than the adopted value (Table \ref{tab:beam}). 
However, we assume that a dominant fraction of the observed J=4--3 
emission originates from the adopted beam, because of the following reasons.
(1) The J=4--3 emission is generally more spatially compact than the lower
J-transition lines owing to the higher critical density in the former
\citep{shi15}. 
(2) Changing originally small beam-sized data to
unnecessarily large beam sizes increases the noise in units of mJy
beam$^{-1}$, and thus the scatter of the spectral data becomes large.  
For J=3--2 data of certain molecular lines for IRAS 12127$-$1412, IRAS
13509$+$0442, and the Superantennae, and J=4--3 data for certain lines
for IRAS 12112$+$0305 and IRAS 22491$-$1808, the major axes of the
synthesized beams are slightly larger than the adopted values (Table
\ref{tab:beam}); however, it is also assumed that the bulk of the
observed J=3--2 and J=4--3 fluxes are emitted within the adopted beams. 

%%%%%%%%%% Table 2 (beam) %%%%%%%%%
\begin{deluxetable}{lllcccccc}[t]
\rotate
%\tabletypesize{\small}
\tabletypesize{\scriptsize}
\tablecaption{Summary of Beam Sizes \label{tab:beam}} 
\tablewidth{0pt}
\tablehead{
\colhead{Object} & \multicolumn{6}{c}{Beam size (arcsec $\times$ arcsec)} 
& \colhead{Adopted beam} \\
\colhead{} & \colhead{J21a} & \colhead{J21b} & \colhead{J32a} 
& \colhead{J32b} & \colhead{J43a} & \colhead{J43b} & \colhead{arcsec (kpc) } \\
\colhead{(1)} & \colhead{(2)} & \colhead{(3)} & \colhead{(4)} &
\colhead{(5)} & \colhead{(6)} & \colhead{(7)}  & \colhead{(8)} 
}
\startdata 
NGC 1614 & 0.55$\times$0.37 (Cy5) & 0.62$\times$0.35
(Cy5) & 1.1$\times$0.58 (Cy2) & 1.1$\times$0.39 (Cy2) & 1.5$\times$1.3
(Cy0) & 1.3$\times$1.3 (Cy0) & 3.1$''$ (1 kpc) \\ 
IRAS 06035$-$7102 & 1.1$\times$0.80 (Cy5) &
0.48$\times$0.34 (Cy5) & 1.1$\times$0.78 (Cy3) & 1.1$\times$0.74 (Cy3) 
& 0.14$\times$0.16 (Cy5) & 0.12$\times$0.09 (Cy5) & 1.1$''$ (1.6 kpc) \\
IRAS 08572$+$3915 & 0.71$\times$0.35 (Cy5) &
0.68$\times$0.33 (Cy5) & 0.89$\times$0.47 (Cy2) & 0.90$\times$0.51
(Cy2) & 1.8$\times$1.1 \tablenotemark{A} (Cy0) & 0.22$\times$0.13
(Cy5) & 0.9$''$ (1 kpc)\\ 
IRAS 12112$+$0305 & 0.87$\times$0.68 (Cy5) &
0.83$\times$0.57 (Cy5) & 0.73$\times$0.55 (Cy2) & 0.81$\times$0.56
(Cy2) & 1.4$\times$0.56 \tablenotemark{B} (Cy2) 
& 1.0$\times$0.65 (Cy2) & 1.1$''$ (1.5 kpc)\\
IRAS 12127$-$1412 & 0.29$\times$0.19 (Cy5) &
0.24$\times$0.16 (Cy5) & 0.94$\times$0.74 \tablenotemark{B} (Cy3) &
0.86$\times$0.74 (Cy3) & 0.60$\times$0.49 (Cy0) &  
0.61$\times$0.52 (Cy0) & 0.86$''$ (2 kpc) \\
IRAS 13509$+$0442 & 0.27$\times$0.19 (Cy5) &
0.29$\times$0.18 (Cy5) & 0.99$\times$0.81 \tablenotemark{B} (Cy3) &
0.92$\times$0.81 \tablenotemark{B} (Cy3) &  
0.70$\times$0.53 (Cy3) & 0.69$\times$0.60 (Cy3) & 0.84$''$ (2 kpc) \\
IRAS 15250$+$3609 & 1.1$\times$0.68 (Cy5) &
0.70$\times$0.39 (Cy5) & 1.2$\times$0.72 (Cy3) & 1.3$\times$0.74 (Cy3)
& 0.87$\times$0.51 (Cy3) &  1.1$\times$0.56 (Cy3) & 1.4$''$ (1.5 kpc) \\
Superantennae & 0.54$\times$0.36 (Cy5) &
0.49$\times$0.34 (Cy5) & 0.87$\times$0.61 \tablenotemark{B} (Cy2) &
1.0$\times$0.61 \tablenotemark{B} (Cy2) & 0.74$\times$0.41 (Cy2) &  
0.61$\times$0.50 (Cy2) & 0.85$''$ (1 kpc)\\
IRAS 20551$-$4250 & 1.1$\times$0.82 (Cy5) &
0.60$\times$0.41 (Cy5) & 0.50$\times$0.46 (Cy1) & 0.52$\times$0.47
(Cy1) & 0.55$\times$0.39 (Cy0) & 0.64$\times$0.40 (Cy0) & 1.2$''$ (1 kpc) \\
IRAS 22491$-$1808 & 0.50$\times$0.33 (Cy5) &
0.41$\times$0.34 (Cy5) & 0.92$\times$0.59 (Cy2) & 0.41$\times$0.36
(Cy2) & 0.56$\times$0.51 (Cy0) & 1.1$\times$0.49 \tablenotemark{B}
(Cy0) & 1.0$''$ (1.5 kpc) \\
\enddata

\tablenotetext{A}{
Data were obtained in Cycle 0 with a beam size larger than the adopted 
one. 
It is assumed that the bulk of the observed J=4--3 flux originates
from the nuclear region within the adopted beam (see $\S$3).
}

\tablenotetext{B}{
The major axis is slightly larger than that of the adopted beam.
We modify only the minor axis to the adopted value. 
We assume that the bulk of the observed J=3--2 and J=4--3 flux
originates from the nuclear region within the adopted beam 
(see $\S$3).  
}

\tablecomments{ 
Col.(1): Object name. 
Cols.(2)--(7): Beam size of the continuum data in arcsec $\times$
arcsec.  
Information on the ALMA observing Cycle when individual data were
taken, is shown in parentheses. 
Col.(2): J21a obtained from HCN and HCO$^{+}$ J=2--1 observations.
Col.(3): J21b obtained from HNC J=2--1 observations.
Col.(4): J32a obtained from HCN and HCO$^{+}$ J=3--2 observations.
Col.(5): J32b obtained from HNC J=3--2 observations.
Col.(6): J43a obtained from HCN and HCO$^{+}$ J=4--3 observations.
Col.(7): J43b obtained from HNC J=4--3 observations.
Col.(8) Adopted circular beam size in arcsec.
Value in kpc is shown in parentheses.
}

\end{deluxetable}
%%%%%%%%%%%%%%%%%%%%%%%%%%%%%%

We obtained 1--2 kpc beam spectra by changing the beam size of 
continuum-subtracted data cube from the original values shown in the
previous publications
\citep{ima13a,ima13b,ima14,ima16a,ima16b,ima18,ima21,ima22}, to 
the adopted ones (Table \ref{tab:beam}). 
However, we noticed that the HCN, HCO$^{+}$, and HNC J=4--3 emission
lines of IRAS 22491$-$1808 obtained in ALMA Cycle 0 were substantially
narrower than J=3--2 and J=2--1 data obtained in later ALMA Cycles. 
This is because (1) the molecular-line-derived redshift of z $=$
0.0776 \citep{ima14} was found to be significantly larger than the
optically derived redshift of z $=$ 0.076 \citep{kim98}, which had
been assumed at the time of the Cycle 0 observation setup, and 
(2) low-frequency (high-velocity) side of the emission tail was
located at the edge of the ALMA spectral windows, which resulted in
continuum oversubtraction owing to the improper inclusion of data
points with significant emission components. 
Subsequently, we reanalyzed the HCN, HCO$^{+}$, and HNC J=4--3 Cycle 0
data of IRAS 22491$-$1808 by redefining a constant continuum flux level.
We found that the revised profiles of the J=4--3 emission lines were
more consistent with those of the J=2--1 and J=3--2 data, and thus
adopt the revised results. 

\section{Results} 

Newly obtained spectra of the two ULIRGs, IRAS 06035$-$7102 and IRAS
08572$+$3915, which were never presented in our previous publications  
\citep{ima13a,ima13b,ima14,ima16a,ima16b,ima18,ima21,ima22}, are shown
in Appendix A. 
Figure \ref{fig:profile} displays the velocity profiles for J=2--1, 
J=3--2, and J=4--3 emission lines of HCN, HCO$^{+}$, and HNC for all
the observed (U)LIRGs.

%\clearpage

%%%%%%%%%% Figure 1 %%%%%%%%%
\begin{figure}
\includegraphics[angle=-90,scale=0.25]{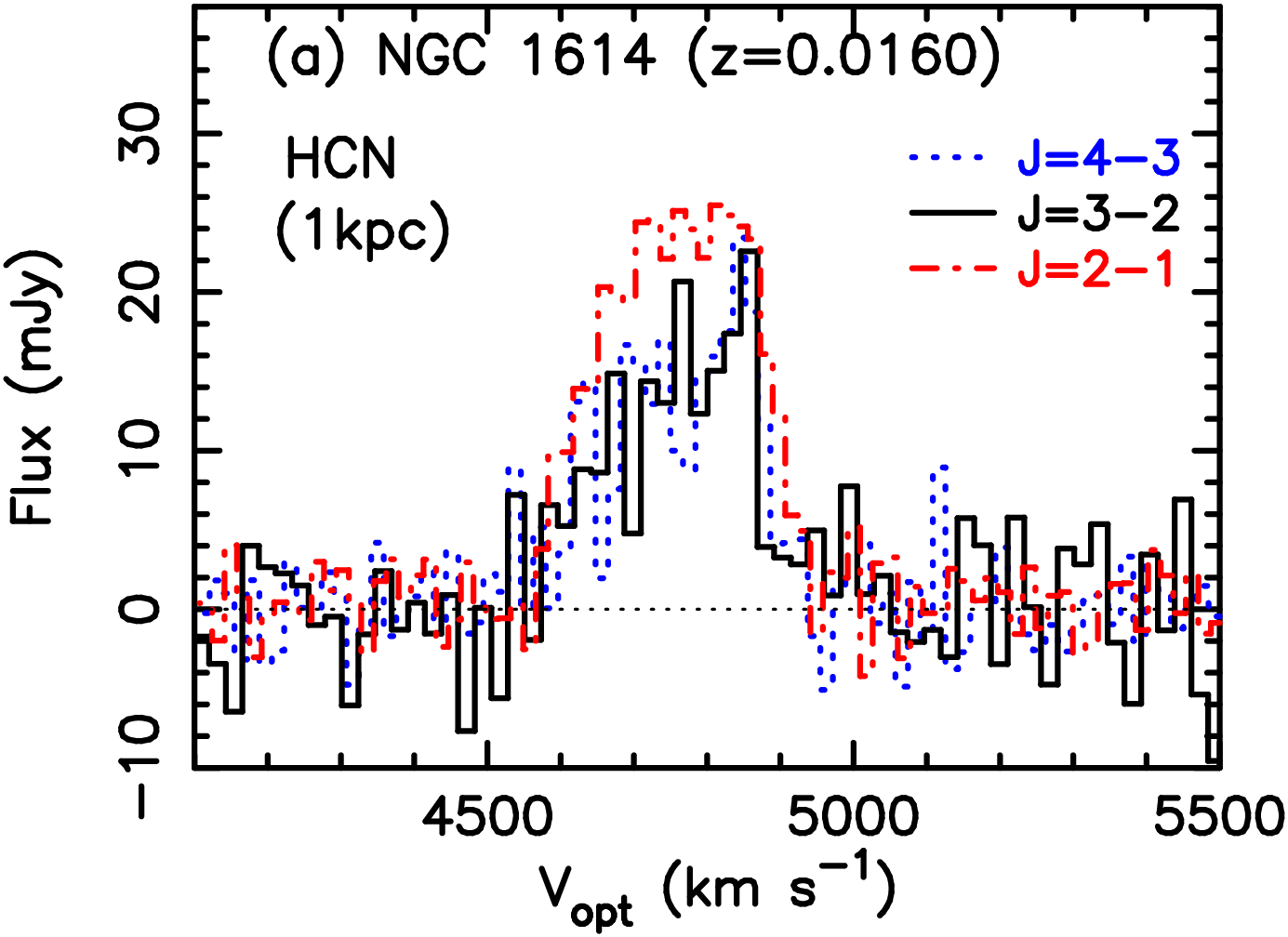} 
\includegraphics[angle=-90,scale=0.25]{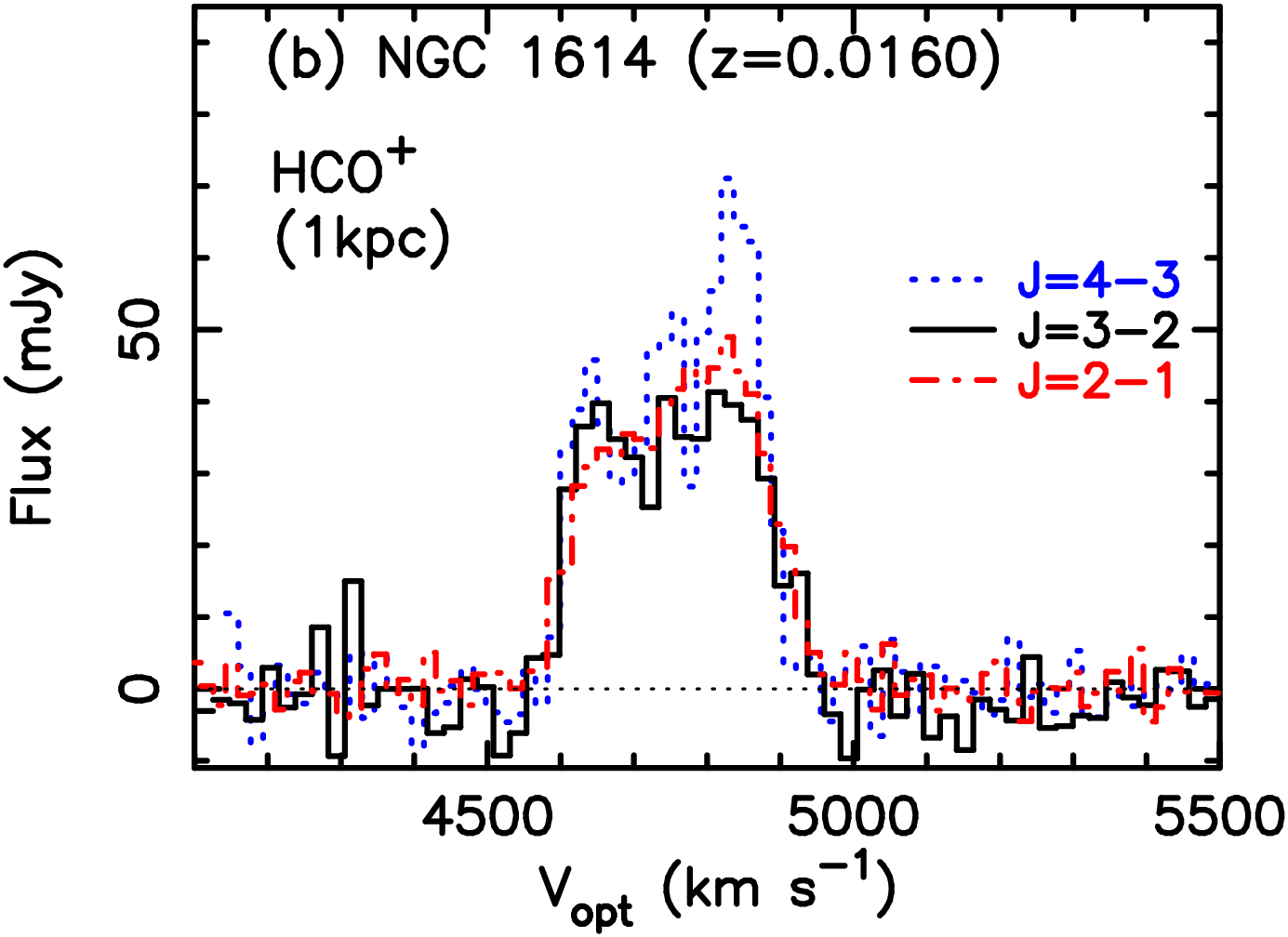} 
\includegraphics[angle=-90,scale=0.25]{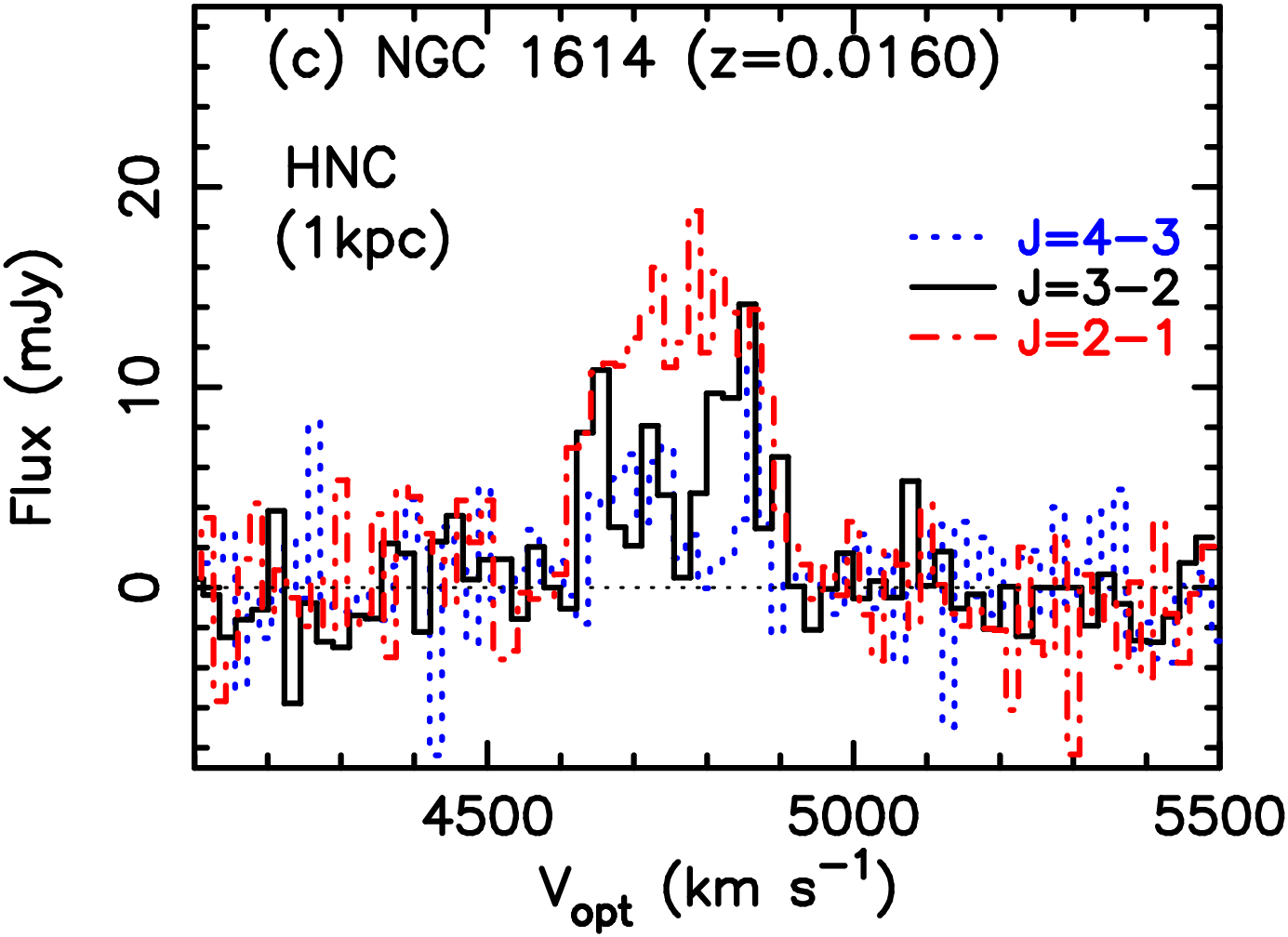} \\
\includegraphics[angle=-90,scale=0.25]{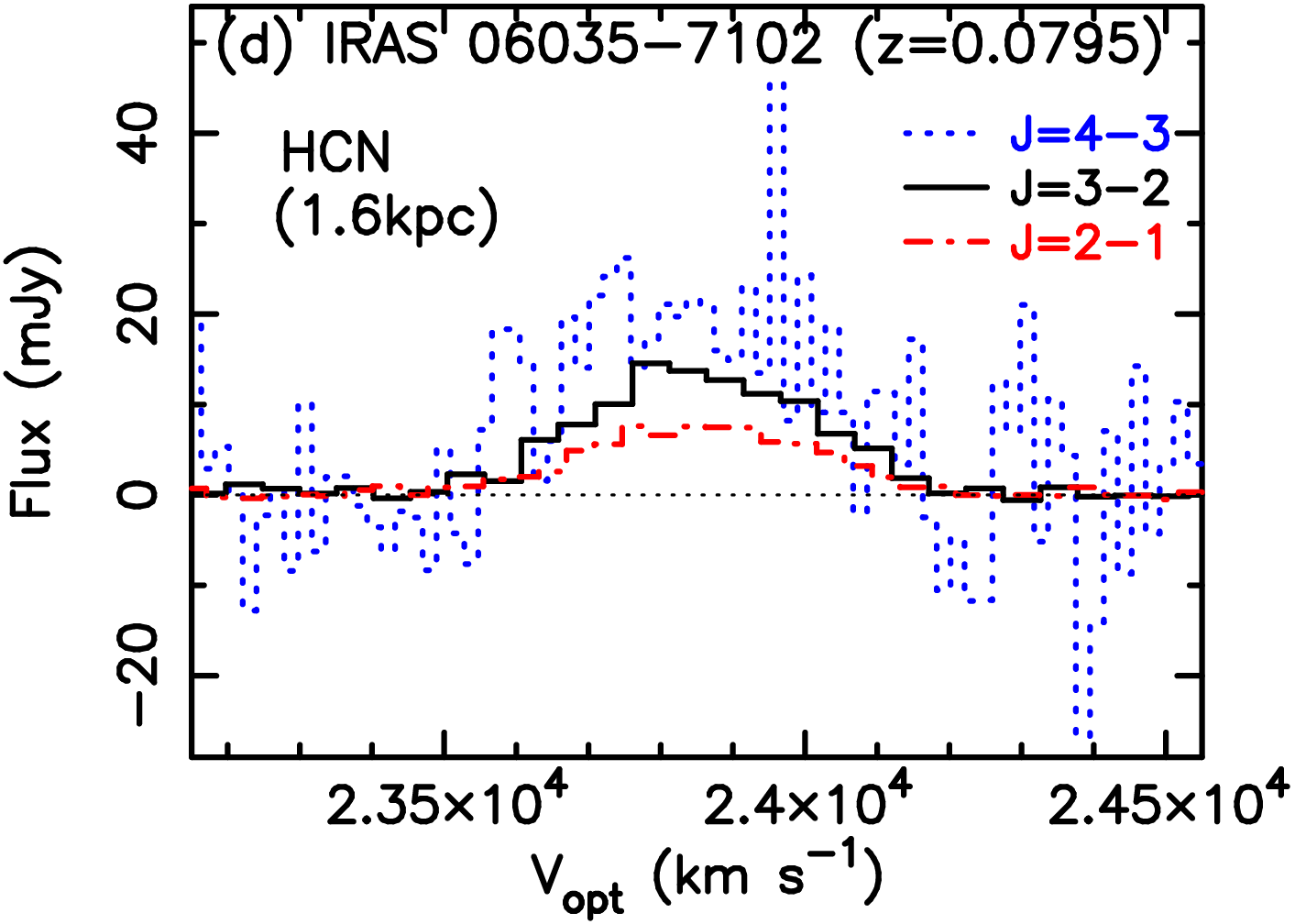} 
\includegraphics[angle=-90,scale=0.25]{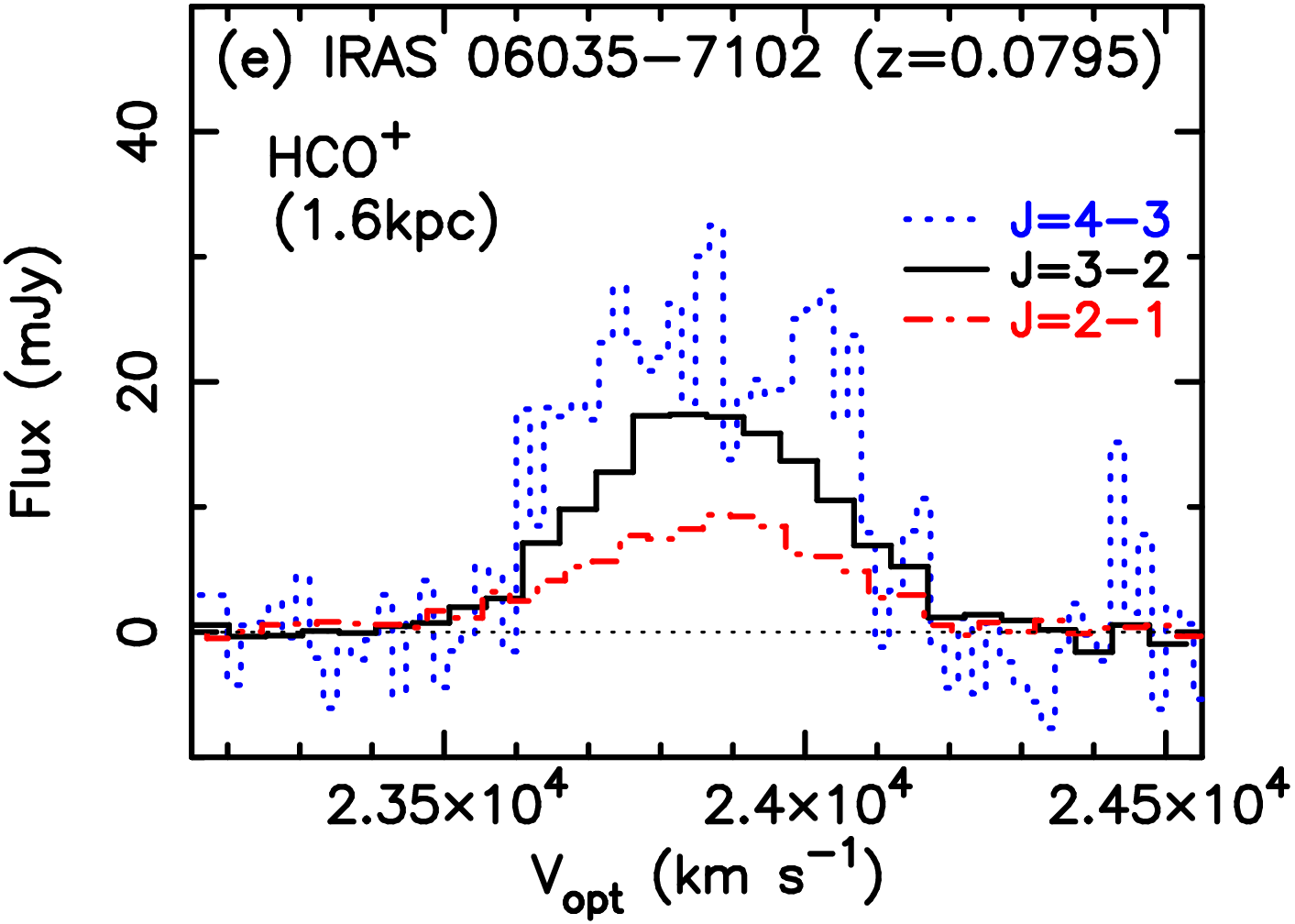} 
\includegraphics[angle=-90,scale=0.25]{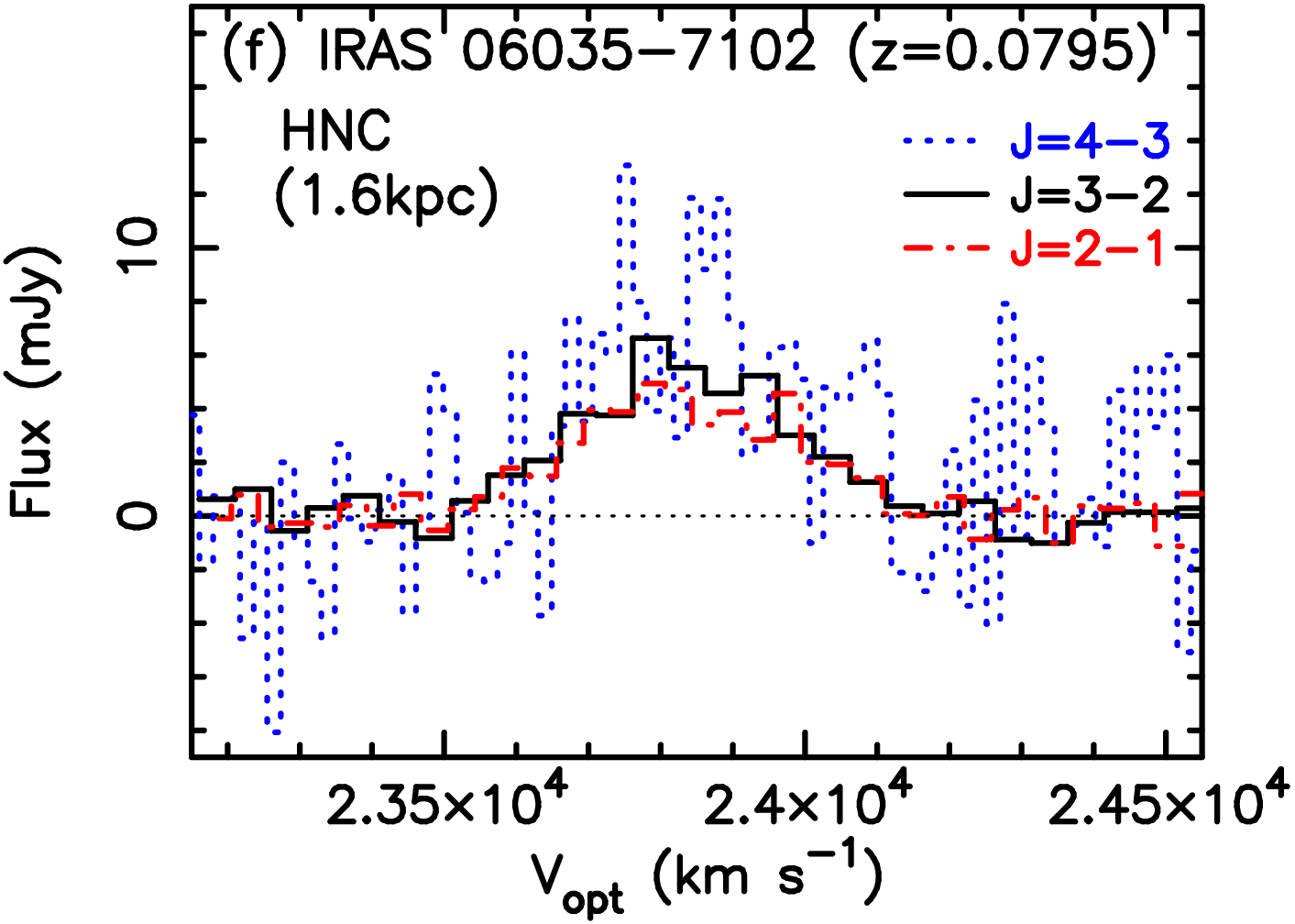} \\
\includegraphics[angle=-90,scale=0.25]{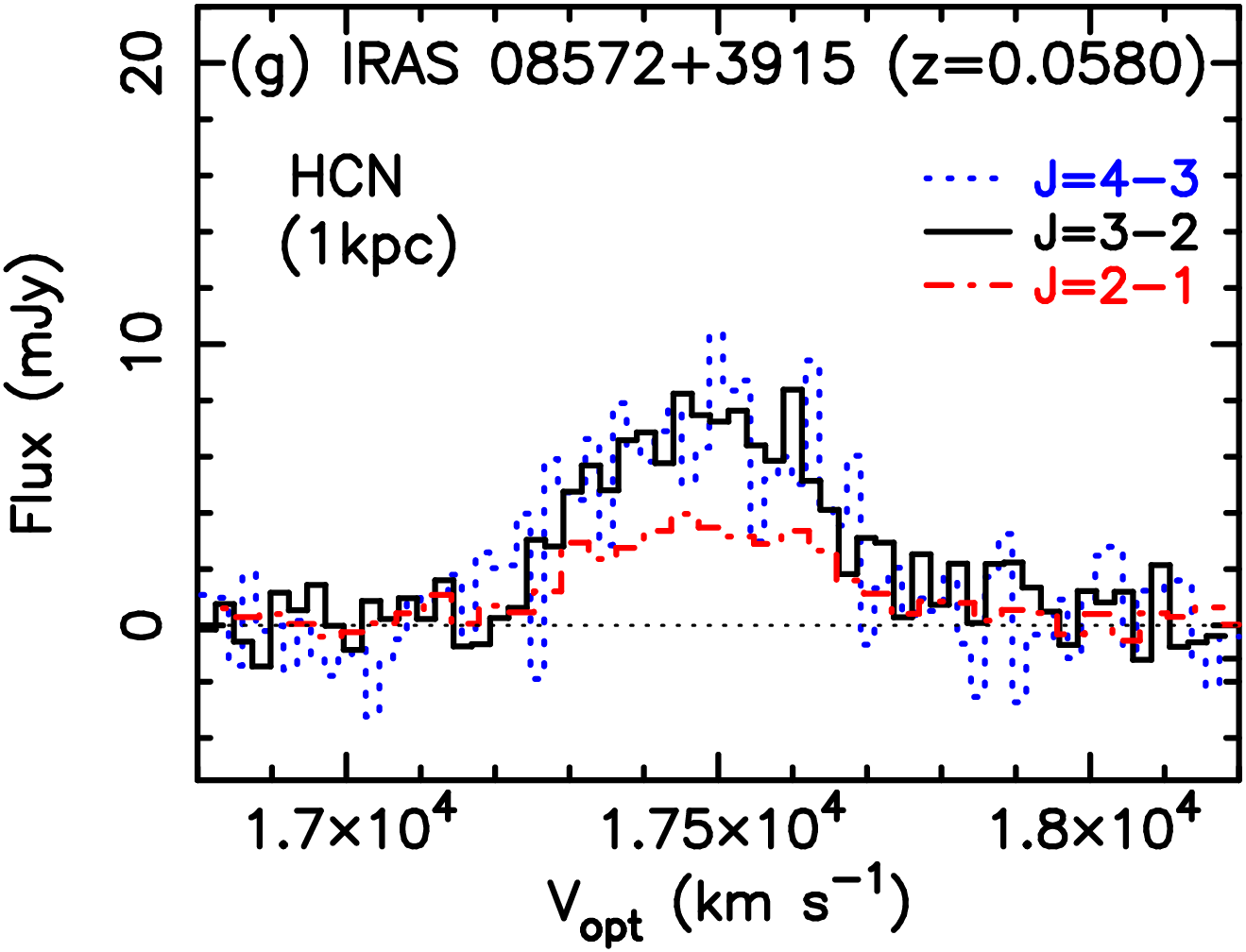} 
\hspace{0.2cm}
\includegraphics[angle=-90,scale=0.25]{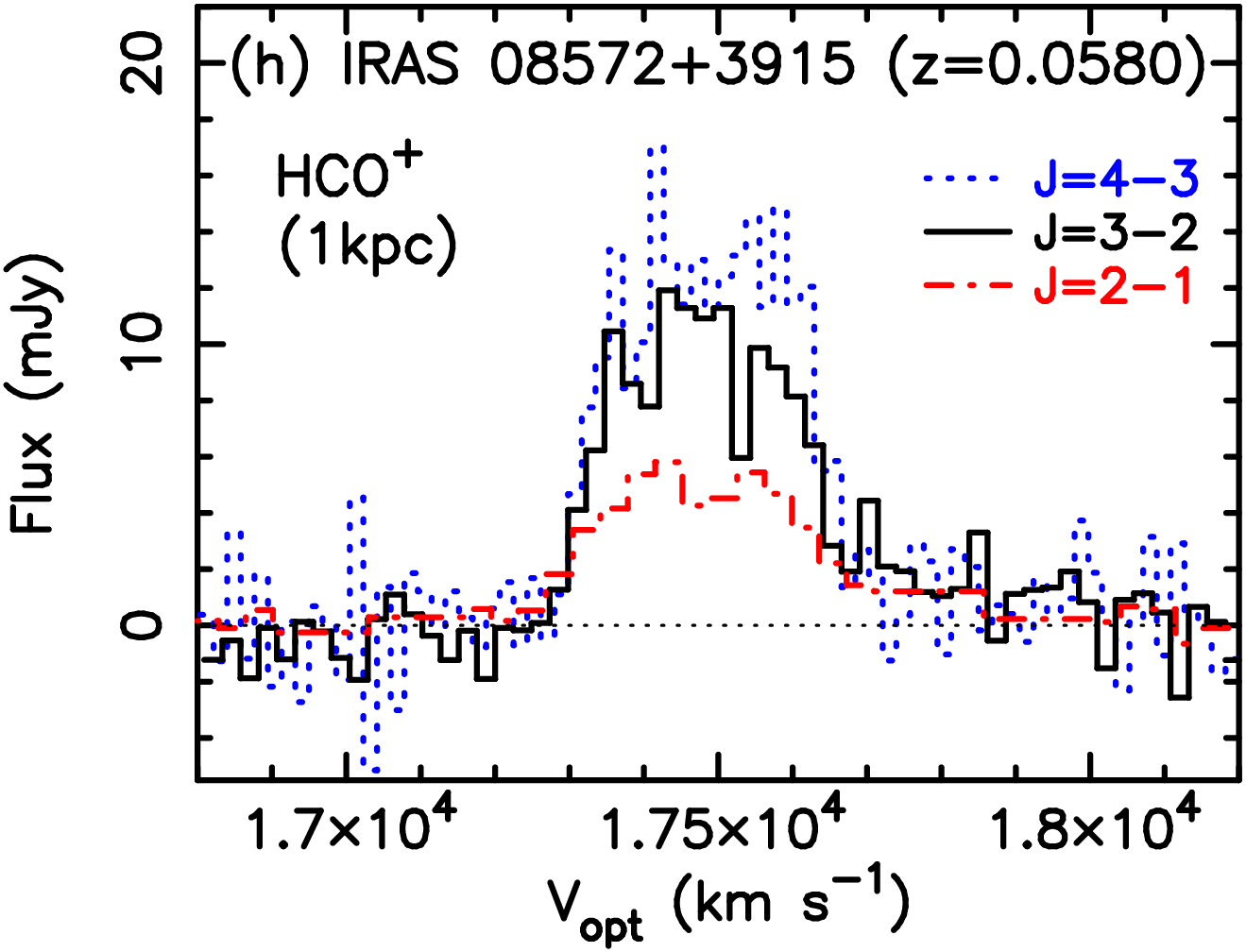} 
\hspace{0.2cm}
\includegraphics[angle=-90,scale=0.25]{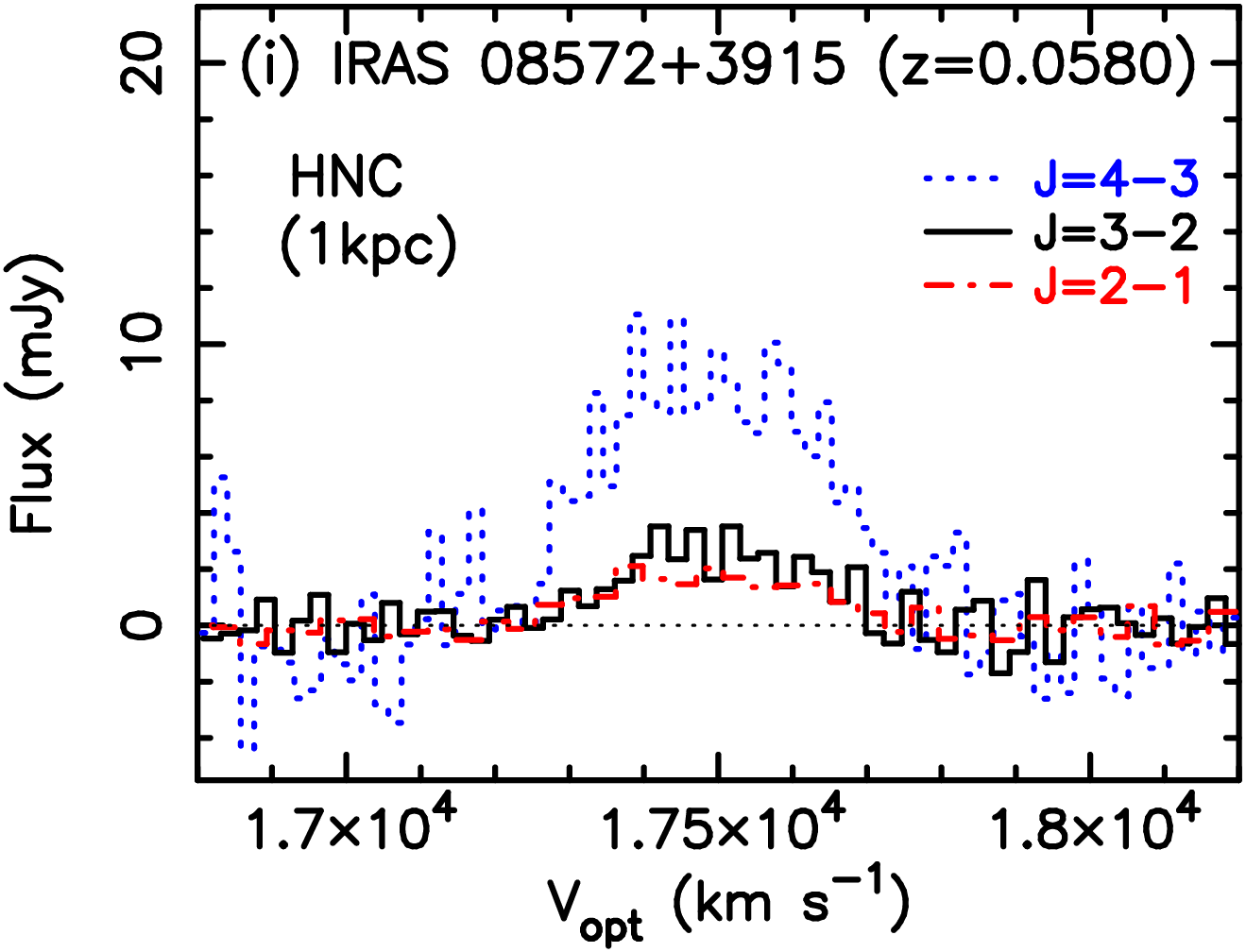} \\
\includegraphics[angle=-90,scale=0.25]{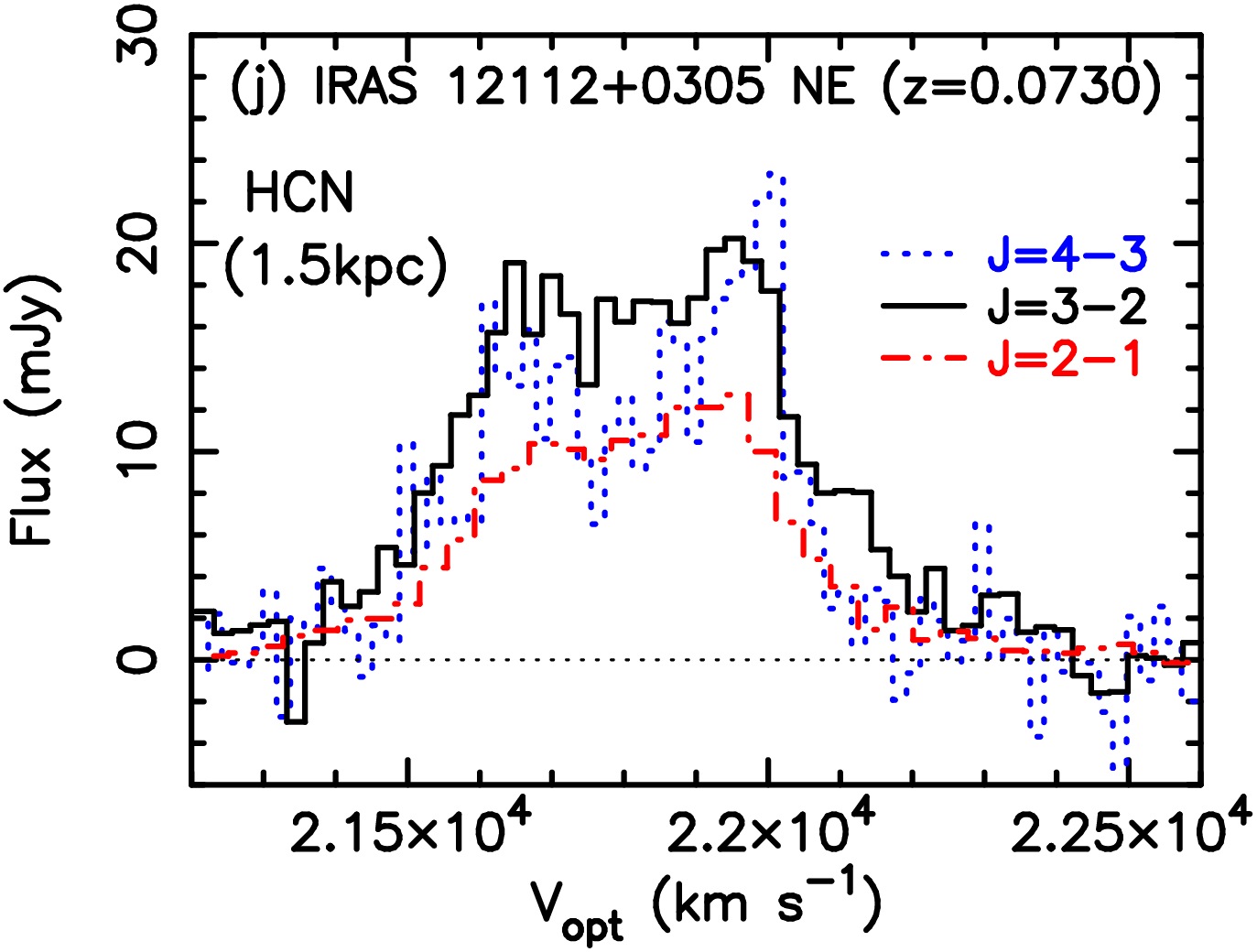} 
\includegraphics[angle=-90,scale=0.25]{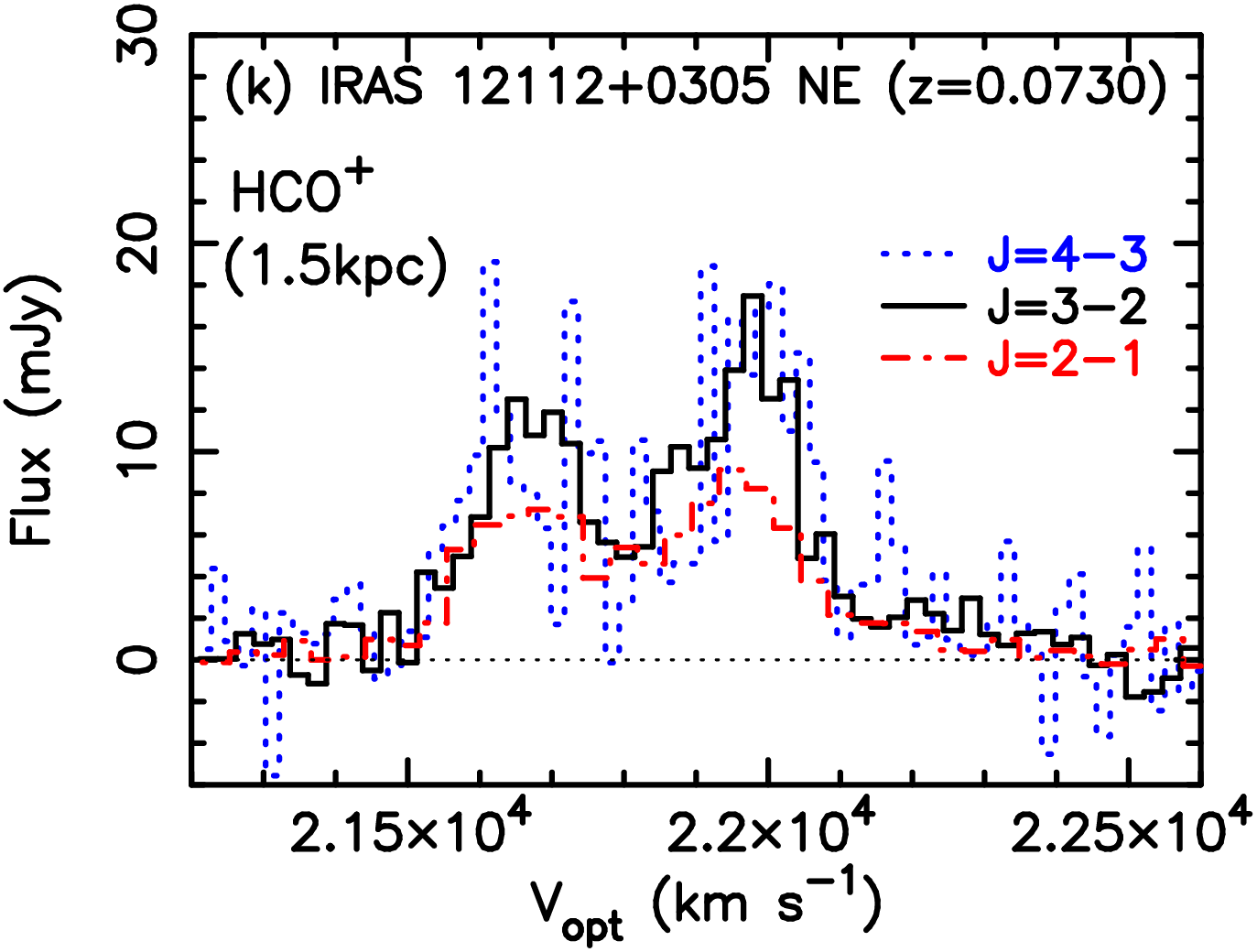} 
\includegraphics[angle=-90,scale=0.25]{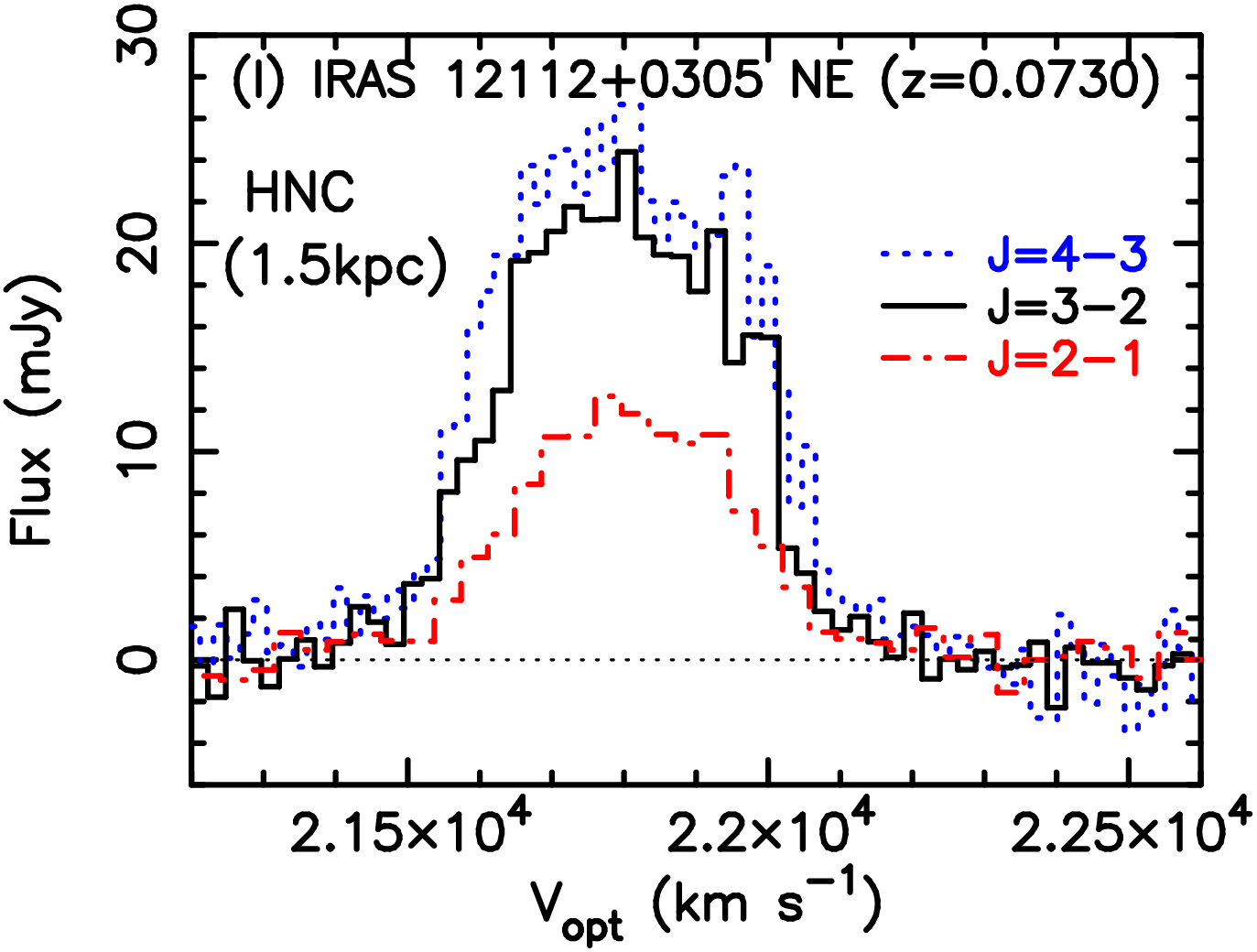} \\
\includegraphics[angle=-90,scale=0.25]{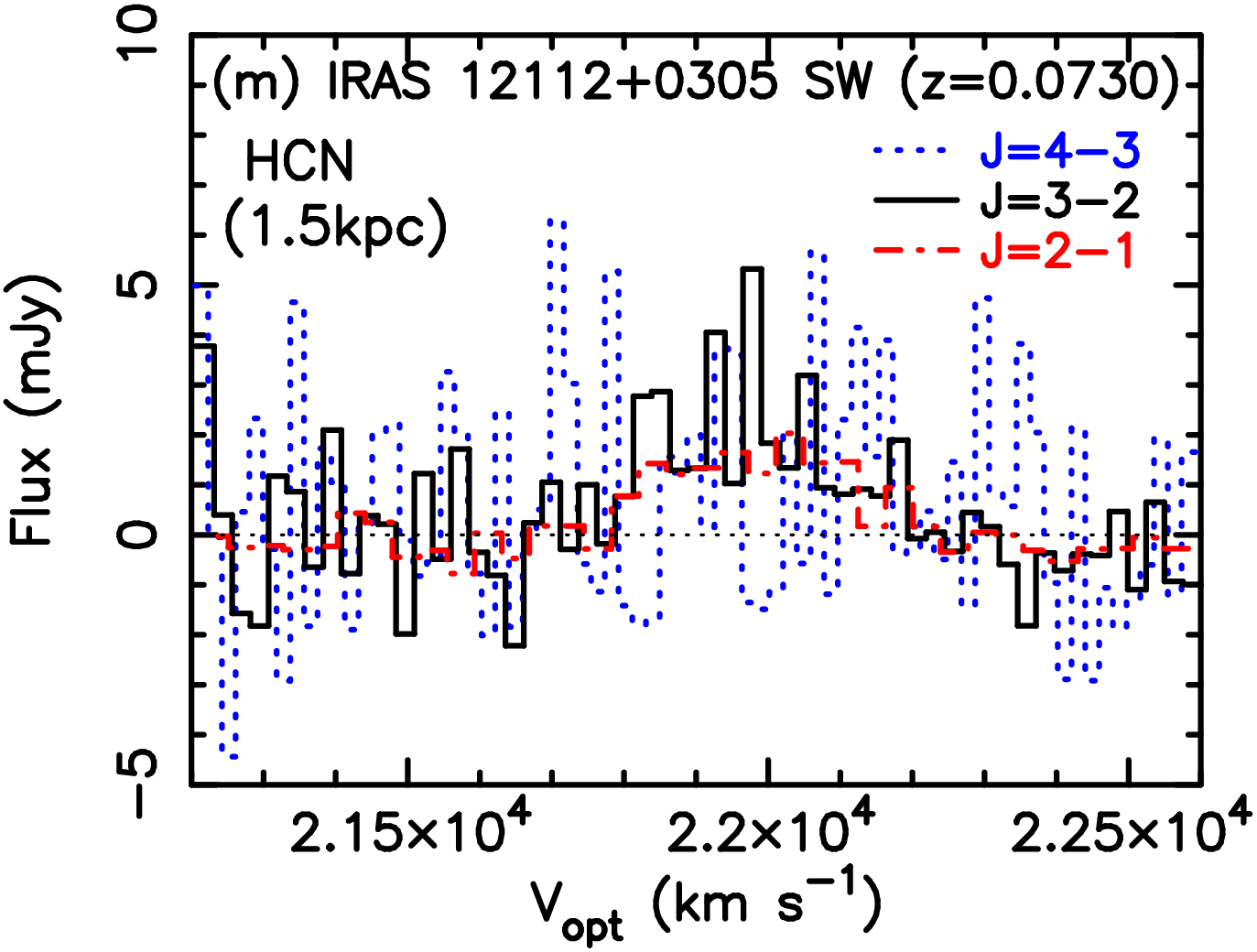} 
\includegraphics[angle=-90,scale=0.25]{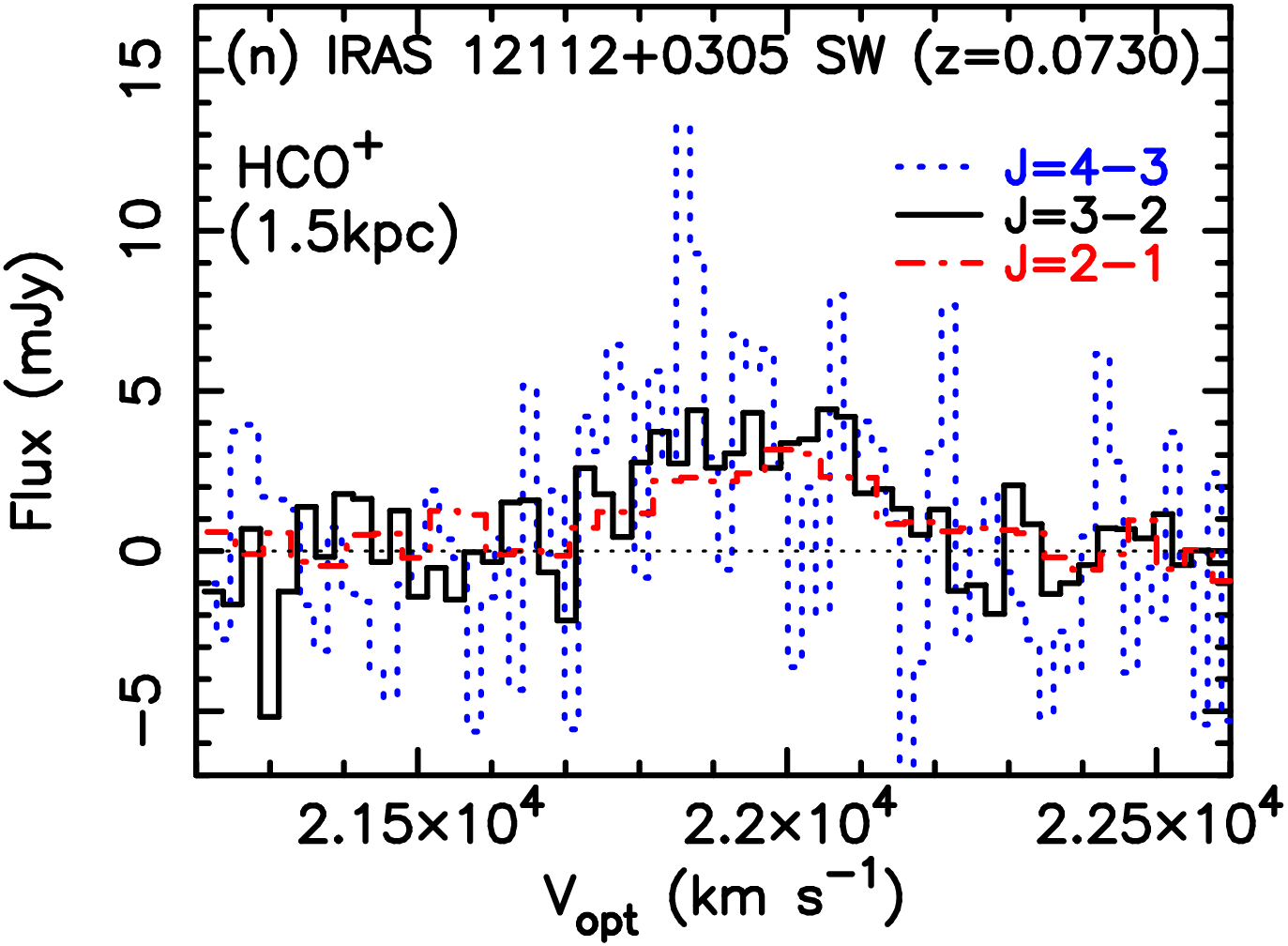} 
\includegraphics[angle=-90,scale=0.25]{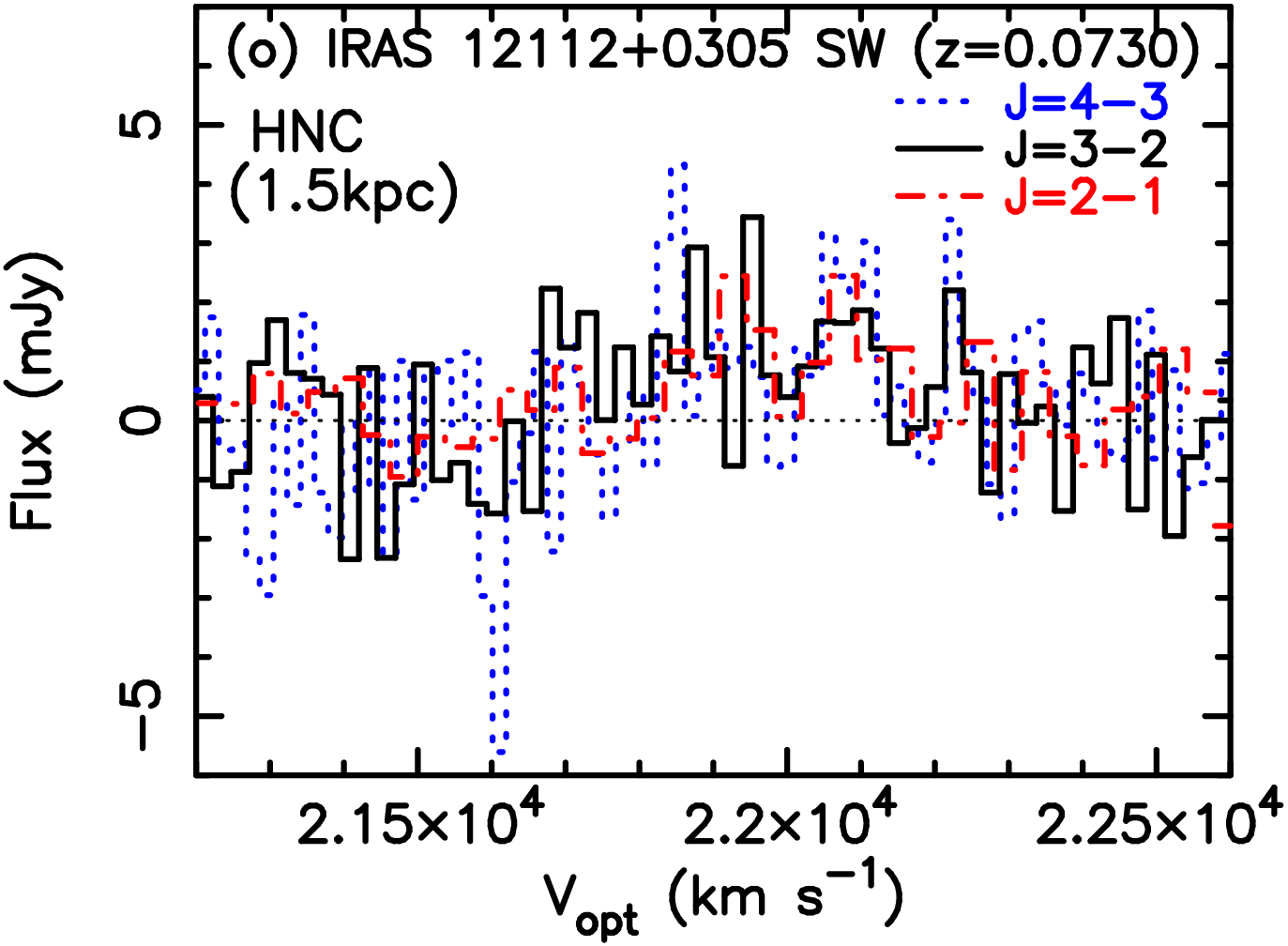} \\
\end{figure}

\clearpage

\begin{figure}
\includegraphics[angle=-90,scale=0.25]{f1p.eps} 
\hspace{0.02cm}
\includegraphics[angle=-90,scale=0.25]{f1q.eps} 
\hspace{0.02cm}
\includegraphics[angle=-90,scale=0.25]{f1r.eps} \\
\includegraphics[angle=-90,scale=0.25]{f1s.eps} 
\includegraphics[angle=-90,scale=0.25]{f1t.eps} 
\includegraphics[angle=-90,scale=0.25]{f1u.eps} \\
\includegraphics[angle=-90,scale=0.25]{f1v.eps} 
\hspace{0.07cm}
\includegraphics[angle=-90,scale=0.25]{f1w.eps} 
\hspace{0.07cm}
\includegraphics[angle=-90,scale=0.25]{f1x.eps} \\
\includegraphics[angle=-90,scale=0.25]{f1y.eps} 
\hspace{0.07cm}
\includegraphics[angle=-90,scale=0.25]{f1z.eps} 
\hspace{0.07cm}
\includegraphics[angle=-90,scale=0.25]{f1aa.eps} \\
\includegraphics[angle=-90,scale=0.25]{f1ab.eps} 
\includegraphics[angle=-90,scale=0.25]{f1ac.eps} 
\includegraphics[angle=-90,scale=0.25]{f1ad.eps} \\
\end{figure}

\clearpage

\begin{figure}
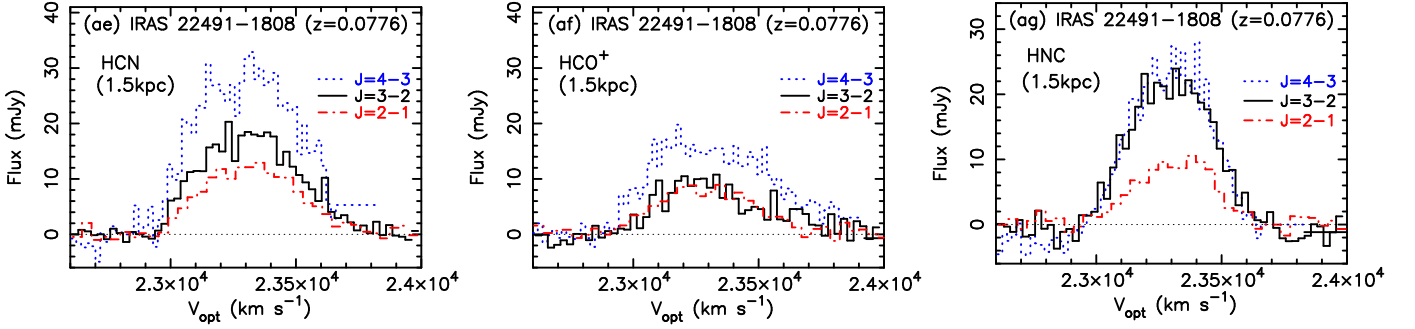

\includegraphics[angle=-90,scale=0.25]{f1ae.eps} 
\includegraphics[angle=-90,scale=0.25]{f1af.eps} 
\includegraphics[angle=-90,scale=0.25]{f1ag.eps} \\
\caption{
Velocity profiles of emission lines detected for matched beam sizes
(1--2 kpc). 
The abscissa represents the optical local standard of rest (LSR)
velocity in km s$^{-1}$. 
The ordinate represents the flux density in mJy. 
{\it (Left)}: HCN. {\it (Middle)}: HCO$^{+}$. {\it (Right)}: HNC.
Red dash-dotted line: J=2--1. Black solid line: J=3--2.
Blue dotted line: J=4--3.
The object name, redshift, and adopted beam size in kpc are shown in
each plot.  
We adopt the rest-frame frequency of 
$\nu_{\rm rest}$ = 177.261 GHz for HCN J=2--1, 
$\nu_{\rm rest}$ = 265.886 GHz for HCN J=3--2,
$\nu_{\rm rest}$ = 354.505 GHz for HCN J=4--3,
$\nu_{\rm rest}$ = 178.375 GHz for HCO$^{+}$ J=2--1,
$\nu_{\rm rest}$ = 267.558 GHz for HCO$^{+}$ J=3--2, 
$\nu_{\rm rest}$ = 356.734 GHz for HCO$^{+}$ J=4--3, 
$\nu_{\rm rest}$ = 181.325 GHz for HNC J=2--1,
$\nu_{\rm rest}$ = 271.981 GHz for HNC J=3--2, and 
$\nu_{\rm rest}$ = 362.630 GHz for HNC J=4--3.
The horizontal black thin dotted straight line indicates the zero flux  
level. 
\label{fig:profile}
}
\end{figure}
%%%%%%%%%%%%%%%%%%%%%%%%%%%

Gaussian fits are applied to estimate the emission line fluxes within
the adopted beam sizes.
Single Gaussian fits are adopted, except for the HCN and HCO$^{+}$
lines of IRAS 12112$+$0305 NE, which clearly exhibit a double-peaked
profile with a strong central dip (Figure 
\ref{fig:profile}j,k) wherein the results of a double Gaussian fit
are adopted.  
For certain emission lines from other sources, certain weak
signatures of double-peaked profiles with modest amounts of central
dips are observed; however, we decide to adopt single Gaussian
fitting results after confirming that the flux estimates between the
single and double Gaussian fits agree within $\sim$10\% for
significantly detected emission lines ($\gtrsim$3$\sigma$). 
This is because we aim to estimate the fluxes of multiple J-transition
emission lines in a consistent manner, using the simplest profile
possible.  
The final Gaussian fitting results are summarized in Appendix B.

The flux density of the continuum emission, simultaneously taken during
individual molecular line observations, is also estimated using the
adopted beam sizes for individual (U)LIRGs. 
Flux density and spectral energy distribution (SED) of the continuum
emission are summarized in Appendix C. 

\begin{figure}[!hbt]
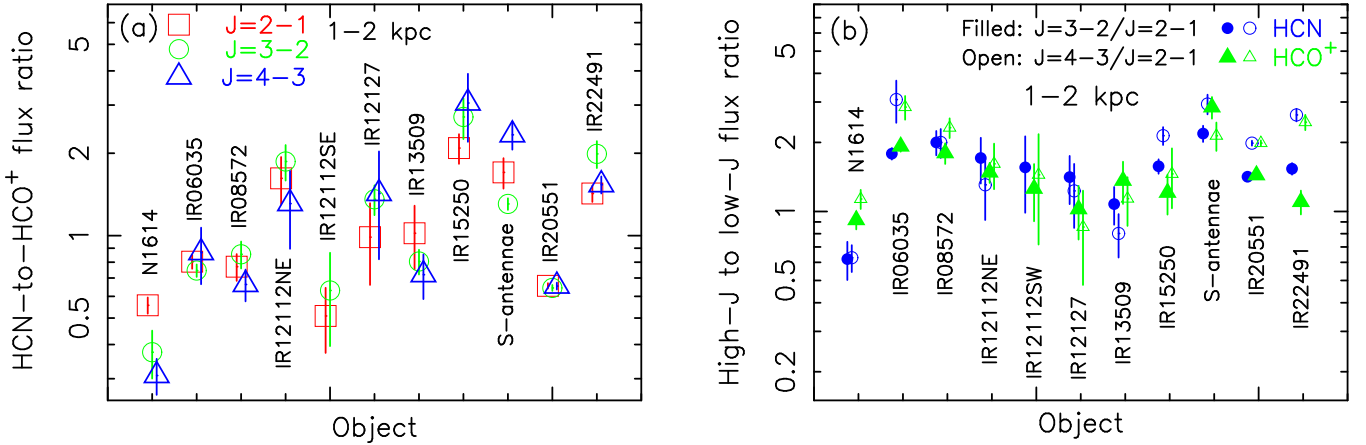

\vspace*{0.5cm}
\begin{center}
\includegraphics[angle=-90,scale=.38]{f2a.eps} \hspace{0.7cm} 
\includegraphics[angle=-90,scale=.38]{f2b.eps} 
\end{center}
\caption{
{\it (a)}: HCN-to-HCO$^{+}$ flux ratio (ordinate) with a 1--2 kpc beam
size at J=2--1 (red open square), 
J=3--2 (green open circle), and J=4--3 (blue open triangle).
J=4--3 data of IRAS 12112$+$0305 SW are not shown because no HCN J=4--3
emission line is detected.
{\it (b)}: J=3--2 to J=2--1 (filled) and J=4--3 to J=2--1 (open) flux
ratio of HCN (blue circle) to HCO$^{+}$ (green triangle), 
calculated from 1--2 kpc beam-sized flux measurements of individual
lines in Jy km s$^{-1}$.
For IRAS 12112$+$0305 SW, the flux ratio of HCN J=4--3 to J=2--1 is
not shown, because the HCN J=4--3 emission line is not detected.
\label{fig:FluxRatio}
}
\end{figure}

Table \ref{tab:flux} summarizes the adopted Gaussian-fit
velocity-integrated emission line fluxes of HCN, HCO$^{+}$, and HNC at 
J=2--1, J=3--2, and J=4--3, together with the adopted line width which
will be used for model calculations in $\S$5.
Table \ref{tab:HCN-to-HCOratio} and Figure \ref{fig:FluxRatio}a show the 
derived HCN-to-HCO$^{+}$ flux ratios at J=2--1, J=3--2, and J=4--3.
High-J to low-J emission line flux ratios for HCN, HCO$^{+}$, and HNC 
are summarized in Table \ref{tab:Jratio}, and those for the HCN and
HCO$^{+}$ are plotted in Figure \ref{fig:FluxRatio}b.

%%%%%%%%%% Table 3 (flux) %%%%%%%%%
\begin{deluxetable}{l|ccc|ccc|ccc|c}
%\rotate
%\tabletypesize{\small}
\tabletypesize{\scriptsize}
\tablecaption{Adopted Nuclear Dense Molecular Line Flux and Width 
\label{tab:flux}} 
\tablewidth{0pt}
\tablehead{
\colhead{Object} & \multicolumn{3}{c}{HCN} & \multicolumn{3}{c}{HCO$^{+}$} 
& \multicolumn{3}{c}{HNC} & \colhead{$\Delta$v} \\  
\colhead{} & \colhead{J=2--1} & \colhead{J=3--2} & \colhead{J=4--3} &
\colhead{J=2--1} & \colhead{J=3--2} & \colhead{J=4--3} &
\colhead{J=2--1} & \colhead{J=3--2} & \colhead{J=4--3} & \colhead{} \\
\colhead{} & \multicolumn{3}{c}{[Jy km s$^{-1}$]} & 
\multicolumn{3}{c}{[Jy km s$^{-1}$]} & \multicolumn{3}{c}{[Jy km s$^{-1}$]} 
& \colhead{[km s$^{-1}$]}    \\ 
\colhead{(1)} & \colhead{(2)} & \colhead{(3)} & \colhead{(4)} & 
\colhead{(5)} & \colhead{(6)} & \colhead{(7)} & \colhead{(8)} & 
\colhead{(9)} & \colhead{(10)} & \colhead{(11)} 
}
\startdata 
NGC 1614 & 6.8$\pm$0.4 & 4.2$\pm$0.8 & 4.3$\pm$0.5 & 
12$\pm$1 & 11$\pm$1 & 14$\pm$1 & 
3.7$\pm$0.4 & 2.0$\pm$0.6 & 1.1$\pm$0.6 & 250 \\
IRAS 06035$-$7102 & 2.9$\pm$0.1 & 5.1$\pm$0.2 & 8.8$\pm$1.8 & 
3.6$\pm$0.1 & 6.8$\pm$0.2 & 10$\pm$1 & 
1.6$\pm$0.1 & 1.9$\pm$0.2 & 2.6$\pm$0.6 & 350 \\
IRAS 08572$+$3915 & 1.4$\pm$0.1 & 2.8$\pm$0.2 & 2.8$\pm$0.3 &
1.8$\pm$0.1 & 3.3$\pm$0.3 & 4.2$\pm$0.3 &
0.61$\pm$0.09 & 0.84$\pm$0.13 & 3.3$\pm$0.3 & 300 \\
IRAS 12112$+$0305 NE & 5.5$\pm$1.0 & 9.4$\pm$1.2 & 7.2$\pm$1.7 &
3.4$\pm$0.3 & 5.0$\pm$0.4 & 5.5$\pm$1.2 &
4.3$\pm$0.2 & 8.6$\pm$0.3 & 11$\pm$1 & 350 \\
IRAS 12112$+$0305 SW & 0.49$\pm$0.10 & 0.76$\pm$0.23 &
$<$1.0 \tablenotemark{A} & 
0.96$\pm$0.16 & 1.2$\pm$0.3 & 1.4$\pm$0.7 & 
0.44$\pm$0.17 & 0.62$\pm$0.30 & 0.49$\pm$0.19 & 300 \\
IRAS 12127$-$1412 & 0.94$\pm$0.21 & 1.3$\pm$0.1 & 1.2$\pm$0.2 &
0.95$\pm$0.23 & 0.97$\pm$0.09 & 0.81$\pm$0.30 & 
0.93$\pm$0.20 & 0.84$\pm$0.15 & 1.0$\pm$0.3 & 500 \\
IRAS 13509$+$0442 & 1.0$\pm$0.2 & 1.1$\pm$0.1 & 0.83$\pm$0.11 &
1.0$\pm$0.2 & 1.4$\pm$0.1 & 1.2$\pm$0.2 &
0.57$\pm$0.14 & 0.91$\pm$0.07 & 0.84$\pm$0.13 & 250 \\
IRAS 15250$+$3609 & 3.5$\pm$0.2 & 5.5$\pm$0.3 & 7.5$\pm$0.5 & 
1.7$\pm$0.2 & 2.0$\pm$0.3 & 2.5$\pm$0.7 &
4.1$\pm$0.3 & 7.1$\pm$0.3 & 10$\pm$1 & 250 \\
Superantennae & 4.9$\pm$0.4 & 11$\pm$1 & 14$\pm$1 &
2.9$\pm$0.3 & 8.2$\pm$0.3 & 6.1$\pm$0.6 &
1.9$\pm$0.2 & 1.7$\pm$0.2 & 1.7$\pm$0.5 & 800 \\
IRAS 20551$-$4250 & 4.8$\pm$0.1 & 6.8$\pm$0.1 & 9.6$\pm$0.2 &
7.4$\pm$0.1 & 11$\pm$1 & 15$\pm$1 &
2.3$\pm$0.1 & 3.7$\pm$0.1 & 5.6$\pm$0.2 & 200 \\
IRAS 22491$-$1808  & 5.5$\pm$0.2 & 8.4$\pm$0.3 & 14$\pm$1 &
3.8$\pm$0.2 & 4.2$\pm$0.4 & 9.4$\pm$0.5 &
3.6$\pm$0.3 & 8.6$\pm$0.4 & 9.2$\pm$0.5 & 400 \\
\enddata

\tablenotetext{A}{This 3$\sigma$ upper limit is estimated from a 
1.5 kpc beam moment 0 map.}

\tablecomments{
Col.(1): Object name.
Cols.(2)--(10): Adopted Gaussian-fit velocity-integrated flux of the
dense molecular emission line in Jy km s$^{-1}$ used in our discussion
($\S$5). 
Only the statistical uncertainty of the Gaussian fit is included.
Col.(2): HCN J=2--1.
Col.(3): HCN J=3--2.
Col.(4): HCN J=4--3.
Col.(5): HCO$^{+}$ J=2--1.
Col.(6): HCO$^{+}$ J=3--2.
Col.(7): HCO$^{+}$ J=4--3.
Col.(8): HNC J=2--1.
Col.(9): HNC J=3--2.
Col.(10): HNC J=4--3.
Col.(11): Adopted line width in km s$^{-1}$ based on our Gaussian fit
(FWHM value in Table \ref{tab:Gaussfit}, column 6).
This line width will be used for our RADEX non-LTE model calculations.
}

\end{deluxetable}
%%%%%%%%%%%%%%%%%%%%%%%%%%%%%%%%%%%

%%%%%%%%%% Table 4 (HCN-to-HCOratio) %%%%%%%%%
\begin{deluxetable}{l|ccc}[!hbt]
%\rotate
%\tabletypesize{\small}
\tabletypesize{\scriptsize}
\tablecaption{HCN-to-HCO$^{+}$ Flux Ratio \label{tab:HCN-to-HCOratio}} 
\tablewidth{0pt}
\tablehead{
\colhead{Object} & \colhead{$\frac{\rm HCN}{\rm HCO^+}$ J=2--1} &
\colhead{$\frac{\rm HCN}{\rm HCO^+}$ J=3--2} & 
\colhead{$\frac{\rm HCN}{\rm HCO^+}$ J=4--3} \\  
\colhead{(1)} & \colhead{(2)} & \colhead{(3)} & \colhead{(4)} 
}
\startdata 
NGC 1614 (1 kpc) & 0.56$\pm$0.04 & 0.38$\pm$0.07 & 0.31$\pm$0.05 \\
IRAS 06035$-$7102 (1.6 kpc) & 0.80$\pm$0.05 & 0.75$\pm$0.04 & 0.87$\pm$0.20 \\
IRAS 08572$+$3915 (1 kpc) & 0.77$\pm$0.09 & 0.85$\pm$0.10 & 0.66$\pm$0.09 \\ 
IRAS 12112$+$0305 NE (1.5 kpc) & 1.6$\pm$0.3 & 1.9$\pm$0.3 & 1.3$\pm$0.4 \\
IRAS 12112$+$0305 SW (1.5 kpc) & 0.51$\pm$0.14 & 0.63$\pm$0.24 &
$<$1.3 \\
IRAS 12127$-$1412 (2 kpc) & 0.99$\pm$0.33 & 1.4$\pm$0.2 & 1.4$\pm$0.6 \\
IRAS 13509$+$0442 (2 kpc) & 1.0$\pm$0.3 & 0.80$\pm$0.08 & 0.72$\pm$0.13 \\
IRAS 15250$+$3609 (1.5 kpc) & 2.1$\pm$0.3 & 2.7$\pm$0.5 & 3.1$\pm$0.9 \\
Superantennae (1 kpc) & 1.7$\pm$0.2 & 1.3$\pm$0.1 & 2.3$\pm$0.3 \\
IRAS 20551$-$4250 (1 kpc) & 0.66$\pm$0.01 & 0.64$\pm$0.02 & 0.65$\pm$0.02 \\
IRAS 22491$-$1808 (1.5 kpc) & 1.4$\pm$0.1 & 2.0$\pm$0.2 & 1.5$\pm$0.1 \\
\enddata

%\vspace{1cm}

\tablecomments{
Col.(1): Object name.
The adopted beam size in kpc is shown in parentheses. 
Cols.(2)--(4): The HCN-to-HCO$^{+}$ emission line flux ratio
calculated in units of Jy km s$^{-1}$. 
Col.(2): J=2--1.
Col.(3): J=3--2.
Col.(4): J=4--3.
Only the Gaussian-fit statistical uncertainty is considered because  
both HCN and HCO$^{+}$ data at each J-transition were obtained 
simultaneously, and so their flux ratio is not affected by the
possible absolute flux calibration uncertainty of individual ALMA
observations. 
}

\end{deluxetable}
%%%%%%%%%%%%%%%%%%%%%%%%%%%%%%%%%%%

%%%%%%%%%% Figure 2 %%%%%%%%%
%%%%%%%%%%%%%%%%%%%%%%%%%%%%%%%%%%%

%%%%%%%%%% Table 5 (Jratio) %%%%%%%%%
\begin{deluxetable}{l|ccc|ccc|ccc}
\vspace*{0.5cm}
%\rotate
%\tabletypesize{\small}
\tabletypesize{\scriptsize}
\tablecaption{High-J to Low-J Molecular Emission Line Flux 
Ratio \label{tab:Jratio}}  
\tablewidth{0pt}
\tablehead{
\colhead{Object} & \multicolumn{3}{c}{HCN} & \multicolumn{3}{c}{HCO$^{+}$} 
& \multicolumn{3}{c}{HNC} \\  
\colhead{} & \colhead{$\frac{J=3-2}{J=2-1}$} &
\colhead{$\frac{J=4-3}{J=2-1}$} & \colhead{$\frac{J=4-3}{J=3-2}$}
& \colhead{$\frac{J=3-2}{J=2-1}$} &
\colhead{$\frac{J=4-3}{J=2-1}$} & \colhead{$\frac{J=4-3}{J=3-2}$}
& \colhead{$\frac{J=3-2}{J=2-1}$} &
\colhead{$\frac{J=4-3}{J=2-1}$} & \colhead{$\frac{J=4-3}{J=3-2}$} \\
\colhead{(1)} & \colhead{(2)} & \colhead{(3)} & \colhead{(4)} & 
\colhead{(5)} & \colhead{(6)} & \colhead{(7)} & \colhead{(8)} & 
\colhead{(9)} & \colhead{(10)}  
}
\startdata 
NGC 1614 (1 kpc) & 0.62$\pm$0.12 & 0.63$\pm$0.08 & 1.0$\pm$0.2 & 
0.92$\pm$0.08 & 1.1$\pm$0.1 & 1.2$\pm$0.1 & 
0.54$\pm$0.16 & 0.30$\pm$0.18 & 0.55$\pm$0.35 \\
IRAS 06035$-$7102 (1.6 kpc) & 1.8$\pm$0.1 & 3.1$\pm$0.6 & 1.7$\pm$0.4 & 
1.9$\pm$0.1 & 2.9$\pm$0.3 & 1.5$\pm$0.2 & 
1.2$\pm$0.1 & 1.6$\pm$0.4 & 1.4$\pm$0.4 \\
IRAS 08572$+$3915 (1 kpc) & 2.0$\pm$0.2 & 2.0$\pm$0.3 & 1.0$\pm$0.1 &
1.8$\pm$0.2 & 2.3$\pm$0.2 & 1.3$\pm$0.1 &
1.4$\pm$0.3 & 5.5$\pm$1.0 & 4.0$\pm$0.7 \\
IRAS 12112$+$0305 NE (1.5 kpc) & 1.7$\pm$0.4 & 1.3$\pm$0.4 & 0.76$\pm$0.21 &
1.5$\pm$0.1 & 1.6$\pm$0.4 & 1.1$\pm$0.2 &
2.0$\pm$0.1 & 2.4$\pm$0.2 & 1.2$\pm$0.1 \\
IRAS 12112$+$0305 SW (1.5 kpc) & 1.6$\pm$0.6 & $<$2.4 & $<$1.8 & 
1.3$\pm$0.4 & 1.4$\pm$0.7 & 1.1$\pm$0.6 & 
1.4$\pm$0.9 & 1.1$\pm$0.6 & 0.79$\pm$0.49 \\
IRAS 12127$-$1412 (2 kpc) & 1.4$\pm$0.3 & 1.2$\pm$0.4 & 0.87$\pm$0.20 &
1.0$\pm$0.3 & 0.86$\pm$0.38 & 0.84$\pm$0.32 & 
0.91$\pm$0.25 & 1.1$\pm$0.4 & 1.2$\pm$0.4 \\
IRAS 13509$+$0442 (2 kpc) & 1.1$\pm$0.2 & 0.80$\pm$0.17 & 0.74$\pm$0.10 &
1.4$\pm$0.3 & 1.1$\pm$0.3 & 0.83$\pm$0.13 &
1.6$\pm$0.4 & 1.5$\pm$0.4 & 0.93$\pm$0.16 \\
IRAS 15250$+$3609 (1.5 kpc) & 1.6$\pm$0.1 & 2.1$\pm$0.2 & 1.4$\pm$0.1 & 
1.2$\pm$0.2 & 1.5$\pm$0.4 & 1.2$\pm$0.4 &
1.8$\pm$0.2 & 2.5$\pm$0.2 & 1.4$\pm$0.1 \\
Superantennae (1 kpc) & 2.2$\pm$0.2 & 2.9$\pm$0.3 & 1.3$\pm$0.1 &
2.8$\pm$0.3 & 2.1$\pm$0.3 & 0.75$\pm$0.08 &
0.89$\pm$0.14 & 0.89$\pm$0.25 & 1.0$\pm$0.3 \\
IRAS 20551$-$4250 (1 kpc) & 1.4$\pm$0.1 & 2.0$\pm$0.1 & 1.4$\pm$0.1 &
1.4$\pm$0.1 & 2.0$\pm$0.1 & 1.4$\pm$0.1 &
1.6$\pm$0.1 & 2.5$\pm$0.1 & 1.5$\pm$0.1 \\
IRAS 22491$-$1808 (1.5 kpc) & 1.5$\pm$0.1 & 2.6$\pm$0.2 & 1.7$\pm$0.1 &
1.1$\pm$0.1 & 2.5$\pm$0.2 & 2.2$\pm$0.3 &
2.4$\pm$0.2 & 2.6$\pm$0.3 & 1.1$\pm$0.1 \\
\enddata

\tablecomments{Col.(1): Object name.
The adopted beam size in kpc is shown in parentheses. 
Cols.(2)--(10): High-J to low-J flux ratio of dense molecular emission 
line measured in Jy km s$^{-1}$. 
Only the statistical uncertainty of the Gaussian fit is considered.
Col.(2): HCN J=3--2 to J=2--1.
Col.(3): HCN J=4--3 to J=2--1.
Col.(4): HCN J=4--3 to J=3--2.
Col.(5): HCO$^{+}$ J=3--2 to J=2--1.
Col.(6): HCO$^{+}$ J=4--3 to J=2--1.
Col.(7): HCO$^{+}$ J=4--3 to J=3--2.
Col.(8): HNC J=3--2 to J=2--1.
Col.(9): HNC J=4--3 to J=2--1.
Col.(10): HNC J=4--3 to J=3--2.
}

\end{deluxetable}
%%%%%%%%%%%%%%%%%%%%%%%%%%%%%%%%%%%

The high-J to low-J (J=2--1) flux ratios of the LIRG NGC 1614 are 
smaller than those of the remaining ULIRGs particularly for HCN 
(by a factor of 2--5 except IRAS 13509$+$0442) and for HCO$^{+}$
as well (by a factor of 1--3)  
(Figure \ref{fig:FluxRatio}b and Table \ref{tab:Jratio}), suggesting
that gas density and temperature of NGC 1614 are much lower and thus
collisional excitation to high-J levels (J=3 and 4) is insufficient.
Furthermore, only NGC 1614 shows (1) a clear systematic
decreasing trend in the HCN-to-HCO$^{+}$ flux ratio from low-J to high-J
transition in Figure \ref{fig:FluxRatio}a and (2) significantly
smaller high-J to low-J flux ratios for HCN than for HCO$^{+}$ in Figure
\ref{fig:FluxRatio}b, while the remaining ten ULIRGs' nuclei do not 
except IRAS 13509$+$0442 which shows a weaker, but similar
decreasing trend in the former. 
These results can also be explained by the lower gas density and
temperature in NGC 1614 because the critical density of HCN is a
factor of $\sim$5 higher than that of HCO$^{+}$ \citep{shi15}; thus,
collisional excitation to a high J-level is less efficient for HCN
than for HCO$^{+}$ in low-density and low-temperature molecular gas.  
We will employ non-LTE model calculations to quantitatively
constrain the nuclear dense molecular gas properties of the LIRG 
NGC 1614 and the remaining ULIRGs in more detail in the next section
($\S$5).  
 
\vspace{1cm}

%\clearpage

\section{Discussion} 

\subsection{General Descriptions of Our RADEX Non-LTE Modeling}

There are now three rotational transition line data (J=2--1, J=3--2,
and J=4--3) of HCN, HCO$^{+}$, and HNC, whose fluxes are 
measured with the same beam sizes (1--2 kpc) for individual (U)LIRGs. 
To investigate nuclear (1--2 kpc) molecular gas properties in detail, 
the observed emission line flux ratios are compared with the
calculated flux ratios using RADEX \citep{RADEX}.
RADEX is a non-LTE radiative transfer code that solves the statistical
equilibrium problems in a one-zone medium based on the escape
probability approximations. 
Molecular emission line flux ratios in the RADEX code are
primarily determined by the following three parameters: 
(i) H$_{2}$ volume number density (n$_{\rm H2}$). 
(ii) H$_{2}$ kinetic temperature (T$_{\rm kin}$).
(iii) Molecular column density divided by line width 
(N$_{\rm mol}$/$\Delta$v), which reflects the line opacity, where
$\Delta$v can be derived from the observed molecular line width (Table 
\ref{tab:flux}, column 11). 
We constrain these three parameters from the three J-transition line data. 
 
In all RADEX non-LTE calculations, molecular gas is assumed to
be static and spherically symmetric homogeneous medium.
The background temperature is set as the cosmic microwave background 
(2.73 K). 
The collision partner for HCN, HCO$^+$, and HNC is regarded as H$_2$ only. 
The emission line flux ratios are calculated after converting the
fluxes from Kelvin (brightness temperature) to Jy units, as adopted in 
Tables \ref{tab:HCN-to-HCOratio}--\ref{tab:Jratio}. 
For implementation,
\texttt{pyradex}\footnote{\url{https://github.com/keflavich/pyradex}}, 
a Python wrapper for RADEX, is used.

To perform the RADEX non-LTE calculations, we use only the
HCN and HCO$^{+}$ data, but exclude the HNC data for the following two
reasons. 
First, in previous observations of the Galactic sources, the HNC
abundance is 
found to be very low in warm molecular gas in the vicinity of a
luminous energy source \citep[e.g.,][]{sch92,hir98,gra14,bub19}.
In fact, in the nearby ($z =$ 0.0038, luminosity distance $\sim$ 14 Mpc), 
well-studied AGN NGC 1068, the HNC emission is found to be extremely
weak, compared with HCN and HCO$^{+}$, in close proximity to the central
luminous AGN \citep{ima20}. 
Therefore, HNC is not considered to be as good as HCN and
HCO$^{+}$ to investigate the physical properties of warm
molecular gas near luminous energy sources (AGN and/or
starburst) at the center of (U)LIRGs' nuclei.
Second, the HNC data were not taken simultaneously with
the HCN and HCO$^{+}$ data at the same J-transition 
\citep{ima13a,ima13b,ima14,ima16a,ima16b,ima18,ima21,ima22}.
Thus, possible absolute flux calibration uncertainty of three more
ALMA observations (HNC J=2--1, J=3--2, and J=4--3) must be considered
as additional free parameters for our RADEX non-LTE calculations (see
$\S$5.3 and 5.5), which makes it extremely difficult to find the
best fit values without systematic uncertainty under the limited
number of observational constraints.

\subsection{HCN-to-HCO$^+$ Emission Line Flux Ratios}

In the RADEX model calculations for the (U)LIRGs' nuclei, we set 
density and temperature in the range 10$^{3}$--10$^{8}$ cm$^{-3}$ and
10--1000 K, respectively. 
We also adopt the HCO$^{+}$ column density with N$_{\rm HCO+}$ = 
1 $\times$ 10$^{16}$ cm$^{-2}$ as the fiducial value, by assuming that
the observed (U)LIRGs are modestly Compton-thick (N$_{\rm H}$ $\sim$
a few $\times$ 10$^{24}$ cm$^{-2}$) and the HCO$^{+}$-to-H$_{2}$
abundance ratio is $\sim$10$^{-8}$ \citep[e.g.,][]{mar06,sai18}. 

\begin{figure}[!hbt]
\begin{center}
%\vspace*{2cm}
%\hspace*{-11.7cm}
\includegraphics[angle=0,scale=.26]{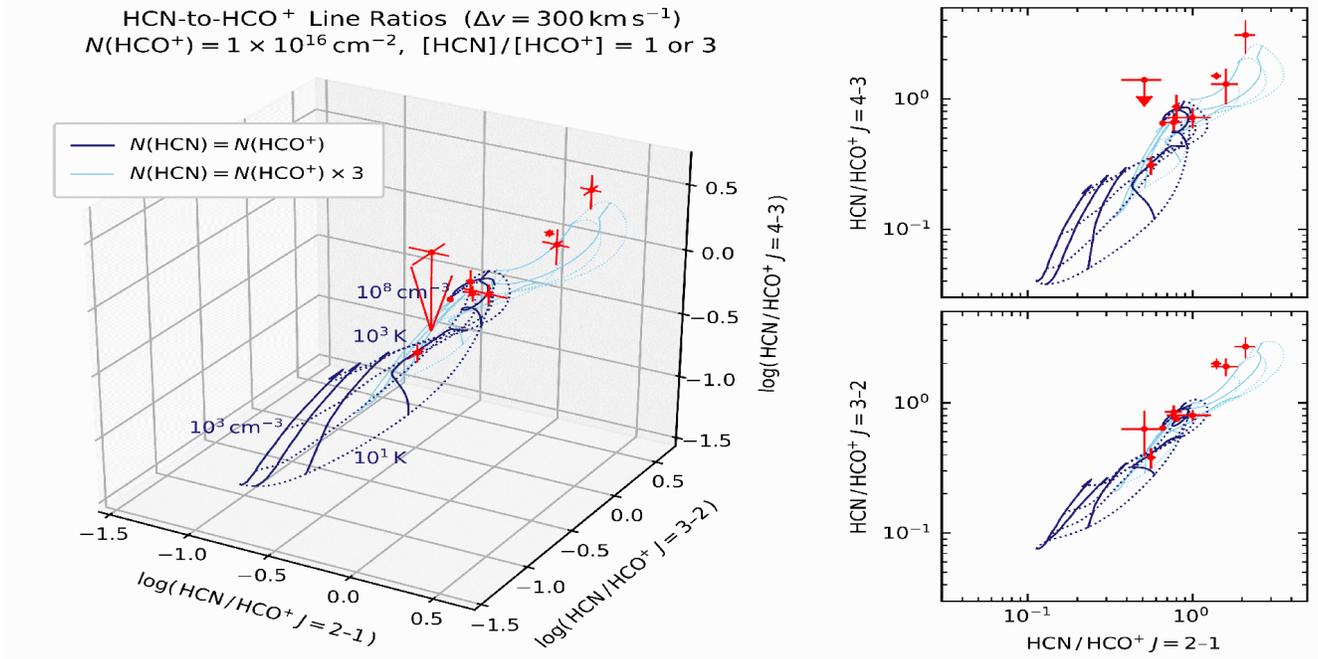} 
\end{center}
%\vspace{-1.0cm}
\caption{
Comparison of the observed and RADEX-calculated HCN-to-HCO$^{+}$ flux
ratios. 
{\it (Left panel)}: 3D plot of HCN-to-HCO$^{+}$ flux ratios at J=2--1,
J=3--2, and J=4--3. 
The red points indicate the observed values.
Only nine (U)LIRGs' nuclei with adopted line widths 
$\Delta$v = 200--400 km s$^{-1}$ (excluding IRAS 12127$-$1412 and
the Superantennae) (Table \ref{tab:flux}) are included here. 
The overlaid blue meshes show the RADEX-calculated results computed
over the range of density log(n$_{\rm H_2}$/cm$^{-3}$) = 3--8 
in steps of 0.1 and 
temperature log(T$_{\rm kin}$/K) = 1--3 in steps of 0.05.
The solid lines represent the iso-density results (log n$_{\rm H_2}$ = 3,
4, 5, 6, 7, 8), whereas the dotted lines represent the iso-temperature
results (log T$_{\rm kin}$ = 1.0, 1.5, 2.0, 2.5, 3.0). 
In the calculations, the line width and HCO$^+$ column density are 
fixed at $\Delta$v = 300 km s$^{-1}$  
and N$_{\rm HCO+}$ = 1 $\times$ 10$^{16}$ cm$^{-2}$, respectively. 
The results of HCN-to-HCO$^{+}$ abundance ratio of [HCN]/[HCO$^{+}$] =
1 or 3 are shown in dark and light blue colors, respectively.
{\it (Right panels)}: Projection of the 3D plot along J=3--2
{\it (upper)} and J=4--3 {\it (lower)} directions. 
\label{fig:3Dplot}
}
\end{figure}

Figure \ref{fig:3Dplot} shows a three-dimensional plot of 
the HCN-to-HCO$^+$ emission line flux ratio at J=2--1, J=3--2, and
J=4--3. 
The overall distribution of the observed HCN-to-HCO$^{+}$ flux ratios
for the (U)LIRGs' nuclei can be naturally reproduced with the
HCN-to-HCO$^{+}$ abundance ratio of [HCN]/[HCO$^{+}$] = 3 rather than 1. 
In Appendix D, we also make the same plot with [HCN]/[HCO$^{+}$]
= 7 and 1, and confirm that the enhanced HCN-to-HCO$^{+}$ abundance
ratio can better fit the observed flux ratios for the majority of the
(U)LIRGs' nuclei.
Two sources with small HCN-to-HCO$^{+}$ flux ratios, located at the
bottom-left part of Figure \ref{fig:3Dplot} (NGC 1614 and IRAS
12112$+$0305 SW) may be better reproduced with [HCN]/[HCO$^{+}$] = 1
(i.e., comparable HCN and HCO$^{+}$ abundances). 
Both of these sources are classified as starburst-dominated ($\S$2).
However, the remaining ULIRGs showing luminous AGN signatures 
(except IRAS 13509$+$0442) are better reproduced with enhanced HCN
abundance relative to HCO$^{+}$.  
In fact, the enhanced HCN-to-HCO$^{+}$ abundance ratio (1) has been
reported in dense molecular gas around luminous AGNs  
\citep[e.g.,][]{kri08,ala15,sai18,nak18,tak19,kam20,ima20,but22}, and 
(2) has been argued to be necessary to reproduce the HCN-to-HCO$^{+}$ flux
ratios with larger than unity, as observed in many (U)LIRGs 
\citep[e.g.,][]{yam07,izu16a}. 
We adopt this enhanced HCN abundance of [HCN]/[HCO$^{+}$] = 3 as the
fiducial value for our model calculations in the next subsection.
As we will show in the calculations for the starburst-dominated LIRG
NGC 1614 (Figure \ref{fig:N1614fit}), adopting [HCN]/[HCO$^{+}$] = 1 
provides comparable gas density and temperature 
%---
\footnote{
Our main argument about gas density and temperature at (U)LIRGs' 
nuclei ($\S$5.3--5.4) also will not change even calculating with
[HCN]/[HCO$^{+}$] = 7.
}. 
%---

%%%%%%%%%% Figure 3 %%%%%%%%%
%%%%%%%%%%%%%%%%%%%%%%%%%%%%%%%%%%%

We also perform calculations using the HCO$^{+}$ column density of
N$_{\rm HCO+}$ = 1 $\times$ 10$^{15}$ cm$^{-2}$ (an order of magnitude
smaller than that adopted in Figure \ref{fig:3Dplot}) and a line
width of $\Delta$v = 800 km s$^{-1}$ (a factor of $>$2 broader).
However, the overall trend of the RADEX-calculated HCN-to-HCO$^{+}$
flux ratios does not change significantly (Appendix D). 

\subsection{High-J to Low-J Emission Line Flux Ratios}

\begin{figure}[h]
\begin{center}
%\vspace*{1.0cm}
%\hspace*{-6.6cm}
\includegraphics[angle=0,scale=.13]{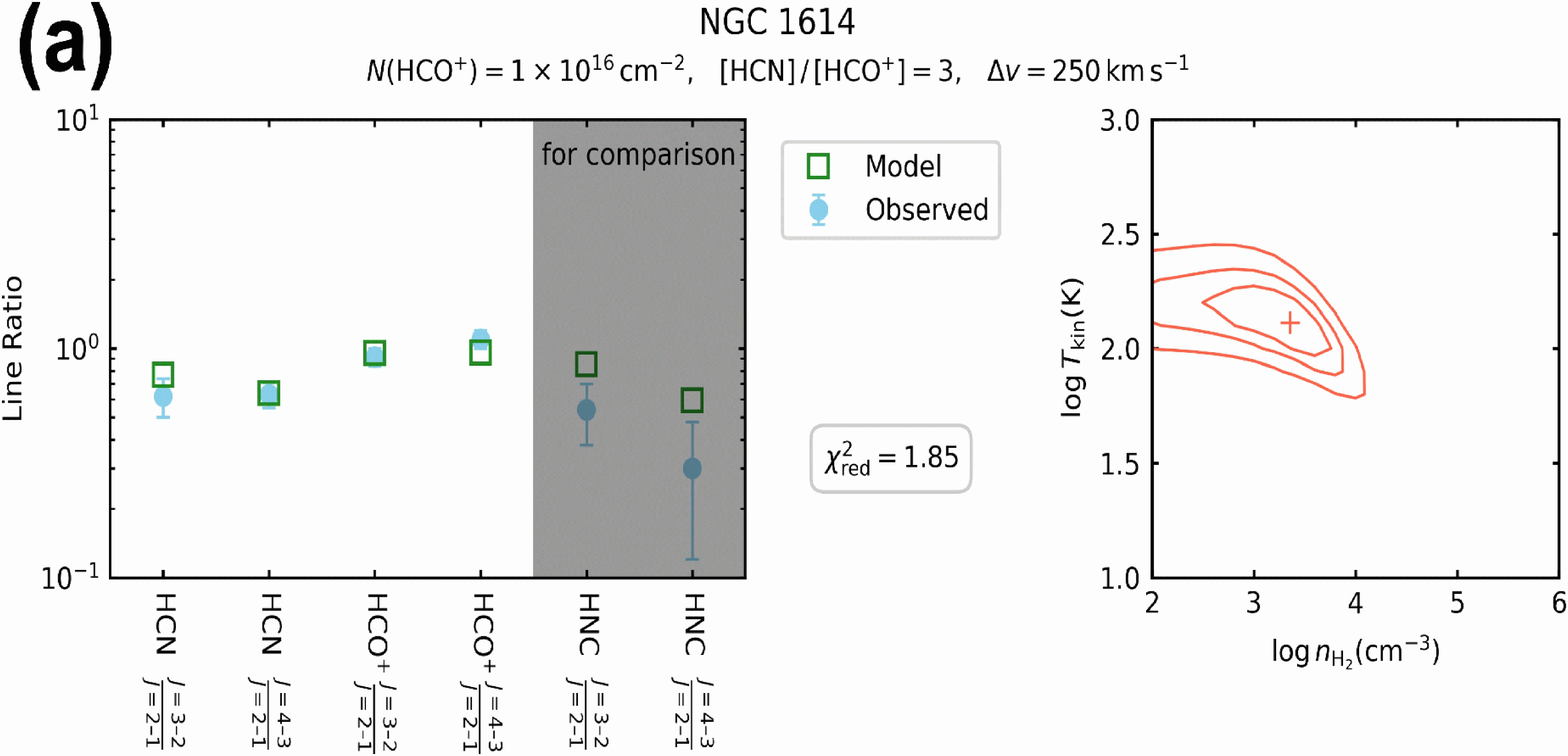} 
\includegraphics[angle=0,scale=.13]{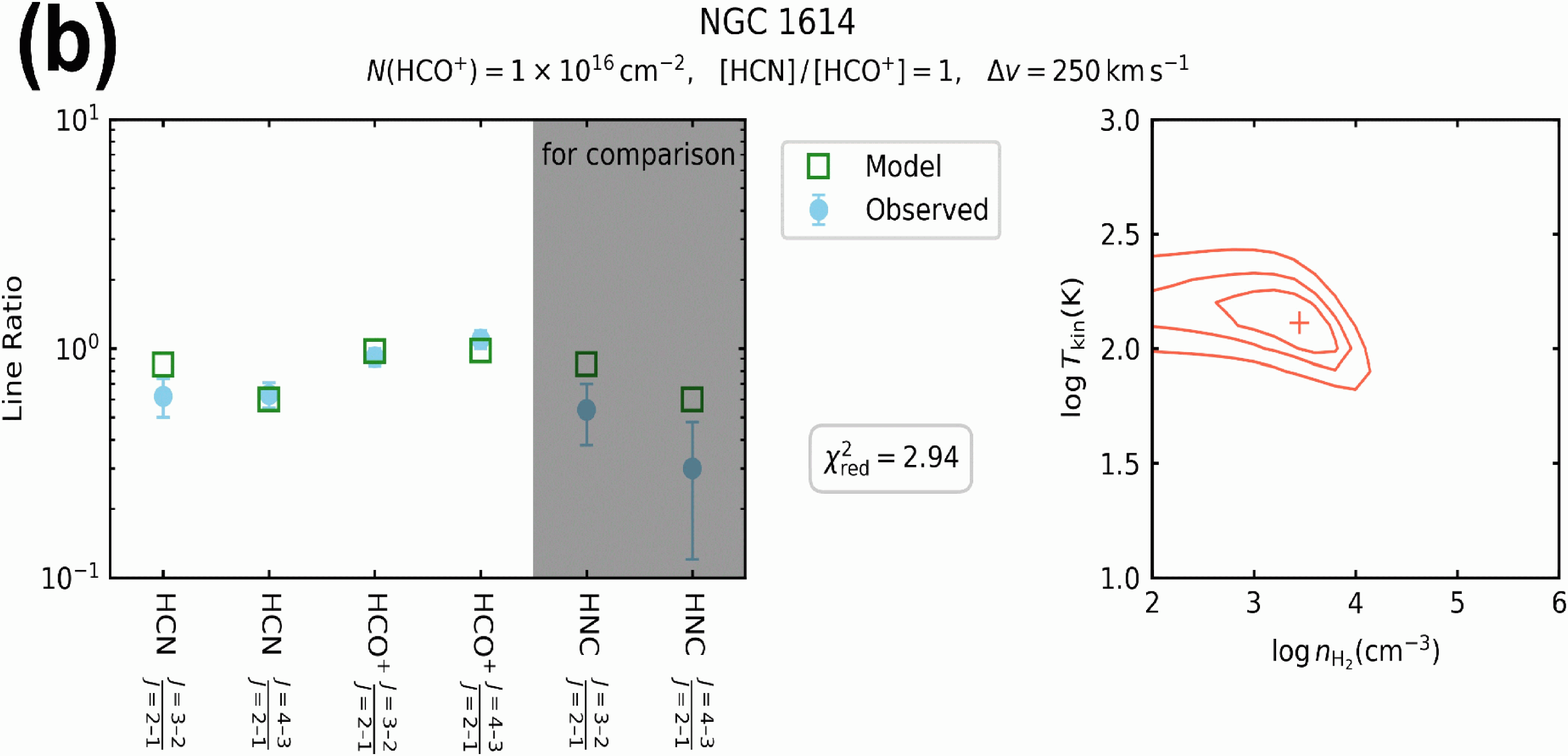} \\
%\hspace*{-6.6cm}
\includegraphics[angle=0,scale=.13]{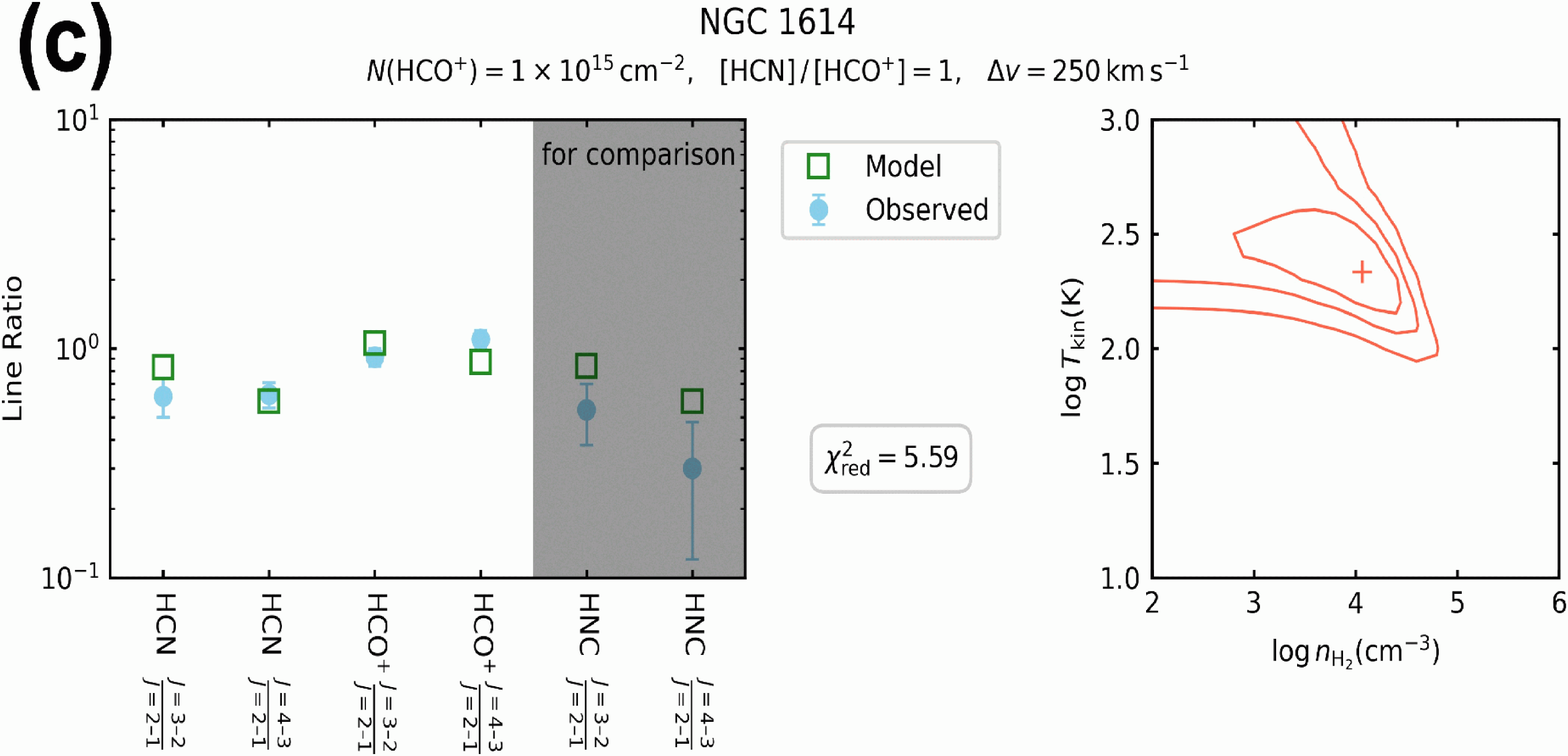} 
\includegraphics[angle=0,scale=.13]{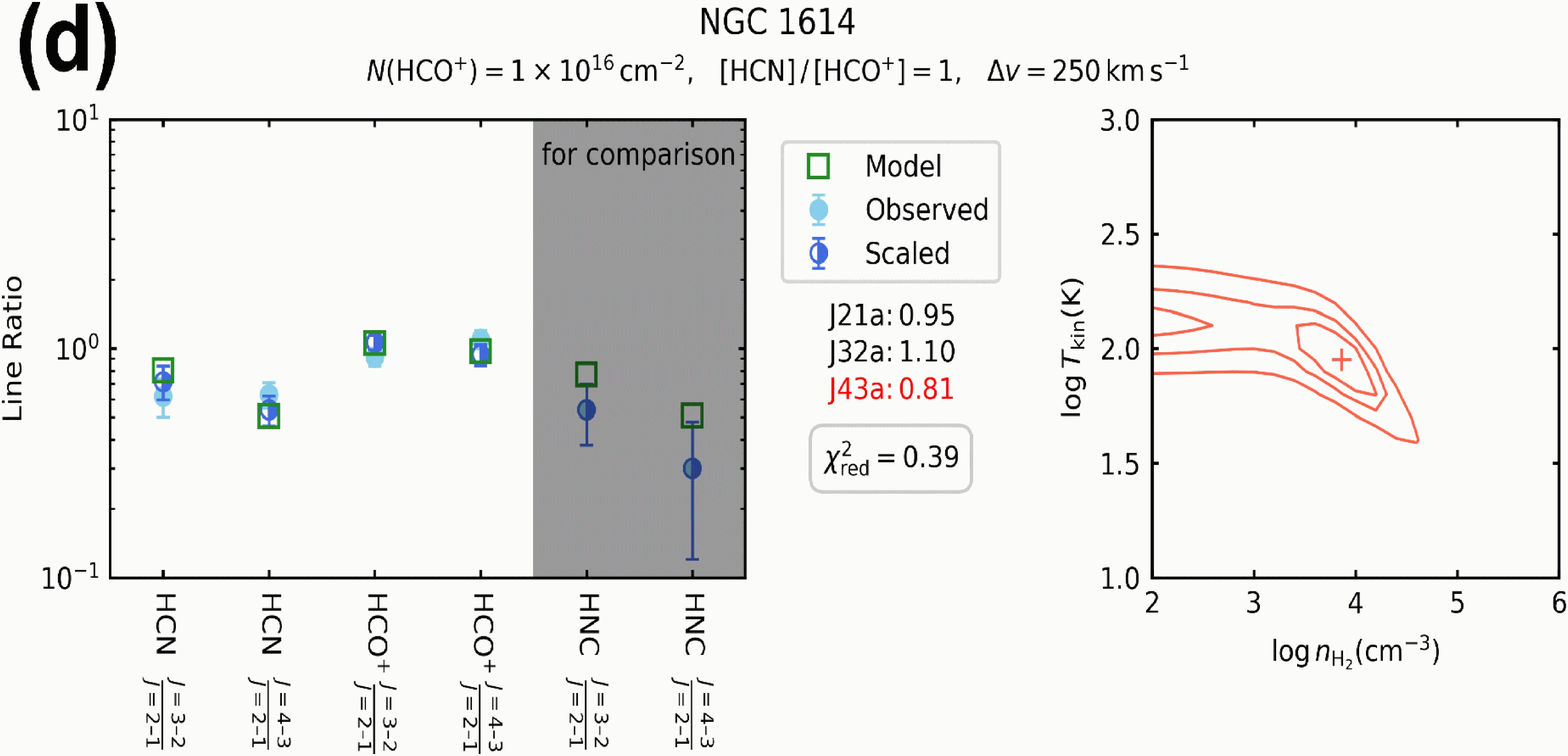} \\ 
\end{center}
%\vspace{-0.6cm}
\caption{
Example of fitting results for the high-J to low-J flux ratios of HCN
and HCO$^+$ for the starburst-dominated LIRG NGC 1614.
HCO$^{+}$ column density and HCN-to-HCO$^{+}$ abundance
ratio are 
{\it (a)}: (N$_{\rm HCO+}$/cm$^{2}$, [HCN]/[HCO$^{+}$]) = (1 $\times$
10$^{16}$, 3), 
{\it (b)}: (1 $\times$ 10$^{16}$, 1), 
{\it (c)}: (1 $\times$ 10$^{15}$, 1), and 
{\it (d)}: (1 $\times$ 10$^{16}$, 1) with scaling factor adjustment
allowed. 
The line width is fixed at $\Delta$v = 250 km s$^{-1}$ for all cases.
{\it (Left panel of each plot)}: Comparison of the observed and
modeled line flux ratios. 
The green open squares represent the best-fit RADEX model.
Its gas density and temperature values are indicated by the plus sign
in the right panel, whereas other fixed parameters (HCO$^{+}$ column
density, HCN-to-HCO$^{+}$ abundance ratio, and line width) are
listed at the top of the figure below the object name. 
The light blue filled circles represent the flux ratios measured from
our observations. 
In (d), the dark blue half-filled circles represent the flux ratios by
allowing scale adjustment within the absolute flux calibration
uncertainty of individual ALMA observations. 
Below the legend, the scaling factors to be multiplied for the line
fluxes are listed.  
If a factor is colored red, it implies that the observation was made in
Cycle 0 (double systematic error). 
For example, in (d), the J=3--2 to J=2--1 flux ratios are multiplied
by a factor of 1.10/0.95 = 1.16. 
In (a)--(d), the reduced $\chi^2$ value for the best-fit model is
listed at the bottom middle part. 
The HNC line flux ratios in the shaded part of the left panel are
shown only for reference and are not used in the fitting; the 
HNC abundance is assumed to be equal to that of HCO$^+$. 
{\it (Right panel)}: Confidence ranges for gas density and temperature. 
The plus sign indicates the position of the best-fit model.
The contours represent 68, 90, and 99\% confidence levels for the two
parameters of interest ($\Delta\chi^2=2.28, 4.61, 9.21$).
\label{fig:N1614fit}
}
\end{figure}

To quantitatively estimate the density and temperature of 
molecular gas at (U)LIRGs' nuclei, a total of four observed
line flux ratios, J=3--2 to J=2--1 and J=4--3 to J=2--1 of HCN and
HCO$^+$ (Table \ref{tab:Jratio}) are fitted using the RADEX model.
Least-squares fitting of log n$_{\rm H_2}$ (density) and log
T$_{\rm kin}$ (temperature) is performed using conventional
Levenberg-Marquardt algorithm through Python package \texttt{lmfit}
\citep{lmfit}. 
The confidence intervals for the parameters are examined 
by grid computing $\Delta\chi^2\equiv\chi^2-\chi^2_{\rm best}$
with log n$_{\rm H_2}$ ranging from 2--6 and log T$_{\rm kin}$
from 1--3.
Figure \ref{fig:N1614fit}a shows the derivation of molecular gas
density and temperature that can best reproduce the observed J=3--2
to J=2--1 and J=4--3 to J=2--1 flux ratios of HCN and HCO$^{+}$ for the
starburst-dominated LIRG NGC 1614, by adopting fiducial values of
N$_{\rm HCO^+}$ = 1 $\times$ 10$^{16}$ cm$^{-2}$, [HCN]/[HCO$^{+}$] =
3, and observed high-J to low-J flux ratios. 
The density and temperature are well constrained.

%%%%%%%%%% Figure 4 %%%%%%%%%
%%%%%%%%%%%%%%%%%%%%%%%%%%%%%%%%%%%

In Figure \ref{fig:3Dplot}, the observed HCN-to-HCO$^{+}$ flux ratio
for the two sources located at the bottom left part, including NGC 1614,
can also be explained by a non-enhanced HCN-to-HCO$^{+}$ abundance
ratio ([HCN]/[HCO$^{+}$] = 1). 
Furthermore, it is possible that the LIRG NGC 1614 has lower
column density than those of ULIRGs. 
Thus, the same least-squares fit is conducted by adopting 
(1) [HCN]/[HCO$^{+}$] = 1 (Figure \ref{fig:N1614fit}b), and 
(2) N$_{\rm HCO^+}$ = 1 $\times$ 10$^{15}$ cm$^{-2}$, in addition to 
[HCN]/[HCO$^{+}$] = 1 (Figure \ref{fig:N1614fit}c).
Only a limited change in the derived H$_{\rm 2}$ molecular gas
density (n$_{\rm H_2}$) and temperature (T$_{\rm kin}$) is observed.
 
In contrast to the HCN-to-HCO$^+$ flux ratios for the same J
transition where both HCN and HCO$^{+}$ data were obtained
simultaneously, the high-J to low-J flux ratios of HCN and
HCO$^{+}$ are affected by the possible absolute flux calibration
uncertainty of individual ALMA observations. 
This is because data for J=4--3, J=3--2, and J=2--1 were obtained at
different times.  
Therefore, this possible systematic uncertainty must be considered.
The absolute flux calibration uncertainty is as high as 5\% for J21a 
%---
\footnote{See Table \ref{tab:beam} caption for the definition of J21a, 
J32a, and J43a.}
%---
(band 5) and 10\% for J32a and J43a (bands 6 and 7), respectively,
according to the ALMA Proposer's Guide. 
However, for Cycle 0 observations, we conservatively choose to adopt
twice the above values, because of possibly large systematic 
uncertainty in ALMA very early phase.
We allow scaling of the absolute fluxes within the above range.
If individual emission line fluxes are scaled, then their flux ratios
will change accordingly. 

We adopt two steps.
First, in addition to gas density (n$_{\rm H_2}$) and temperature 
(T$_{\rm kin}$), the scaling factors for J21a, J32a, and J43a are
treated as free parameters that move within the systematic uncertainty. 
Least-squares fitting with box constraints is performed with
L-BFGS-B algorithm \citep{L-BFGS-B}. 
Subsequently, the scaling factors are fixed at the determined values. 
Then, gas density and temperature are derived using the
Levenberg-Marquardt algorithm in the same manner as before.
Figure \ref{fig:N1614fit}d shows the results obtained by allowing the
flux scale adjustment.  
The best fit values based on the four models in 
Figure \ref{fig:N1614fit} are summarized in Table \ref{tab:bestfit}.  
The derived gas density and temperature largely agree in
the range n$_{\rm H_2}$ = 10$^{3.4-4.1}$ cm$^{-3}$ and T$_{\rm kin}$ =
10$^{2.0-2.3}$ K for all the methods in Figure \ref{fig:N1614fit}.

\begin{figure}[h]
\begin{center}
%\vspace*{0.2cm}
%\hspace*{-6.6cm}
\includegraphics[angle=0,scale=.125]{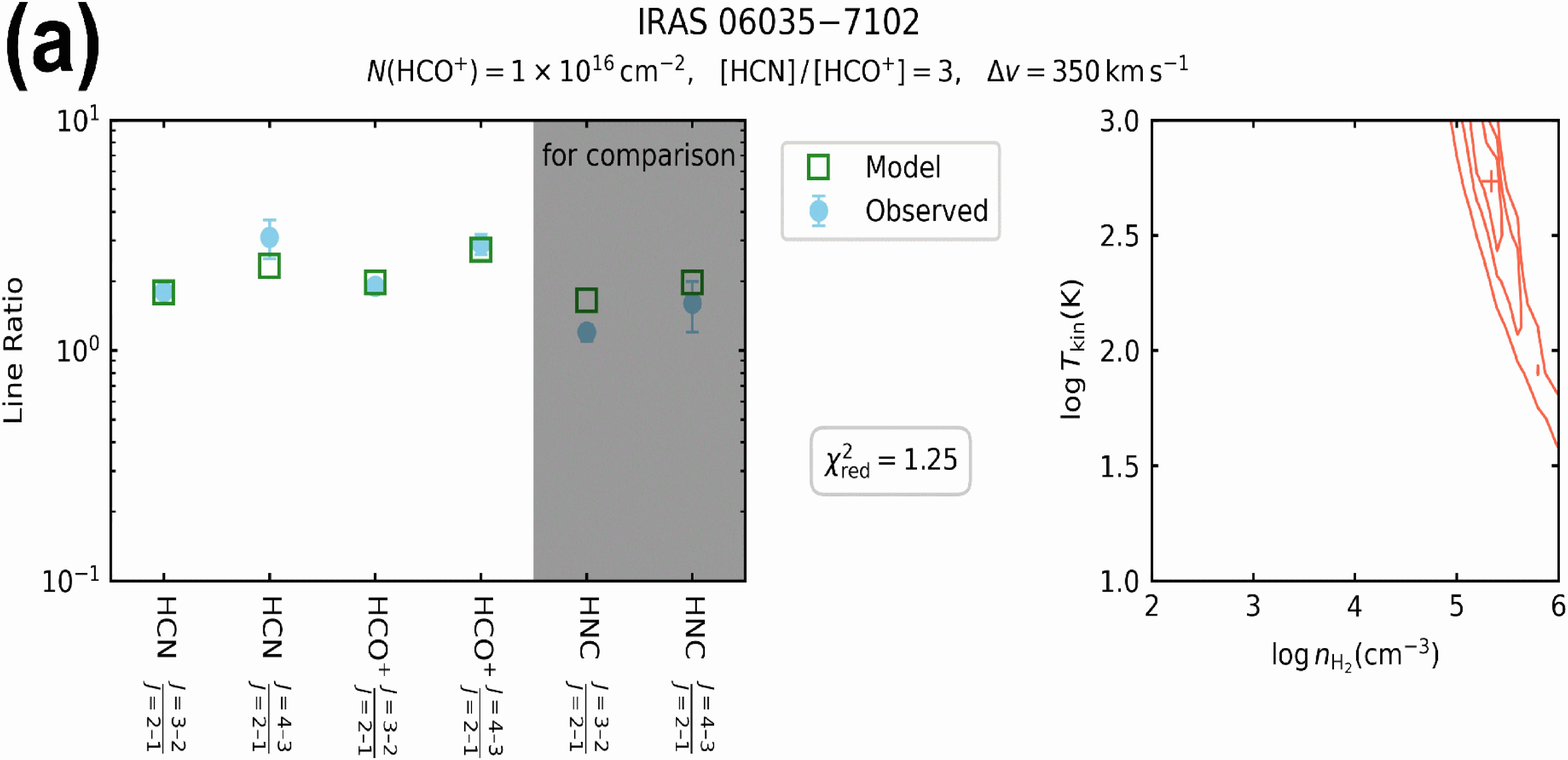} \hspace{0.5cm}
\includegraphics[angle=0,scale=.125]{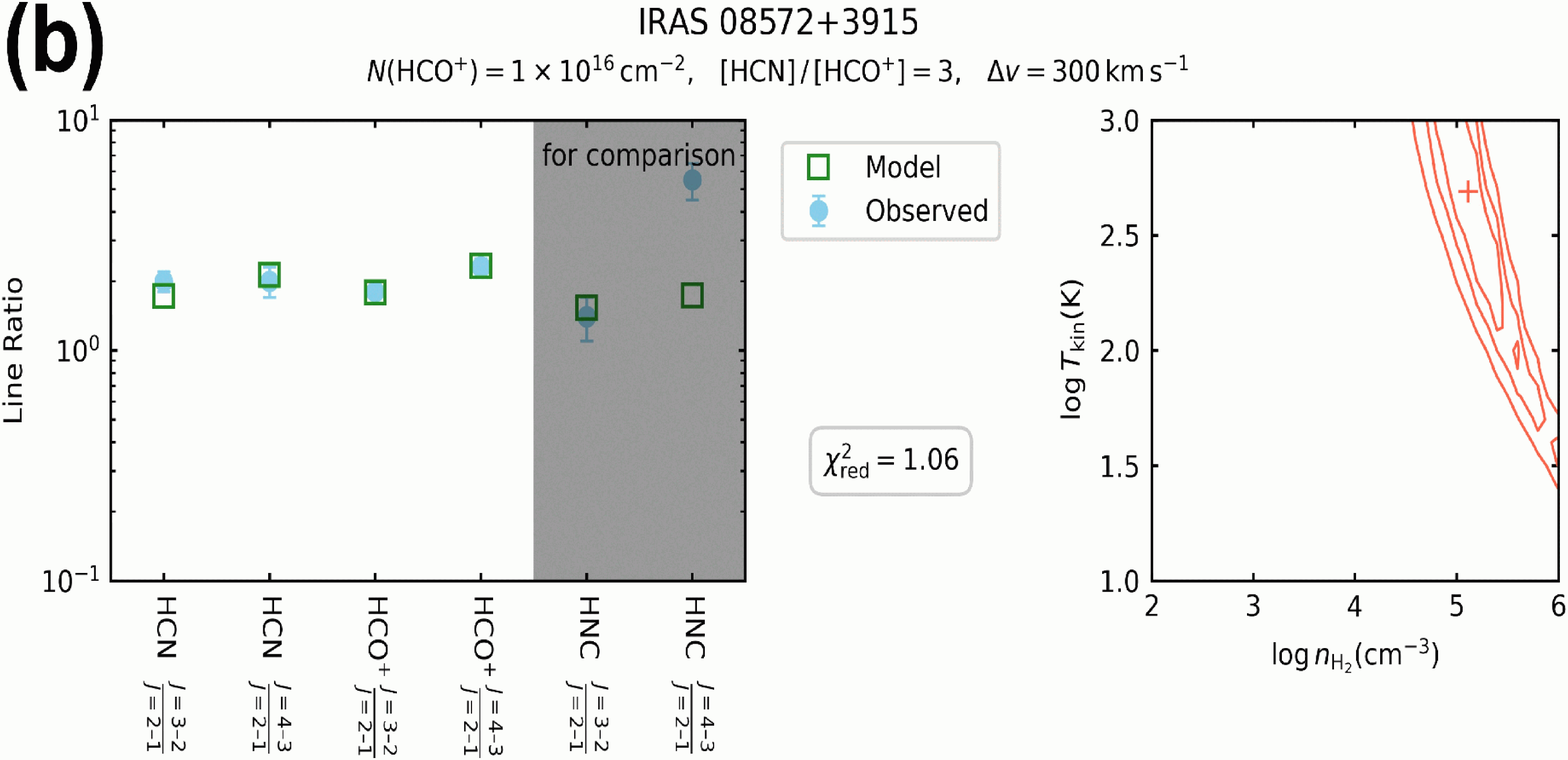} \\
%\hspace*{-6.6cm}
\includegraphics[angle=0,scale=.125]{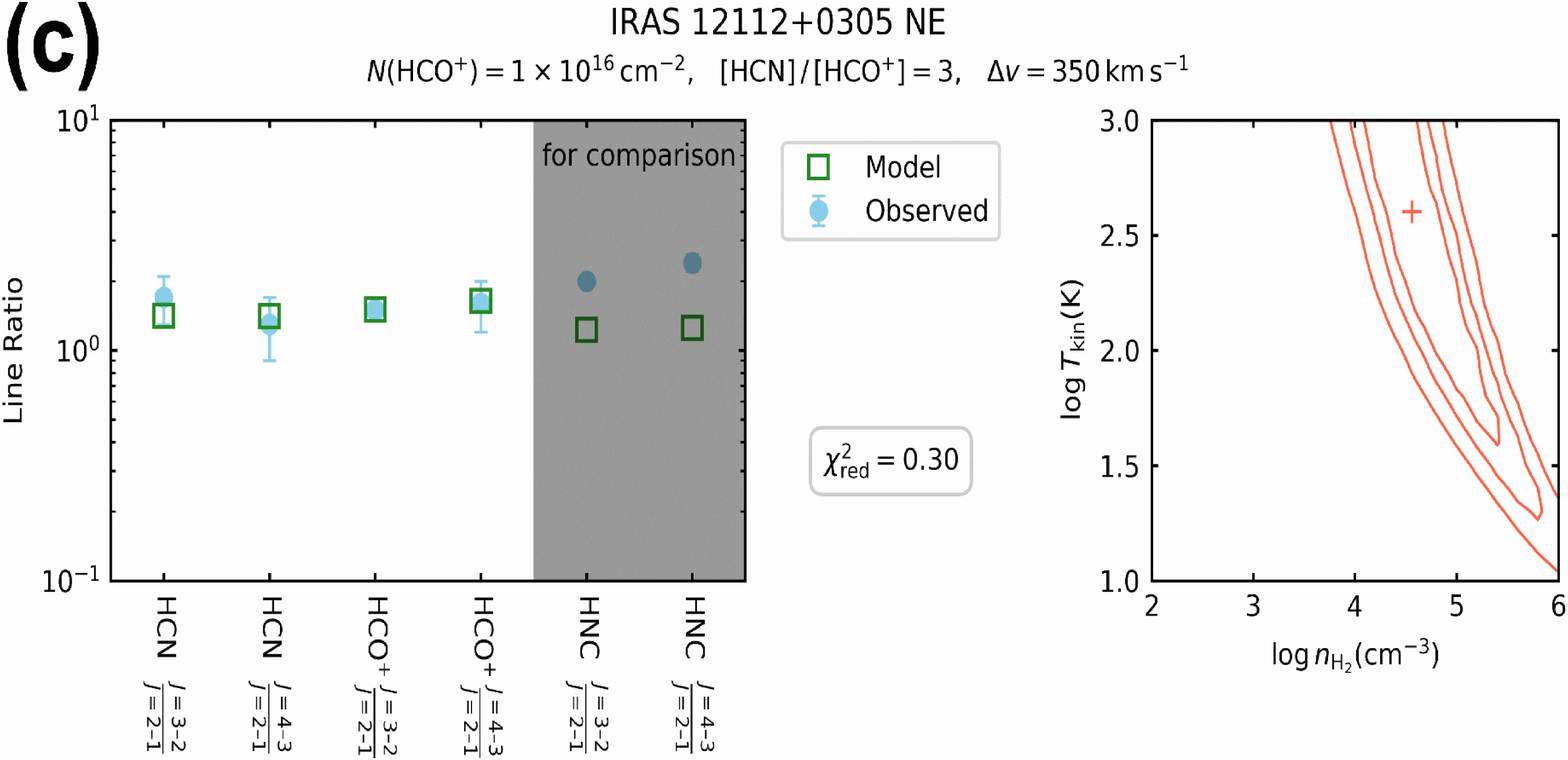} \hspace{0.5cm} 
\includegraphics[angle=0,scale=.125]{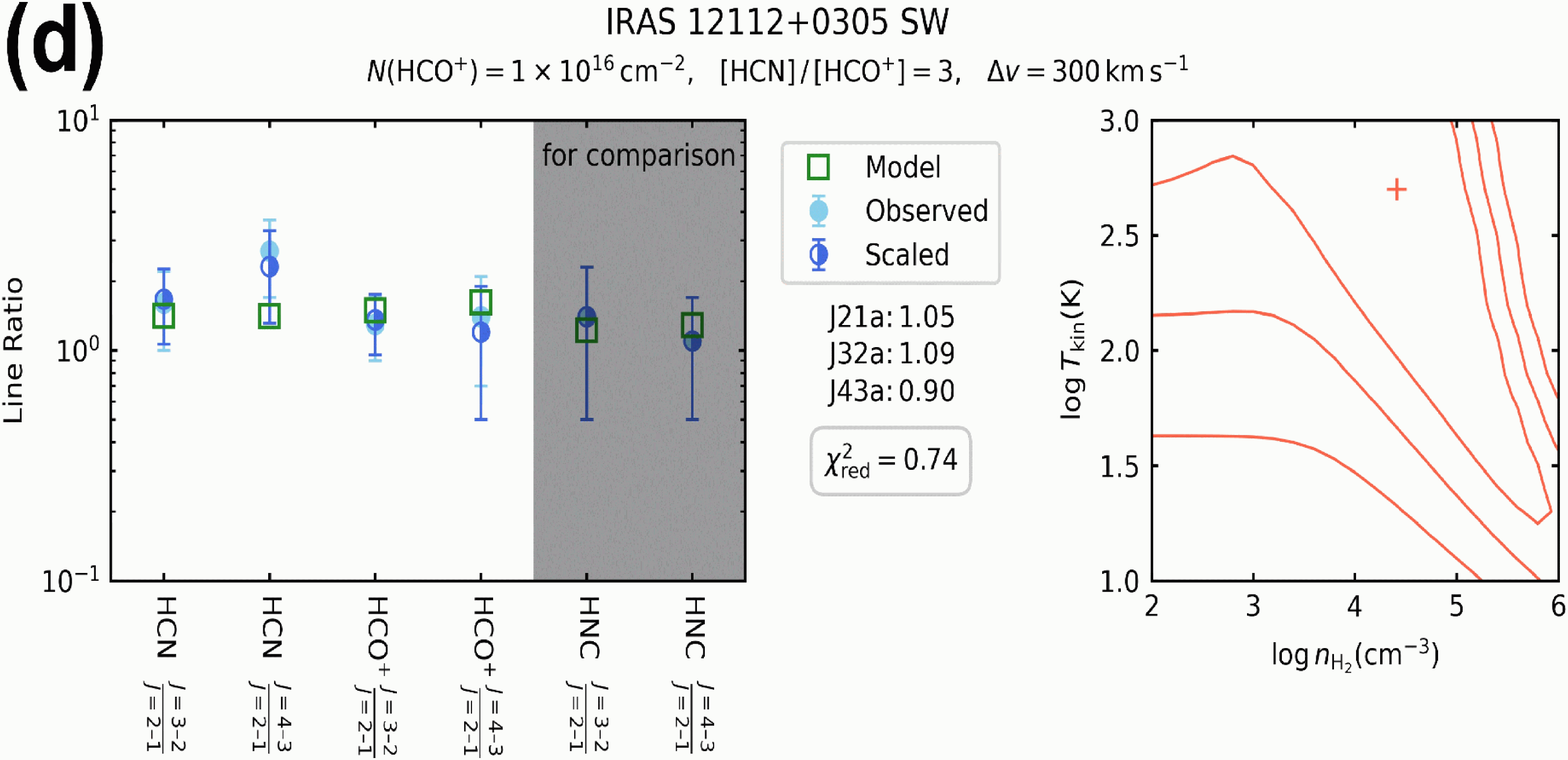} \\
%\hspace*{-6.6cm}
\includegraphics[angle=0,scale=.125]{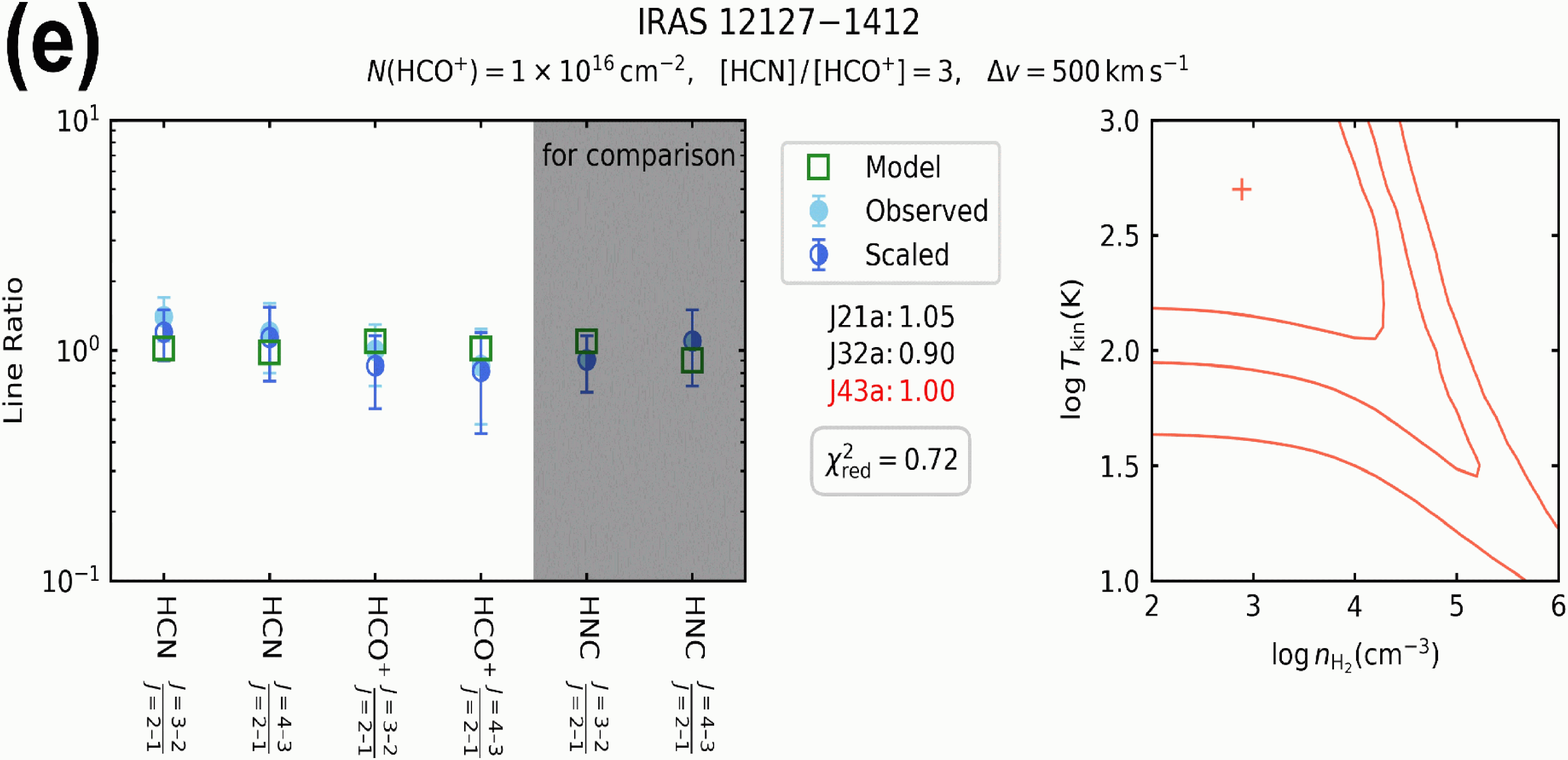} \hspace{0.5cm}
\includegraphics[angle=0,scale=.125]{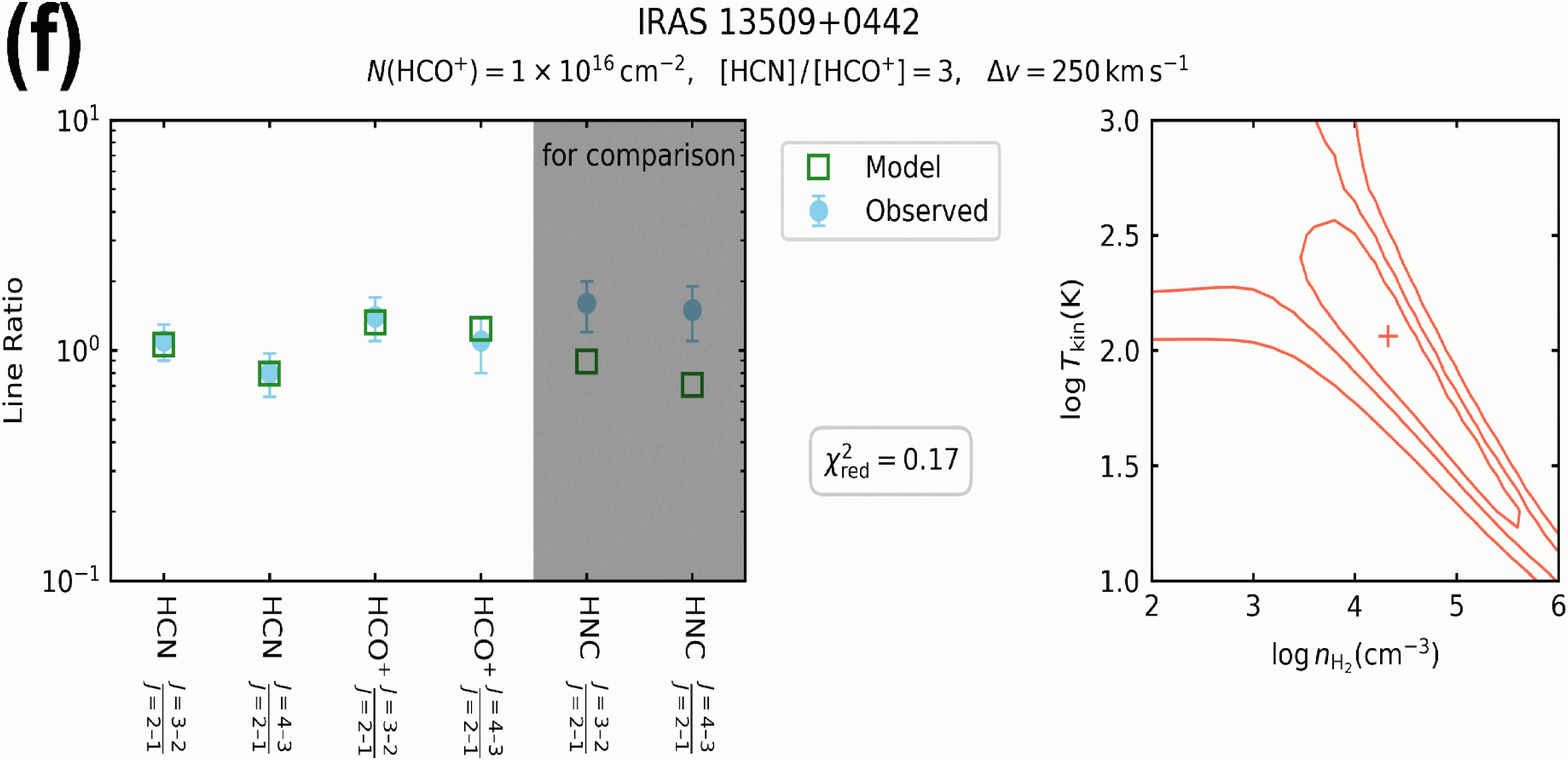} \\
%\hspace*{-6.6cm}
\includegraphics[angle=0,scale=.125]{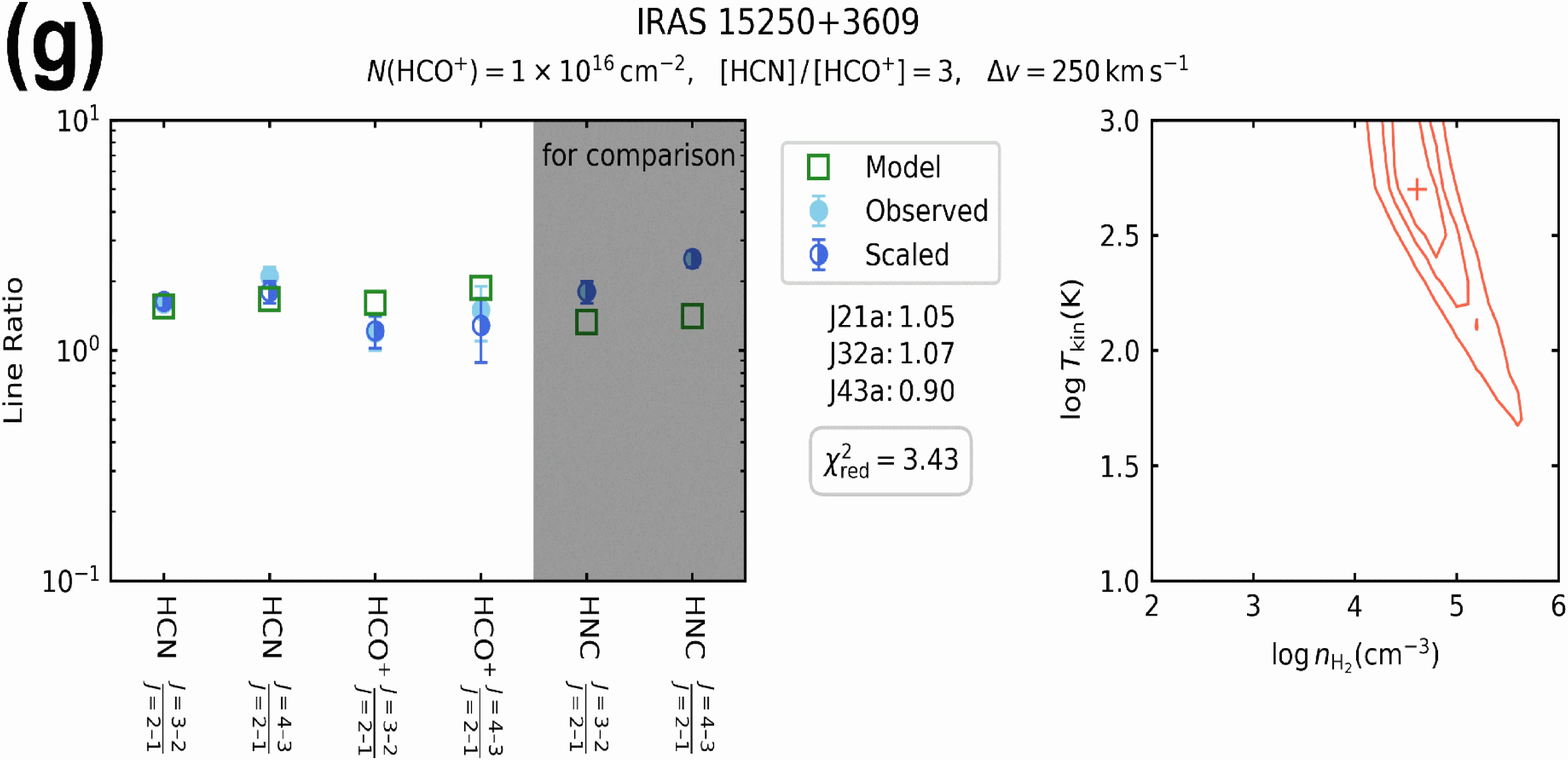} \hspace{0.5cm}
\includegraphics[angle=0,scale=.125]{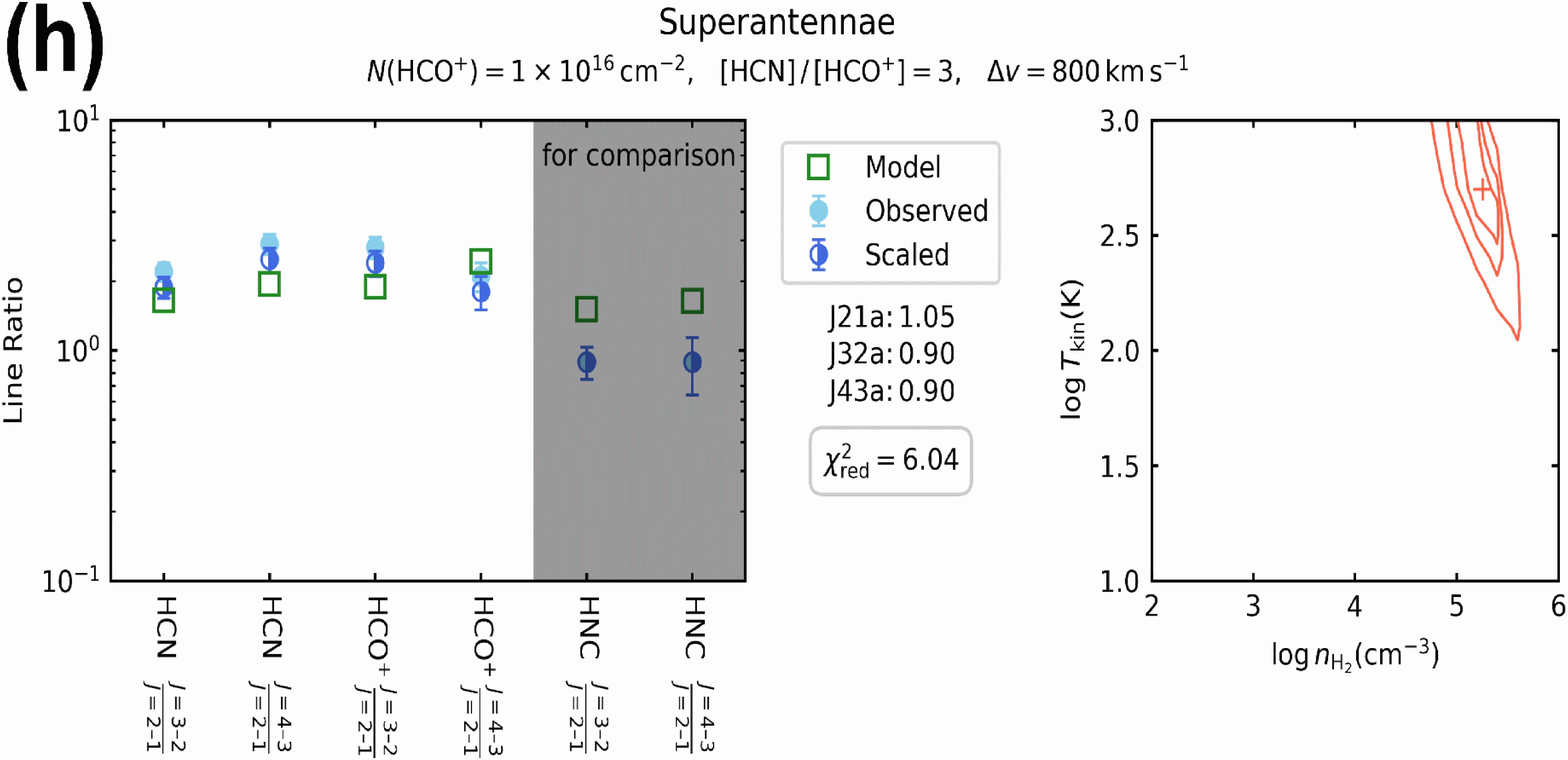} \\
%\hspace*{-6.6cm}
\includegraphics[angle=0,scale=.125]{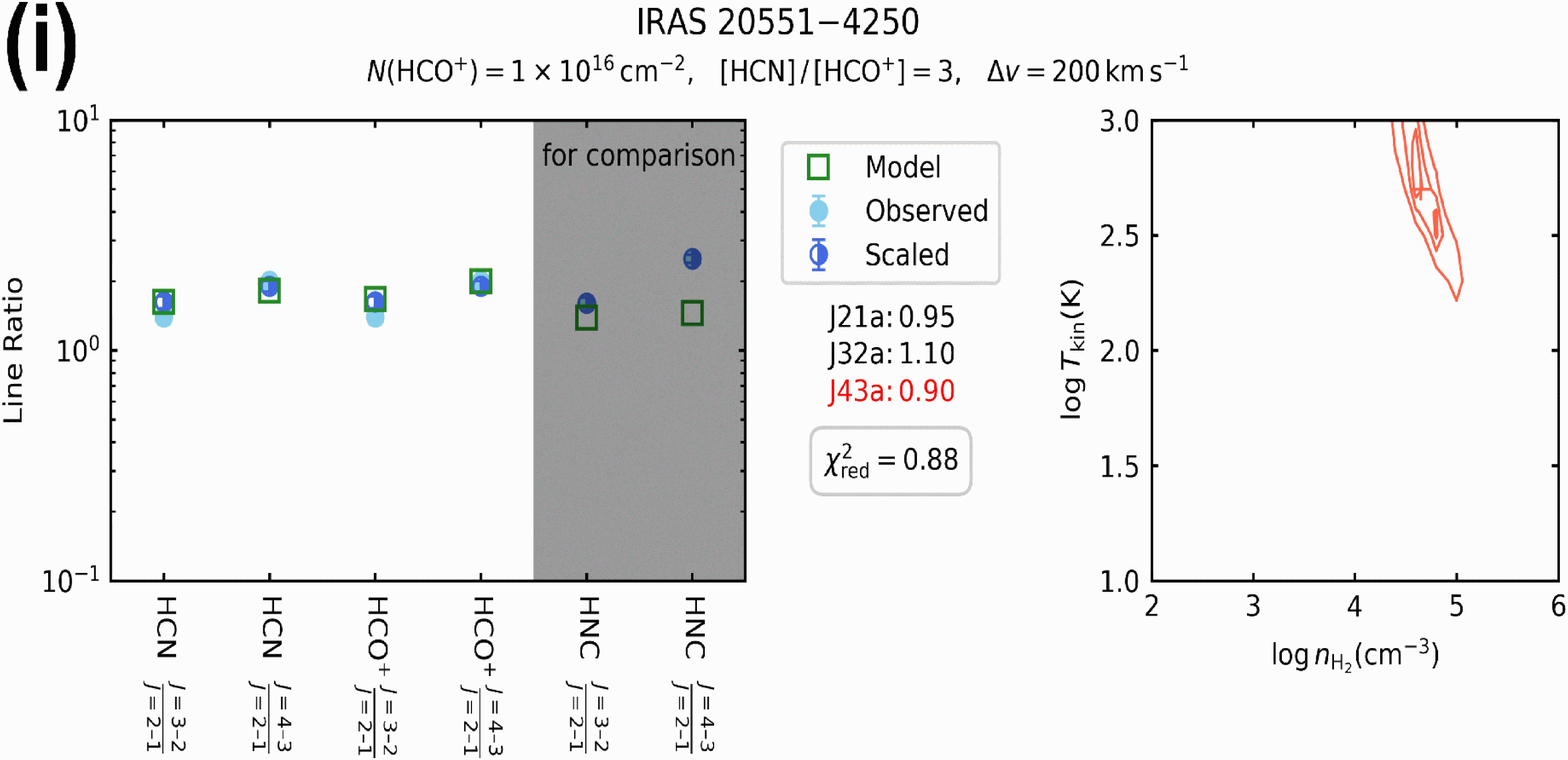} \hspace{0.5cm}
\includegraphics[angle=0,scale=.125]{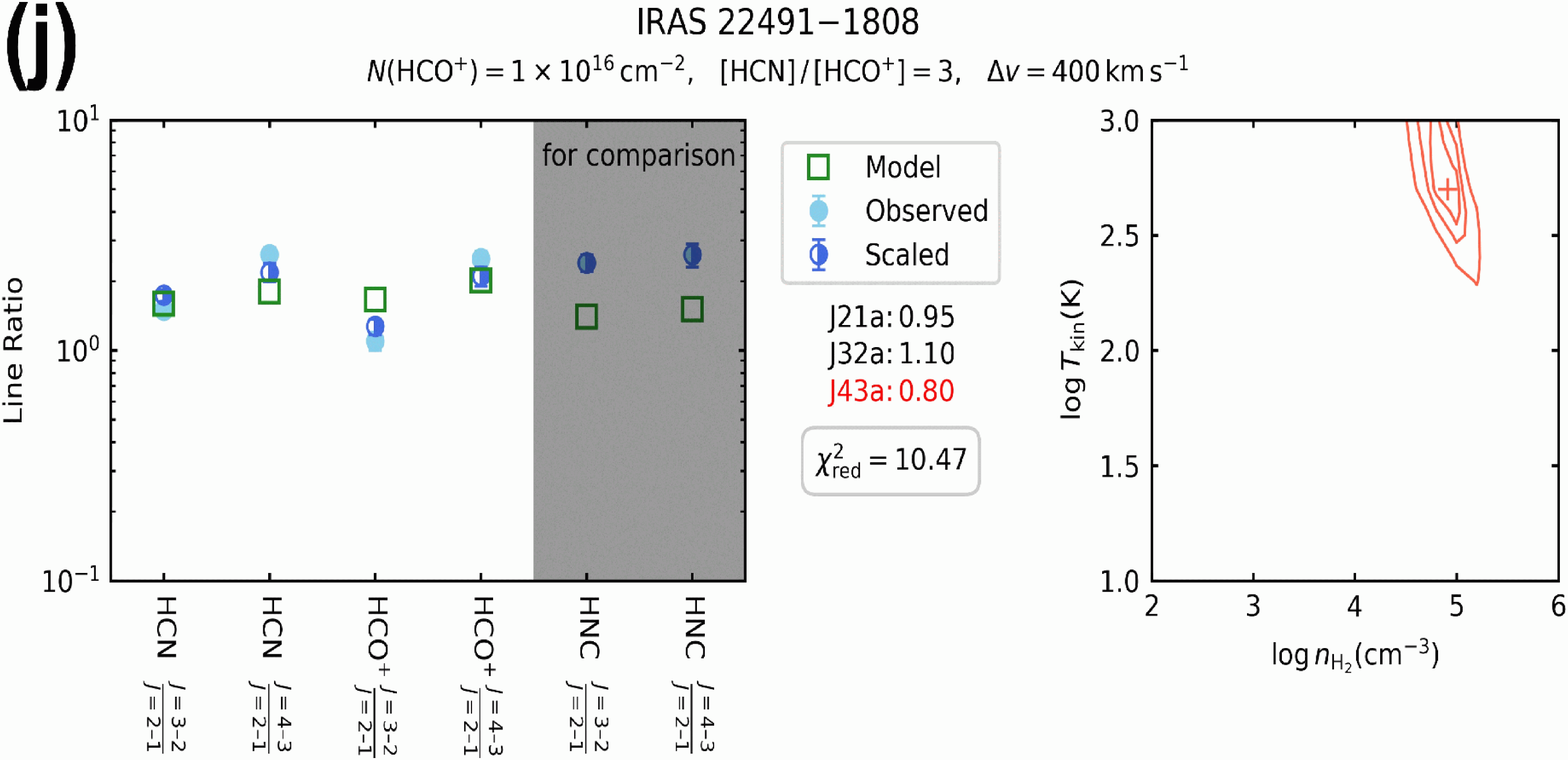} \\
\end{center}
%\end{figure}
%
%\begin{figure}[h]
\vspace{-0.6cm}
\caption{
Same as Figure \ref{fig:N1614fit}, but for ULIRGs' nuclei. 
HCO$^{+}$ column density of N$_{\rm HCO^{+}}$ = 1 $\times$ 10$^{16}$ cm$^{-2}$
and HCN-to-HCO$^{+}$ abundance ratio of [HCN]/[HCO$^{+}$] = 3 are
assumed for all the sources.
Line widths ($\Delta$v) are different among different ULIRGs and are
presented in Table \ref{tab:flux} (Column 11). 
Symbols are the same as those in Figure \ref{fig:N1614fit}.
We adopt the second fitting result (that is, flux-scaling adjustment
allowed) for IRAS 12112$+$0305 SW, IRAS 12127$-$1412, IRAS
15250$+$3609, the Superantennae, IRAS 20551$-$4250, and IRAS
22491$-$1808. 
\label{fig:ULIRGfit}
}
\end{figure}

Figure \ref{fig:N1614fit} shows that the derived gas density and
temperature are not strongly dependent on the choice of the 
HCN-to-HCO$^{+}$ abundance ratio ([HCN]/[HCO$^{+}$]) and HCO$^{+}$
column density (N$_{\rm HCO^+}$).
Thus, we perform the same fit for the remaining ULIRGs by adopting the
fiducial values of [HCN]/[HCO$^{+}$] = 3 and N$_{\rm HCO^+}$ = 1 
$\times$ 10$^{16}$ cm$^{-2}$. 
We adopt the line widths ($\Delta$v) listed in Table \ref{tab:flux}
(column 11) which are not the same among individual ULIRGs.
As in the case of NGC 1614, we 
(1) use the observed high-J to low-J flux ratios as they are, and 
(2) allow scale adjustment for individual J=4--3, J=3--2, and J=2--1 
data to consider the possible absolute flux calibration
uncertainty of individual ALMA observations.
The derived gas density and temperature largely agree between the
first and second fitting results, although the second one 
usually provides a smaller reduced $\chi^{2}$ value than
the former.  
We basically adopt the first fitting result, but do the second one 
only if the first one cannot determine the most likely value or
gives a very large reduced $\chi^{2}$ value. 
Figure \ref{fig:ULIRGfit} shows our adopted result for each
ULIRG's nucleus. 
Table \ref{tab:bestfit} summarizes the best fit values for the
ULIRGs.   

%%%%%%%%%% Figure 5 %%%%%%%%%
%%%%%%%%%%%%%%%%%%%%%%%%%%%%%%%%%%%

%%%%%%%%%% Table 6 (bestfit) %%%%%%%%%
\begin{deluxetable}{cccccccc}[!hbt]
%\tabletypesize{\normalsize}
\tablecaption{Summary of the Best Fit Values \label{tab:bestfit}}
\tablewidth{0pt}
\tablehead{\colhead{Object} & \colhead{log N$_{\rm HCO+}$} &
\colhead{[HCN]/[HCO$^{+}$]} & \colhead{Scaling} & 
\colhead{log n$_{\rm H_2}$} & \colhead{log T$_{\rm kin}$} & 
\colhead{Reduced} & \colhead{Remark} \\
\colhead{} & \colhead{[cm$^{-2}$]} & \colhead{} & \colhead{} 
& \colhead{[cm$^{-3}$]} & \colhead{[K]} & \colhead{$\chi^{2}$}
& \colhead{} \\ 
\colhead{(1)} & \colhead{(2)} & \colhead{(3)} & \colhead{(4)} &
\colhead{(5)} & \colhead{(6)} & \colhead{(7)} & \colhead{(8)}  
} 
\startdata
NGC 1614  & 16 & 3 & off & 3.4$^{+0.3}_{-0.5}$ & 2.1$\pm$0.1 & 1.9
& Figure \ref{fig:N1614fit}a \\ 
          & 16 & 1 & off & 3.4$^{+0.3}_{-0.4}$ & 2.1$\pm$0.1 & 2.9
& Figure \ref{fig:N1614fit}b \\ 
          & 15 & 1 & off & 4.1$^{+0.3}_{-0.6}$ & 2.3$^{+0.2}_{-0.1}$ & 5.6
& Figure \ref{fig:N1614fit}c \\ 
          & 16 (16.0$^{+0.4}_{-0.5}$) & 1 (3.5$\pm$0.5) & on (on) &
3.9$^{+0.2}_{-0.3}$ (3.3$\pm$0.8) & 2.0$\pm$0.1 (2.0$\pm$0.2) &
0.39 (1.0) & Figure \ref{fig:N1614fit}d,\ref{fig:MCMC}a \\  
IRAS 06035$-$7102 & 16 (16.6$^{+0.3}_{-0.4}$) & 3
(4.7$^{+1.1}_{-1.4}$) & off (on) & 5.3$\pm$0.2 (5.5$^{+0.3}_{-0.4}$) &
2.7$^{+\infty}_{-0.3}$ (2.0$^{+0.6}_{-0.3}$) & 1.3
(4.2) & Figure \ref{fig:ULIRGfit}a,\ref{fig:MCMC}b \\  
IRAS 08572$+$3915 & 16 (16.4$^{+0.4}_{-0.5}$) & 3
(4.9$^{+1.3}_{-1.5}$) & off (on) & 5.1$\pm$0.3 (5.2$^{+0.4}_{-0.5}$) &
2.7$^{+\infty}_{-0.5}$ (2.0$^{+0.6}_{-0.4}$) & 1.1 (1.7) & Figure
\ref{fig:ULIRGfit}b,\ref{fig:MCMC}c \\   
IRAS 12112$+$0305 NE & 16 (15.7$^{+0.8}_{-1.0}$) & 3
(8.7$^{+1.0}_{-1.4}$) & off (on) & 4.6$^{+0.5}_{-0.4}$ (4.6$^{+0.4}_{-0.7}$) & 
2.6$^{+\infty}_{-0.7}$ (2.7$^{+0.2}_{-0.3}$) & 0.30 (2.1) & Figure
\ref{fig:ULIRGfit}c,\ref{fig:MCMC}d \\  
IRAS 12112$+$0305 SW & 16 & 3 & on & 4.4$^{+0.6}_{-0.7}$ &
2.7$^{+\infty}_{-0.7}$ & 0.74 & Figure \ref{fig:ULIRGfit}d \\ 
IRAS 12127$-$1412 & 16 & 3 & on & 2.9$^{+1.1}_{-\infty}$ &
2.7$^{+\infty}_{-0.4}$ & 0.72 & Figure \ref{fig:ULIRGfit}e \\ 
IRAS 13509$+$0442 & 16 (15.7$\pm$0.8) & 3 (6.3$^{+1.9}_{-2.2}$) & off
(on) & 4.3$^{+1.0}_{-0.5}$ (4.3$^{+0.9}_{-1.2}$) & 2.1$^{+0.3}_{-0.7}$
(2.1$\pm$0.6) & 0.17 (1.3) & Figure \ref{fig:ULIRGfit}f,\ref{fig:MCMC}e \\  
IRAS 15250$+$3609 & 16 & 3 & on & 4.6$\pm$0.2 & 2.7$^{+0.3}_{-0.2}$ &
3.4 & Figure \ref{fig:ULIRGfit}g \\ 
Superantennae & 16 & 3 & on & 5.3$^{+0.1}_{-0.2}$ & 2.7$^{+0.3}_{-0.1}$ &
6.0 & Figure \ref{fig:ULIRGfit}h \\ 
IRAS 20551$-$4250 & 16 (15.4$^{+0.6}_{-0.2}$) & 3 (2.9$\pm$0.2) & on
(on) & 4.6$\pm$0.1 (4.5$^{+0.3}_{-0.2}$) & 2.7$^{+0.2}_{-0.1}$
(2.8$^{+0.1}_{-0.2}$) & 0.88 (8.3) & Figure
\ref{fig:ULIRGfit}i,\ref{fig:MCMC}f \\   
IRAS 22491$-$1808 & 16 & 3 & on & 4.9$\pm$0.1 &
2.7$^{+0.3}_{-0.1}$ & 10.5 & Figure \ref{fig:ULIRGfit}j \\  
\enddata

\tablecomments{
Col.(1): Object name. 
Col.(2): Decimal logarithm of HCO$^{+}$ column density in units of cm$^{-2}$.
Col.(3): HCN-to-HCO$^{+}$ abundance ratio. 
Col.(4): Scaling on or off.
Col.(5): Decimal logarithm of H$_{2}$ gas density in units of cm$^{-3}$. 
Col.(6): Decimal logarithm of gas kinetic temperature in units of K. 
In Cols.(2)--(6), the best fit values based on the Bayesian
method ($\S$5.5) are displayed in parentheses for selected (U)LIRGs.
For very high HCO$^{+}$ column density and HCN-to-HCO$^{+}$
abundance ratio, HCN emission line becomes very optically thick and
its flux does not change. 
Thus, some of the derived HCO$^{+}$ column density and/or HCN-to-HCO$^{+}$
abundance ratio could be lower limits.
Col.(7): Reduced $\chi^{2}$ value.
Col.(8): Remark. Corresponding figure information is added.
}

\end{deluxetable}
%%%%%%%%%%%%%%%%%%%%%%%%%%%%%%%%%%%

\clearpage

\subsection{Overall Trend of Molecular Gas Properties at (U)LIRGs' Nuclei}

In Figure \ref{fig:ULIRGfit}, the uncertainty in the derived gas density 
and temperature is large for IRAS 12112$+$0305 SW and IRAS 12127$-$1412
simply because molecular line emission is fainter than other
(U)LIRGs' nuclei (Figure \ref{fig:profile}).  
Excluding these two ULIRGs' nuclei, we see the following two trends in 
Figures \ref{fig:N1614fit} and \ref{fig:ULIRGfit}; 
(1) The only one LIRG NGC 1614 contains nuclear molecular gas with
modestly high density (10$^{3.4-4.1}$ cm$^{-3}$) and high temperature
(100--200 K or 10$^{2.0-2.3}$ K). 
(2) Nuclear molecular gas probed with HCN and HCO$^{+}$ at J=2--1 to 
J=4--3 in all ULIRGs but IRAS 13509$+$0442 is significantly 
denser (10$^{4.5-5.5}$ cm$^{-3}$) and warmer ($\gtrsim$300 K or
$\gtrsim$10$^{2.5}$ K) than that in the LIRG NGC 1614.  
These two results constitute our first main arguments.
The gas density and temperature at these ULIRGs' nuclei are 
also systematically higher than those of other LIRGs in the
literature, derived from larger beam-sized CO multiple J-transition 
(J=1--0 to J=6--5) line data and RADEX modeling 
(mostly 10$^{2.0-4.0}$ cm$^{-3}$ and 10--90 K) \citep[e.g.,][]{pap12b}. 
IRAS 13509$+$0442 has relatively low density ($\sim$10$^{4.3}$
cm$^{-3}$) and temperature ($\sim$130 K or $\sim$10$^{2.1}$ K) nuclear
molecular gas compared to other ULIRGs. 
Although there are certain difference among the observed sources, the 
derived gas density and temperature of all (U)LIRGs' nuclei are
substantially higher than those of quiescently star-forming, less
infrared luminous galaxies at $\lesssim$kpc scale, suggested from
the observed small high-J to low-J emission line flux ratios of HCN
and HCO$^{+}$ \citep{gar23} or derived from CO multiple J-transition 
(J=1--0 to J=3--2) line data and RADEX modeling with fixed
N$_{\rm CO}$/$\Delta$v ratio ($\lesssim$10$^{4.0}$ cm$^{-3}$ and
$\lesssim$50 K) \citep[e.g.,][]{ler22}.  

\begin{figure}[h]
\begin{center}
%\hspace*{-8cm}
\includegraphics[angle=0,scale=.26]{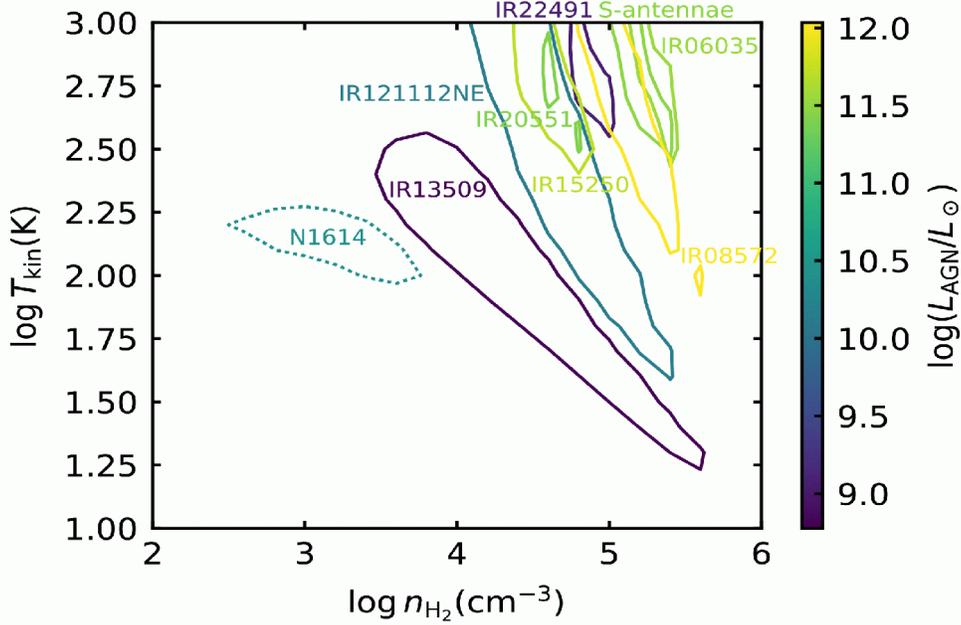} 
\end{center}
%\vspace{-1.5cm}
\caption{
Summary of contours of the RADEX-derived molecular gas density in
cm$^{-3}$ (abscissa) and temperature in K (ordinate) for the observed
(U)LIRGs, excluding the two faint molecular emission line ULIRGs (IRAS
12112$+$0305 SW and IRAS 12127$-$1412). 
Color corresponds to infrared-derived AGN luminosity in units of solar
luminosity (L$_{\odot}$), as listed in Table \ref{tab:objects}
(columns 9 and 11).
NGC 1614 is shown as a dotted line because the upper limit of AGN
luminosity is looser than that of other ULIRGs.
For all the ULIRGs, the AGN luminosity is derived by \citet{nar10} in
a consistent manner, whereas that of NGC 1614 is estimated in a
different method by a different group \citep{per15}. 
\label{fig:AGNvsSB}
}
\end{figure}

NGC 1614 and IRAS 13509$-$0442, together with IRAS 12112$+$0305 SW,
are classified as starburst-dominated with no luminous AGN signature 
at any wavelength, whereas the remaining ULIRGs are regarded as
containing luminous AGNs based on previous observations ($\S$2).
In other words, there is a trend that nuclear molecular gas is denser
and warmer in ULIRGs with luminous AGN signatures than in (U)LIRGs
without.  
This is our second main argument.
Figure \ref{fig:AGNvsSB} shows the contours of the derived gas density
and temperature for (U)LIRGs with small uncertainty (excluding IRAS 
12112$+$0305 SW and IRAS 12127$-$1412).
It has previously been suggested that AGN activity is enhanced in
galaxy nuclei with a larger amount of and/or higher fraction of dense
molecular gas \citep{jun09,izu16b}. 
This appears reasonable because the mass accretion rate onto a SMBH is 
enhanced by a larger amount of dense molecular gas near the SMBH.
Owing to the high SMBH mass accretion, luminous AGN activity can
emerge and make the surrounding dense molecular gas warm.
The second argument can be explained naturally by the combination of
these two effects. 

%%%%%%%%%% Figure 6 %%%%%%%%%
%%%%%%%%%%%%%%%%%%%%%%%%%%%%%%%%%%%

For the three ULIRGs, IRAS 08572$+$3915, IRAS 12112$+$0305, and IRAS
22491$-$1808, molecular gas temperature was estimated to be 
60--150 K, based on large beam-sized CO multiple J-transition line
observations and RADEX modeling \citep{pap12a}.
Our ALMA HCN and HCO$^{+}$ line study provides significantly  
higher temperature for nuclear molecular gas in all the three ULIRGs 
(see Figure \ref{fig:ULIRGfit} and Table \ref{tab:bestfit}).
This suggests that our ALMA 1--2 kpc resolution dense
molecular line observations selectively probe warm nuclear molecular
gas components around luminous energy sources with minimum 
contamination from spatially extended ($\gtrsim$a few kpc) cool
components.
This advantage was also demonstrated in ALMA high-spatial-resolution
HCN and HCO$^{+}$ observations of the nearby well-studied, optically 
identified AGN NGC 1068, to probe molecular gas only in the vicinity
of a luminous energy source \citep[e.g.,][]{vit14}. 
\citet{bab18} revealed the presence of high temperature ($\gtrsim$200 K)
molecular gas at the innermost part of obscuring material around 
luminous AGNs in nearby ULIRGs, based on the observed properties of 
$\sim$4.6 $\mu$m CO ro-vibrational absorption lines detected in
the infrared 4--5 $\mu$m spectra. 
It is likely that HCN and HCO$^{+}$ rotational emission at J=2--1 to
J=4--3 detected in our ALMA high-angular-resolution observations
largely originates from such innermost high temperature molecular gas
at nearby ULIRGs' nuclei. 

\subsection{Bayesian Analysis of Molecular Emission Line Flux Ratios}

For selected (U)LIRGs' nuclei exhibiting bright molecular emission 
lines with high detection significance, molecular gas
parameters are estimated by simultaneously fitting HCN-to-HCO$^+$
and high-J to low-J emission line flux ratios using RADEX
calculations, by making all parameters free, based on a Bayesian
approach. 
The aim is to verify whether the Bayesian approach provides 
consistent results regarding gas density and temperature with those
derived from the previous Levenberg-Marquardt method, where the
HCO$^{+}$ column density and HCN-to-HCO$^{+}$ abundance ratios were
fixed to fiducial values ($\S$5.3).
There are five independent emission line flux ratios 
(HCN-to-HCO$^+$ flux ratios at J=2--1, J=3--2, and J=4--3, and HCO$^{+}$
J=3--2 to J=2--1 and J=4--3 to J=2--1 flux ratios), which are lower
than the total number of parameters, including the absolute flux
calibration scaling factors. 
However, Bayesian estimation facilitates the sampling of the posterior
probability distribution of parameters, considering indeterminacy
of the solution. 

Using the five emission line flux ratios, we explore the parameter
space using a Markov Chain Monte Carlo (MCMC) sampler; \texttt{emcee}
package \citep{emcee} is used in \texttt{lmfit}. 
Flat priors with upper and lower bounds presented in Table
\ref{tab:bounds} are adopted. 
The first guess is made by running the L-BFGS-B solver in $\S$5.3, with
the HCO$^{+}$ column density and HCN-to-HCO$^{+}$ abundance ratio
unfixed, and then 100 walkers are initialized to be distributed around
that guess. 
The chain is run for 100$\tau$ steps, where $\tau$ is the maximum number of
integrated autocorrelation time for each parameter
%---
\footnote{
Ideally, $\tau$ should be derived from a chain of infinite lengths.
In our framework, $\tau$ generally increases with the number of steps 
in the chain ($N$) and converges asymptotically to a reliable value.
Therefore, we lengthen the chain while monitoring $\tau$
and terminate the calculation when $N/\tau$ exceeds 100.
This is a more conservative option than the minimum chain length 
recommended in the \texttt{emcee} documentation ($N/\tau=50$).
}.
%---
The first 5$\tau$ steps are discarded as a ``burn-in'' phase.
Subsequently, the chain is thinned out in 0.5$\tau$ steps to create
independent samples. 
Hence, the effective number of sampling of the posterior probability
distribution is $100\times(100-5)/0.5=19,\!000$.

\begin{figure}[h]
\begin{center}
%\vspace*{1cm}
%\hspace*{-4.6cm}
\includegraphics[angle=0,scale=.234]{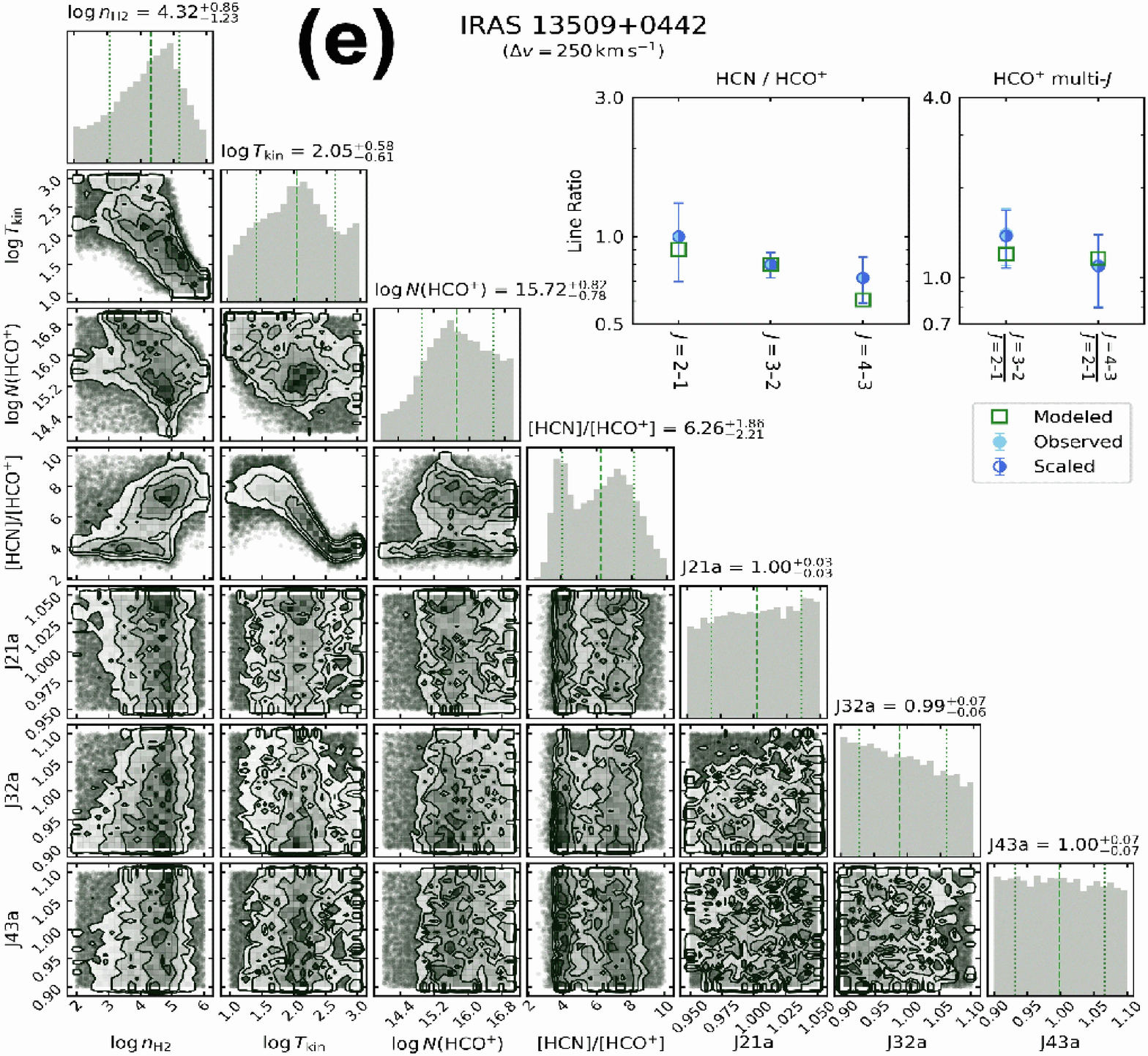} 
\includegraphics[angle=0,scale=.234]{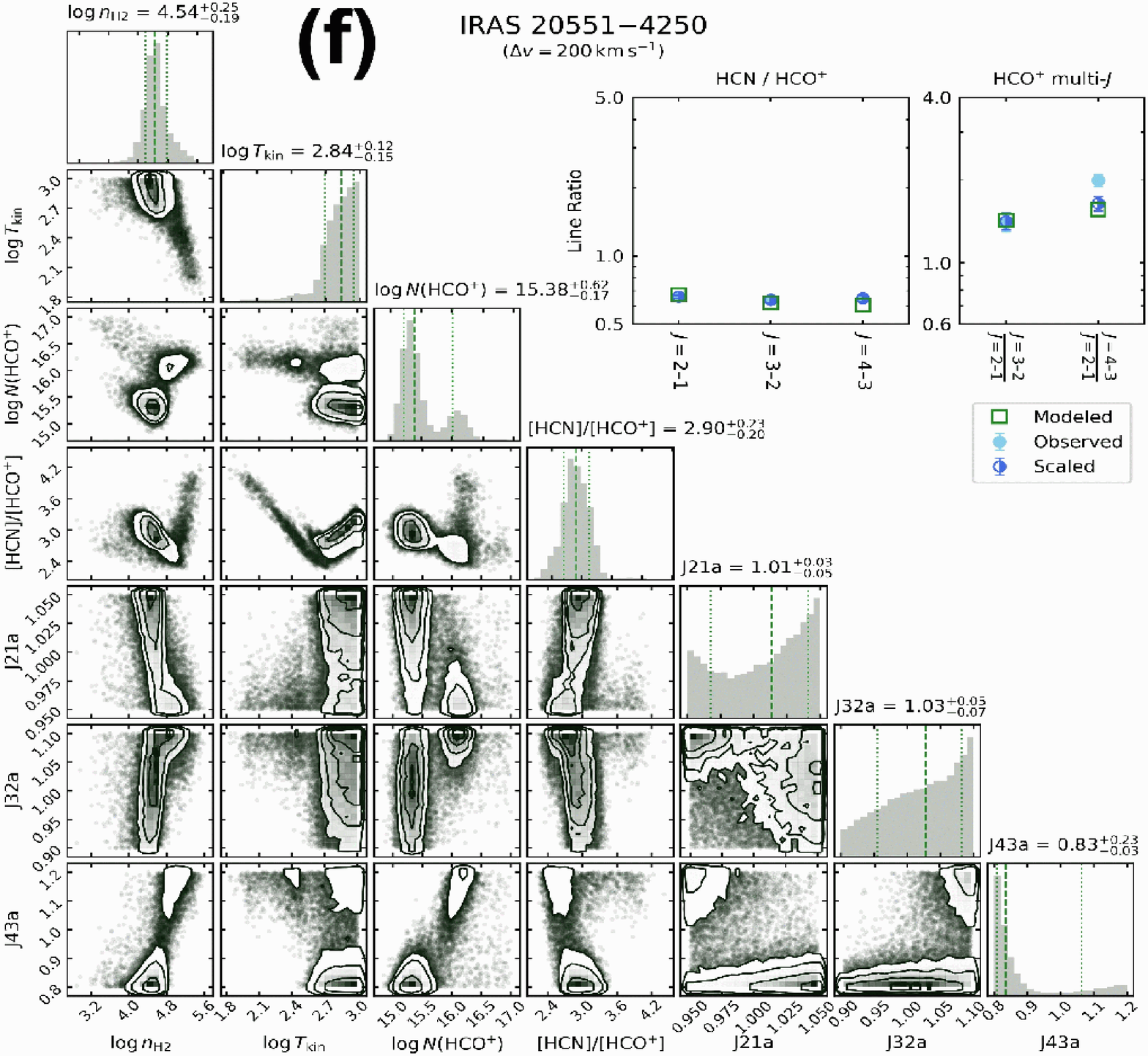} \\
\end{center}
%\vspace{-1.0cm}
\caption{
Example results of the MCMC analysis of the HCN-to-HCO$^+$ line flux ratios
and HCO$^+$ high-J to low-J line flux ratios. 
Shown in the left is a corner plot which displays the posterior
probability distribution of each parameter and the covariance of each
parameter pair as 1D and 2D histograms, respectively.
The median and 68\% credible bounds for each parameter posterior are
indicated by the vertical dashed and dotted lines, respectively, and
are also given in the panel title. 
A comparison of the line flux ratios is shown in the upper right panel.
The green open squares show the model flux ratios calculated at the 
medians of log n$_{\rm H_2}$, log T$_{\rm kin}$, log N$_{\rm HCO^+}$, and
[HCN]/[HCO$^+$].  
The light blue filled circles indicate the flux ratios as observed and
the dark blue half-filled circles indicate line flux ratios scaled 
from the observed values using the medians of the scaling factors for
J21a, J32a, and J43a. 
The residuals used for the likelihood are the differences between the
model and scaled line flux ratios, normalized by the statistical
uncertainty in the latter ratio. 
\label{fig:MCMC}
}
\end{figure}

For our MCMC analysis, IRAS 12112$-$1412 SW and IRAS 12127$-$1412 are
excluded, because of their faint molecular emission lines.  
Examples of the MCMC analysis results for six selected (U)LIRGs' 
nuclei are shown in Figure \ref{fig:MCMC} and the best fit values are
summarized in Table \ref{tab:bestfit}. 
Results for IRAS 15250$+$3609, the Superantennae, and IRAS
22491$-$1808 are not shown because we are unable to provide 
model-calculated results which reproduce the observed results very
well with small systematic discrepancy.
This independent MCMC analysis supports the previous arguments  
that 
(1) molecular gas at the (U)LIRGs' nuclei is dense ($\gtrsim$10$^{3-4}$
cm$^{-3}$) and warm ($\gtrsim$100 K) compared to quiescently
star-forming normal galaxies with less infrared luminosity and 
(2) the density of nuclear molecular gas is higher in ULIRGs
($\gtrsim$10$^{4.3}$ cm$^{-3}$) than in the LIRG NGC 1614 
($\sim$10$^{3.3}$ cm$^{-3}$).  

%%%%%%%%%% Figure 7 %%%%%%%%%
\begin{figure}[h]
\begin{center}
%\vspace*{1cm}
%\hspace*{-4.6cm}
\includegraphics[angle=0,scale=.234]{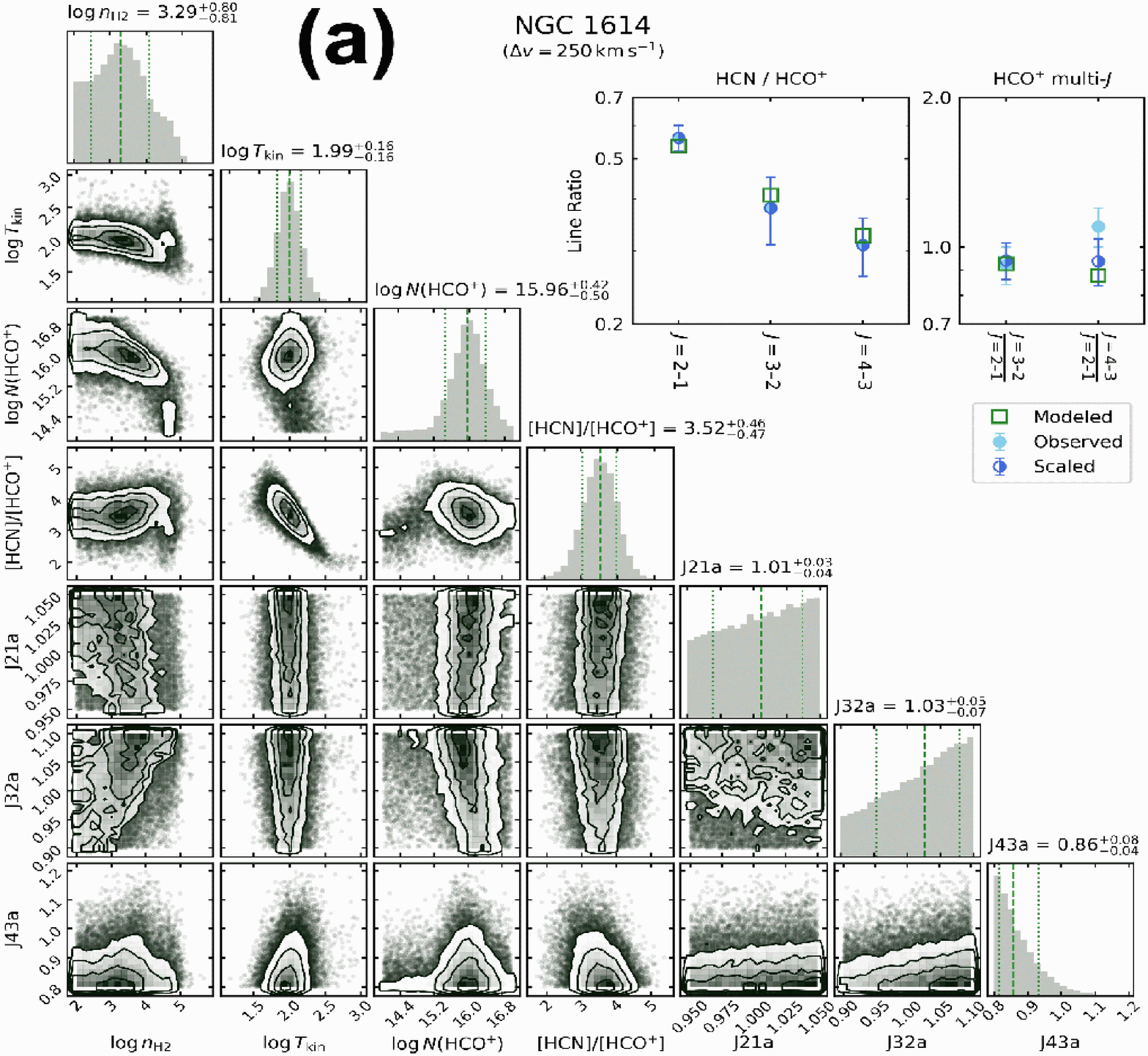} 
\includegraphics[angle=0,scale=.234]{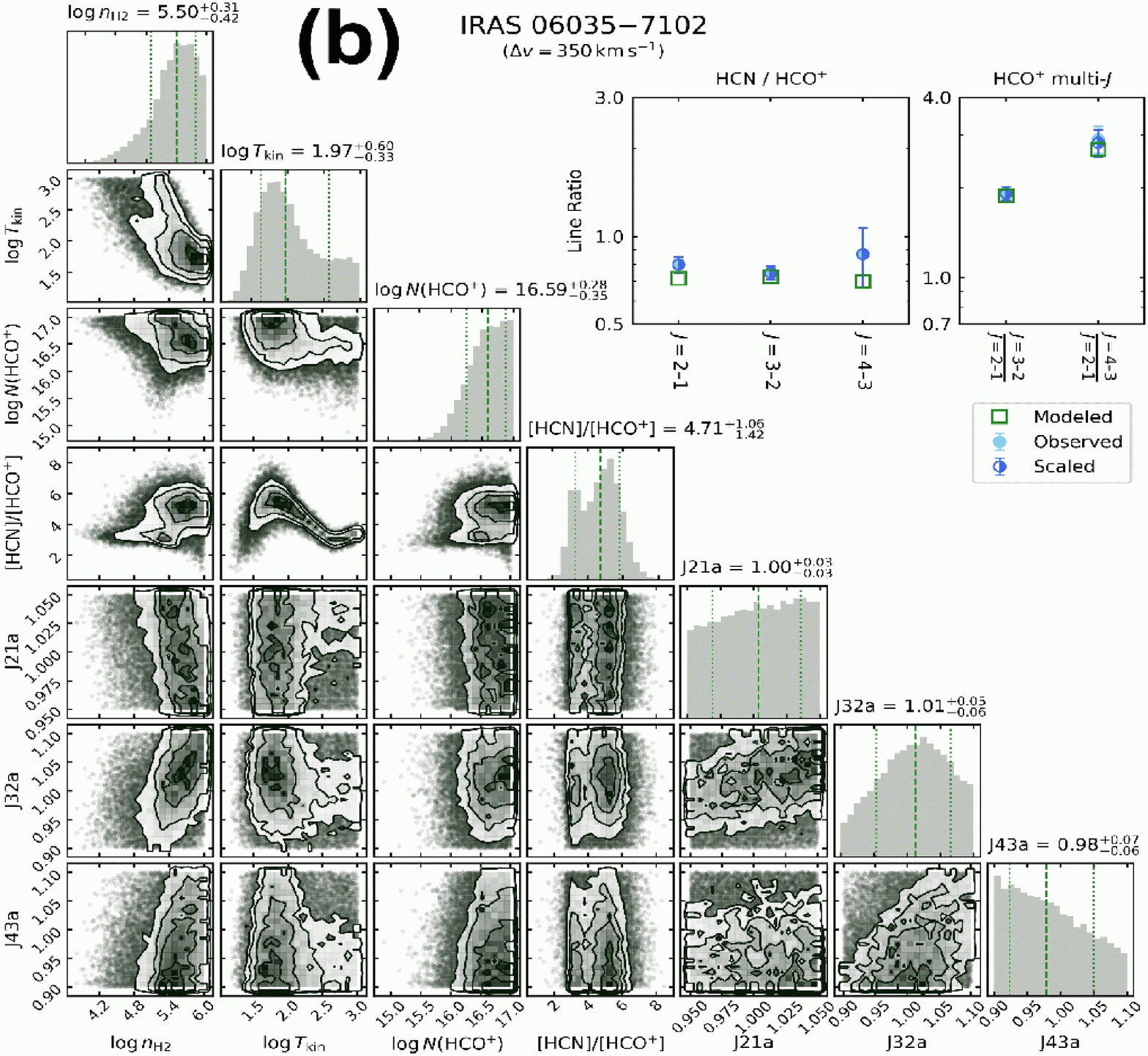} \\
\vspace*{1cm}
%\hspace*{-4.6cm}
\includegraphics[angle=0,scale=.234]{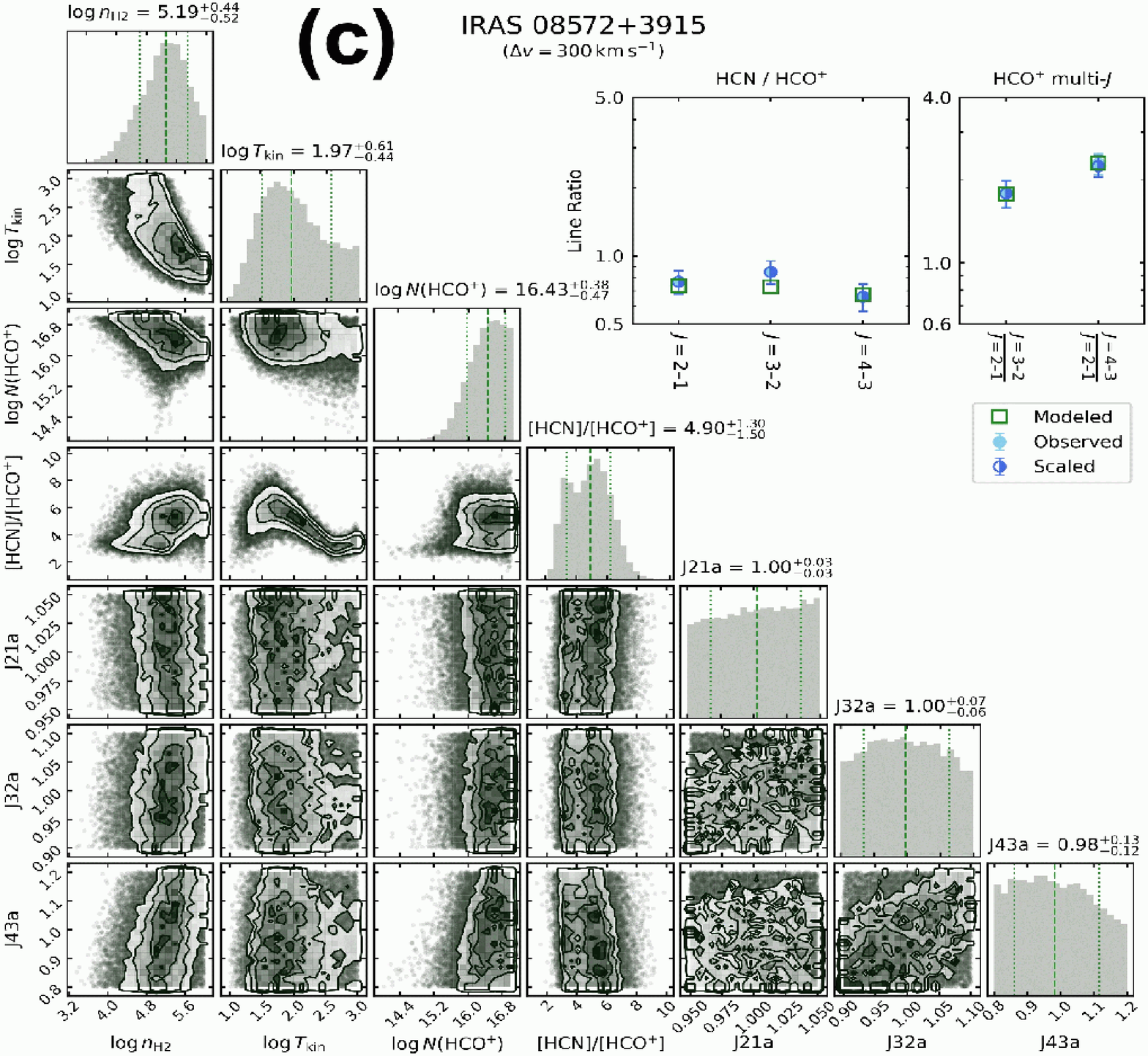} 
\includegraphics[angle=0,scale=.234]{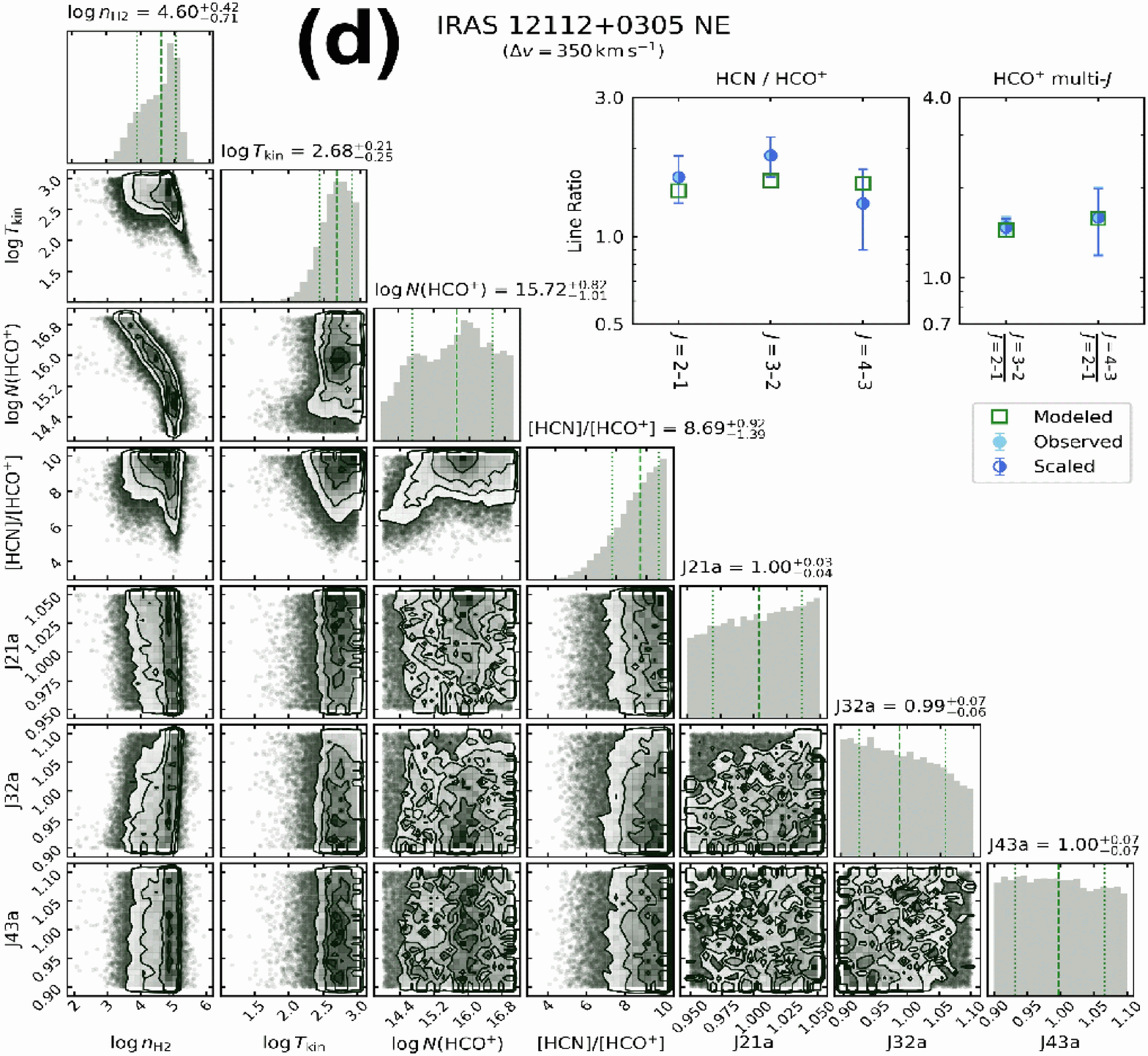} \\
\end{center}
\end{figure}

%%%%%%%%%%%%%%%%%%%%%%%%%%%%%%%%%%%

%%%%%%%%%% Table 7 (bounds) %%%%%%%%%
\begin{deluxetable}{ccc}[!hbt]
\tabletypesize{\small}
\tablecaption{Bounds of the Flat Priors\label{tab:bounds}}
\tablewidth{0pt}
\tablehead{\colhead{Parameter} & \colhead{Lower} & \colhead{Upper}}
\startdata
log(n$_{\rm H_2}$/cm$^{-3}$)    &  2    &  6    \\
log(T$_{\rm kin}$/K)          &  1    &  3    \\
log(N$_{\rm HCO^+}$/cm$^{-2}$) & 14    & 17    \\
% [HCN]/[HCO$^+$]                            &  0.1  & 10    \\
$\mathrm{[HCN]/[HCO^+]}$ & 0.1 & 10 \\
J21a scaling                               &  0.95 &  1.05 \\
J32a scaling                               &  0.9  &  1.1  \\
J43a scaling                               &  0.9  &  1.1  \\
\enddata

\end{deluxetable}
%%%%%%%%%%%%%%%%%%%%%%%%%%%%%%%%%%%

\clearpage

\section{Summary} 

We presented our ALMA $\lesssim$1--2 kpc-resolution observational
results of dense molecular gas tracers (HCN, HCO$^{+}$, and HNC) at
multiple rotational transition lines (J=2--1, 3--2, and 4--3)
for one LIRG NGC 1614 (L$_{\rm IR}$ $\sim$ 10$^{11.7}$L$_{\odot}$) and
nine ULIRGs (L$_{\rm IR}$ $=$ 10$^{12.0-12.3}$L$_{\odot}$).
The beam sizes of all the molecular J-transition lines obtained
with different beams (different array configurations) were matched to
the same value (1--2 kpc) for each (U)LIRG.  
Thereafter, the high-J to low-J emission line flux ratios for each molecule
and emission line flux ratios among the different molecules at each
J-transition were derived.
Based primarily on the HCN and HCO$^{+}$ data which can probe the
properties of dense and warm molecular gas near luminous energy sources, 
the observational and RADEX non-LTE modeling results were compared to
constrain H$_{2}$ gas volume number density and kinetic
temperature in these (U)LIRGs' nuclear 1--2 kpc regions.
We obtained the following main results. 

\begin{enumerate}

\item 
HCN and HCO$^{+}$ line emission was significantly detected at up
to J=4--3, suggesting that dense and warm molecular gas is abundant in
the observed (U)LIRGs' nuclei. 

\item 
The overall distribution of the observed HCN-to-HCO$^{+}$ flux
ratios at J=2--1, J=3--2, and J=4--3 for the (U)LIRGs' nuclei were
better reproduced by enhanced HCN abundance, relative to HCO$^{+}$,
although the ratios in a few sources could be explained by comparable
HCN and HCO$^{+}$ abundance. 

\item 
We applied the Levenberg-Marquardt method for all the (U)LIRGs' nuclei by
adopting fiducial values for molecular column density and
HCN-to-HCO$^{+}$ abundance ratio. 
Consequently, it was quantitatively derived that molecular gas in the
observed (U)LIRGs' nuclei was generally dense ($\gtrsim$10$^{3-4}$
cm$^{-3}$) and warm ($\gtrsim$100 K). 
This conclusion was confirmed to be insensitive to the choice of the
fiducial values. 

\item 
The only one observed LIRG NGC 1614 showed molecular gas density and
temperature distinctly lower than those of the remaining nine ULIRGs. 
However, the derived gas density and temperature in all the observed
(U)LIRGs' nuclei, including NGC 1614, were substantially higher than
those in quiescently star-forming normal galaxies with even less
infrared luminosity.   

\item 
For selected (U)LIRGs' nuclei with bright molecular line emission, we
also applied a Bayesian approach to constrain molecular gas
properties, with all parameters set as free, by using the Markov Chain
Monte Carlo (MCMC) method. 
The above arguments 3 and 4 were supported.

\item 
The derived density and temperature of nuclear molecular gas in 
starburst-dominated sources tended to be lower than those of the
majority of ULIRGs that exhibited signatures of luminous AGNs from
infrared, hard X-ray ($>$10 keV), and (sub)millimeter spectroscopic
observations. 
This conformed to previous suggestions that luminous AGN activity was
associated with a larger amount of and/or higher fraction of nuclear
dense molecular gas. 

\end{enumerate}

This study demonstrated that high-spatial-resolution ($\lesssim$1--2 kpc)
ALMA observations of multiple dense molecular gas tracers at multiple
J-transition lines are an effective tool for quantitatively constraining
molecular gas properties of (U)LIRGs' nuclei, by minimizing possible
contamination from spatially extended ($\gtrsim$a few kpc), more
diffuse and cooler molecular gas emission in the host galaxies.  

\acknowledgments

We thank the anonymous referee for valuable comment which helped
improve the clarity of this manuscript.
This paper made use of the following ALMA data:
ADS/JAO.ALMA\#2011.0.00020.S, \#2013.1.00032.S, \#2013.1.00033.S, 
\#2015.1.00027.S, \#2017.1.00022.S and \#2017.1.00023.S.
ALMA is a partnership of ESO (representing its member states), NSF (USA) 
and NINS (Japan), together with NRC (Canada), NSC and ASIAA
(Taiwan), and KASI (Republic of Korea), in cooperation with the Republic
of Chile. The Joint ALMA Observatory is operated by ESO, AUI/NRAO, and
NAOJ. 
M.I., K.N., and T.I. are supported by JP21K03632, JP19K03937, and
JP20K14531, respectively.  
S.B. is supported by JP19J00892 and JP21H04496.
Data analysis was in part carried out on the open use data analysis
computer system at the Astronomy Data Center, ADC, of the National
Astronomical Observatory of Japan. 
This research has made use of NASA's Astrophysics Data System and the
NASA/IPAC Extragalactic Database (NED) which is operated by the Jet
Propulsion Laboratory, California Institute of Technology, under
contract with the National Aeronautics and Space Administration. 


\vspace{5mm}
\facilities{ALMA}
\software{RADEX \citep{RADEX},
CASA \citep{CASA},
pyradex (\url{https://github.com/keflavich/pyradex}),
IPython \citep{IPython},
Jupyter Notebook \citep{JupyterNotebook},
NumPy \citep{NumPy},
SciPy \citep{SciPy},
Pandas \citep{Pandas},
Matplotlib \citep{Matplotlib},
Astropy \citep{Astropy_v5},
lmfit \citep{lmfit},
emcee \citep{emcee},
corner \citep{corner}}


%\software{astropy \citep{2013A&A...558A..33A},  
%          Cloudy \citep{2013RMxAA..49..137F}, 
%          SExtractor \citep{1996A&AS..117..393B}
%          }



%\section{Appendix information}

\clearpage

\appendix

%\restartappendixnumbering

%%%%% Appendix A %%%%%

\section{Previously Unpublished Observation Details and Obtained
Spectra of Two Ultraluminous Infrared Galaxies} 

\begin{figure}
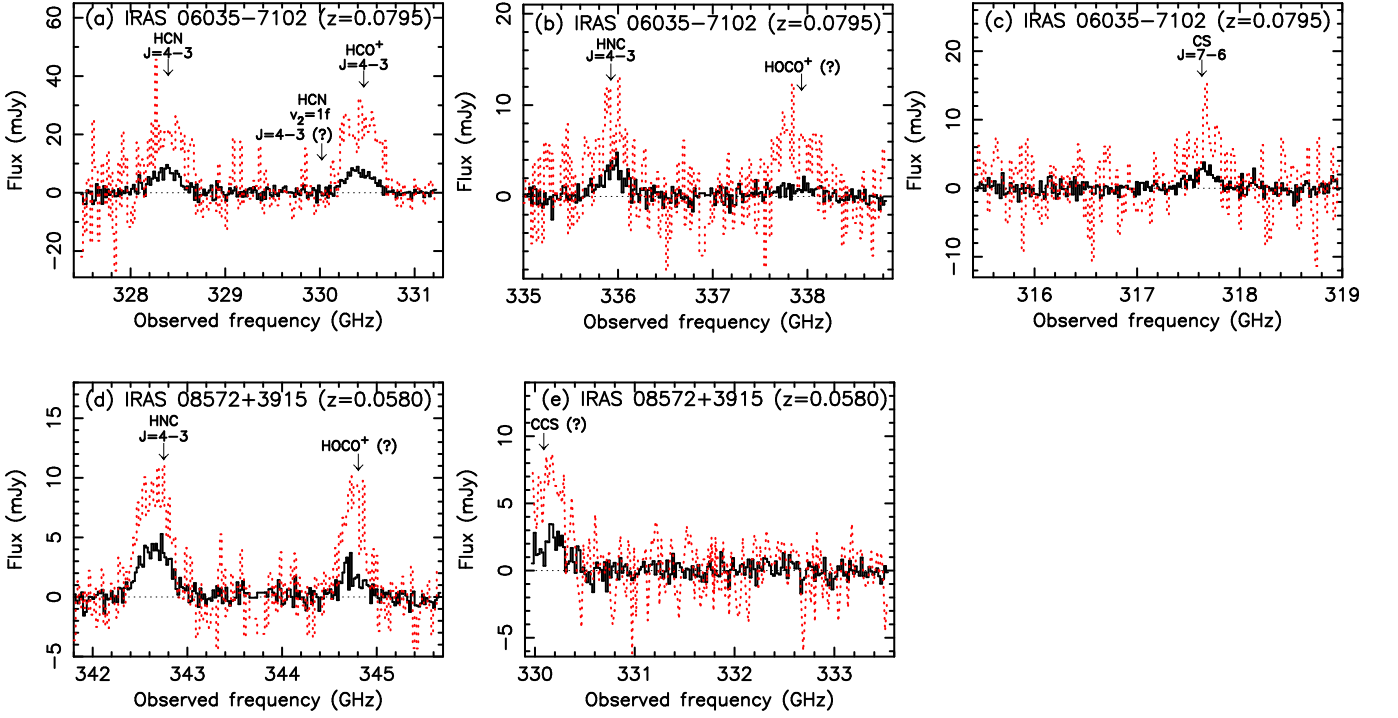

\includegraphics[angle=-90,scale=.263]{fA1a.eps} 
\includegraphics[angle=-90,scale=.263]{fA1b.eps} 
\includegraphics[angle=-90,scale=.263]{fA1c.eps} 
\vspace{0.1cm} 
\\
\includegraphics[angle=-90,scale=.263]{fA1d.eps} 
\includegraphics[angle=-90,scale=.263]{fA1e.eps} 
\vspace{0.2cm}
\caption{
ALMA band 7 (275--373 GHz) spectra obtained in Cycle 5.
The abscissa is the observed frequency in GHz, and the ordinate is the
flux density in mJy.  
IRAS 06035$-$7102 spectra of 
(a) J43a USB (including HCN and HCO$^{+}$ J=4--3), 
(b) J43b LSB (including HNC J=4--3), and 
(c) J43a LSB (including CS J=7--6). 
IRAS 08572$+$3915 spectra of (d) J43b USB (including HNC J=4--3) and 
(e) J43b LSB.
Black solid line: original beam-sized spectrum (Table \ref{tab:beam}).
Red dotted line: 1.6 kpc and 1 kpc beam spectra for (a--c) IRAS 06035$-$7102 
and (d--e) IRAS 08572$+$3915, respectively.
The horizontal black thin dotted straight line indicates the zero flux
level. 
In (b) and (d), a serendipitously detected modestly bright emission line 
is tentatively identified as HOCO$^{+}$ 17(1,16)--16(1,15) at $\nu_{\rm rest}$ 
= 364.804 GHz.
In (e), a serendipitously detected modestly bright emission line 
is tentatively identified as CCS N=27--26, J=26--25 at 
$\nu_{\rm rest}$ = 349.234 GHz. 
IRAS 06035$-$7102 J43b USB spectrum is not shown because no emission line 
signature is seen. \label{fig:spec}
}
\end{figure}

Table \ref{tab:obs} summarizes the log of previously unpublished
ALMA Cycle 5 observations for the two ULIRGs, IRAS 06035$-$7102 and
IRAS 08572$+$3915. 
The obtained spectra are shown in Figure \ref{fig:spec}.

%%%%%%%%%% Table A1 %%%%%%%%%
\begin{deluxetable}{lllccc|ccc}[!hbt]
%\tabletypesize{\small}
%\rotate
\tabletypesize{\scriptsize}
\tablecaption{Observation Log of Two ULIRGs \label{tab:obs}} 
\tablewidth{0pt}
\tablehead{
\colhead{Object} & \colhead{Line} & \colhead{Date} & \colhead{Antenna} & 
\colhead{Baseline} & \colhead{Integration} & \multicolumn{3}{c}{Calibrator} \\ 
\colhead{} & \colhead{} & \colhead{[UT]} & \colhead{Number} & \colhead{[m]} &
\colhead{[min]} & \colhead{Bandpass} & \colhead{Flux} & \colhead{Phase}  \\
\colhead{(1)} & \colhead{(2)} & \colhead{(3)} & \colhead{(4)} &
\colhead{(5)} & \colhead{(6)} & \colhead{(7)}  & \colhead{(8)} 
& \colhead{(9)}
}
\startdata 
IRAS 06035$-$7102 & HCN and HCO$^{+}$ J=4--3 & 2017 December 26 & 46 &
15--2517 & 15 
& J0522$-$3627 & J0522$-$3627 & J0529$-$7245 \\
 & HNC J=4--3 & 2017 December 12 & 45 & 15--3321 & 12
& J0522$-$3627 & J0522$-$3627 & J0529$-$7245 \\
IRAS 08572$+$3915 & HNC J=4--3 & 2017 December 26 & 46 & 15--2517 & 19 & 
J0854$+$2006 & J0854$+$2006 & J0916$+$3854 \\
 & & 2017 December 31 & 47 & 15--2517 & 19 & 
J0854$+$2006 & J0854$+$2006 & J0916$+$3854 \\
 & & 2018 January 22 & 45 & 15--1398 & 19 & 
J0854$+$2006 & J0854$+$2006 & J0916$+$3854 \\ 
\enddata

\tablecomments{ 
Col.(1): Object name. 
Col.(2): Observed molecular line. 
Col.(3): Observation date in UT. 
Col.(4): Number of antennas used for observations. 
Col.(5): Baseline length in meters. The minimum and maximum baseline
lengths are shown.  
Col.(6): Net on source integration time in minutes.
Cols.(7), (8), and (9): Bandpass, flux, and phase calibrator for the 
target source, respectively.
}

\end{deluxetable}
%%%%%%%%%%%%%%%%%%%%%%%%%

%%%%%%%%%% Figure A1 %%%%%%%%%
%%%%%%%%%%%%%%%%%%%%%%%%%%%%%%%%%%%

\clearpage

%%%%% Appendix B %%%%%

\section{Gaussian Fitting Results of Molecular Emission Lines}

\begin{figure*}
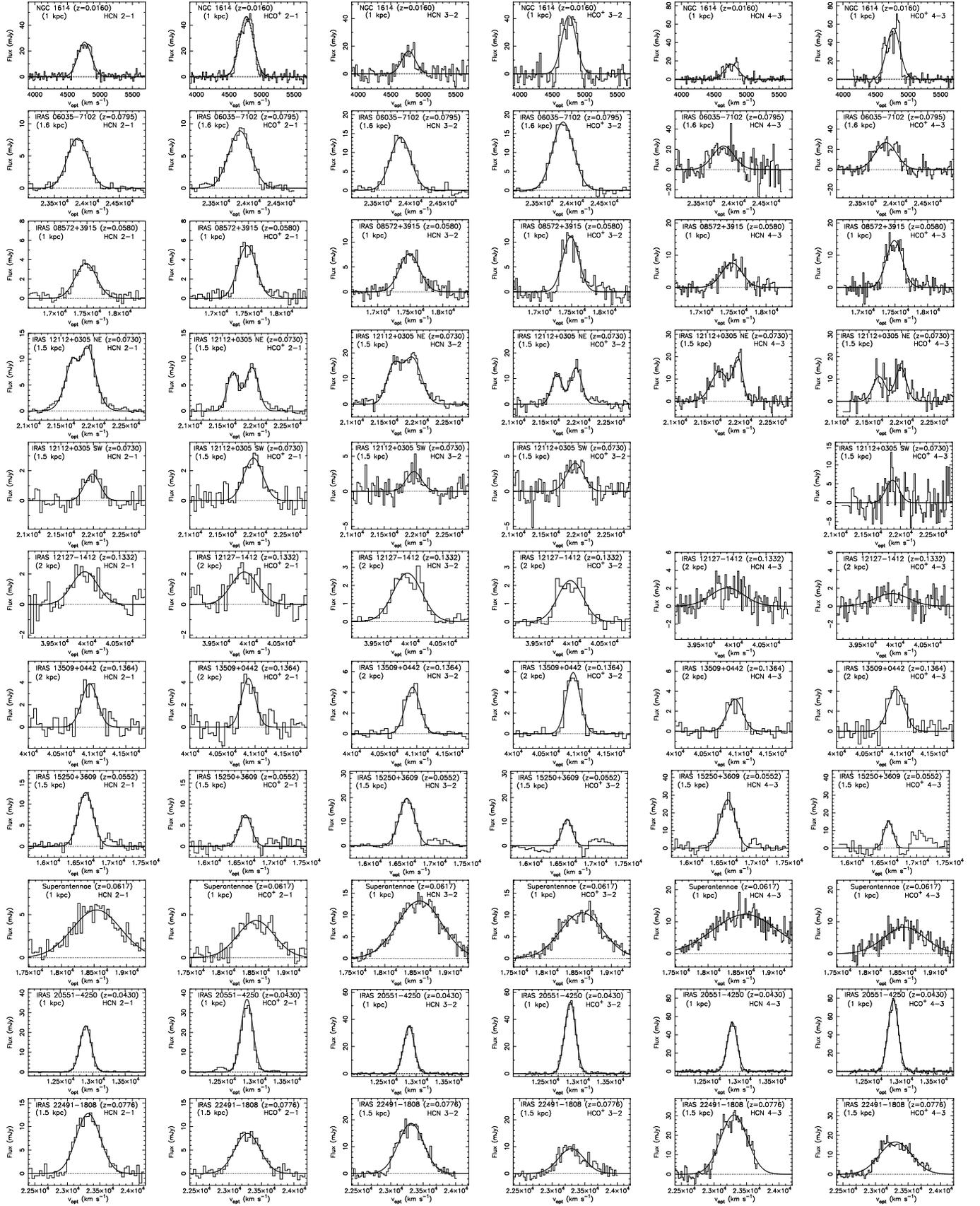

%\begin{center}
%\hspace*{-4.7cm}
\includegraphics[angle=-90,scale=.118]{fB1a.eps} \hspace{0.15cm}
\includegraphics[angle=-90,scale=.118]{fB1b.eps} \hspace{0.15cm}
\includegraphics[angle=-90,scale=.118]{fB1c.eps} \hspace{0.15cm}
\includegraphics[angle=-90,scale=.118]{fB1d.eps} \hspace{0.15cm}
\includegraphics[angle=-90,scale=.118]{fB1e.eps} \hspace{0.15cm}
\includegraphics[angle=-90,scale=.118]{fB1f.eps} \\
\includegraphics[angle=-90,scale=.118]{fB1g.eps} \hspace{0.15cm}
\includegraphics[angle=-90,scale=.118]{fB1h.eps} \hspace{0.15cm}
\includegraphics[angle=-90,scale=.118]{fB1i.eps} \hspace{0.15cm}
\includegraphics[angle=-90,scale=.118]{fB1j.eps} \hspace{0.15cm}
\includegraphics[angle=-90,scale=.118]{fB1k.eps} \hspace{0.15cm}
\includegraphics[angle=-90,scale=.118]{fB1l.eps} \\
\includegraphics[angle=-90,scale=.118]{fB1m.eps} \hspace{0.15cm}
\includegraphics[angle=-90,scale=.118]{fB1n.eps} \hspace{0.15cm}
\includegraphics[angle=-90,scale=.118]{fB1o.eps} \hspace{0.15cm}
\includegraphics[angle=-90,scale=.118]{fB1p.eps} \hspace{0.15cm}
\includegraphics[angle=-90,scale=.118]{fB1q.eps} \hspace{0.15cm}
\includegraphics[angle=-90,scale=.118]{fB1r.eps} \\
\includegraphics[angle=-90,scale=.118]{fB1s.eps} \hspace{0.15cm}
\includegraphics[angle=-90,scale=.118]{fB1t.eps} \hspace{0.15cm}
\includegraphics[angle=-90,scale=.118]{fB1u.eps} \hspace{0.15cm}
\includegraphics[angle=-90,scale=.118]{fB1v.eps} \hspace{0.15cm}
\includegraphics[angle=-90,scale=.118]{fB1w.eps} \hspace{0.15cm}
\includegraphics[angle=-90,scale=.118]{fB1x.eps} \\
\includegraphics[angle=-90,scale=.118]{fB1y.eps} \hspace{0.15cm}
\includegraphics[angle=-90,scale=.118]{fB1z.eps} \hspace{0.15cm}
\includegraphics[angle=-90,scale=.118]{fB1aa.eps} \hspace{0.15cm}
\includegraphics[angle=-90,scale=.118]{fB1ab.eps} \hspace{3.17cm}
\includegraphics[angle=-90,scale=.118]{fB1ac.eps} \\
\includegraphics[angle=-90,scale=.118]{fB1ad.eps} \hspace{0.15cm}
\includegraphics[angle=-90,scale=.118]{fB1ae.eps} \hspace{0.15cm}
\includegraphics[angle=-90,scale=.118]{fB1af.eps} \hspace{0.15cm}
\includegraphics[angle=-90,scale=.118]{fB1ag.eps} \hspace{0.15cm}
\includegraphics[angle=-90,scale=.118]{fB1ah.eps} \hspace{0.15cm}
\includegraphics[angle=-90,scale=.118]{fB1ai.eps} \\
\includegraphics[angle=-90,scale=.118]{fB1aj.eps} \hspace{0.15cm}
\includegraphics[angle=-90,scale=.118]{fB1ak.eps} \hspace{0.15cm}
\includegraphics[angle=-90,scale=.118]{fB1al.eps} \hspace{0.15cm}
\includegraphics[angle=-90,scale=.118]{fB1am.eps} \hspace{0.15cm}
\includegraphics[angle=-90,scale=.118]{fB1an.eps} \hspace{0.15cm}
\includegraphics[angle=-90,scale=.118]{fB1ao.eps} \\
\includegraphics[angle=-90,scale=.118]{fB1ap.eps} %\hspace{0.01cm}
\includegraphics[angle=-90,scale=.118]{fB1aq.eps} %\hspace{0.01cm}
\includegraphics[angle=-90,scale=.118]{fB1ar.eps} %\hspace{0.01cm}
\includegraphics[angle=-90,scale=.118]{fB1as.eps} %\hspace{0.01cm}
\includegraphics[angle=-90,scale=.118]{fB1at.eps} %\hspace{0.01cm}
\includegraphics[angle=-90,scale=.118]{fB1au.eps} \\
\includegraphics[angle=-90,scale=.118]{fB1av.eps} \hspace{0.15cm}
\includegraphics[angle=-90,scale=.118]{fB1aw.eps} \hspace{0.15cm}
\includegraphics[angle=-90,scale=.118]{fB1ax.eps} \hspace{0.15cm}
\includegraphics[angle=-90,scale=.118]{fB1ay.eps} \hspace{0.15cm}
\includegraphics[angle=-90,scale=.118]{fB1az.eps} \hspace{0.15cm}
\includegraphics[angle=-90,scale=.118]{fB1ba.eps} \\
\includegraphics[angle=-90,scale=.118]{fB1bb.eps} \hspace{0.15cm}
\includegraphics[angle=-90,scale=.118]{fB1bc.eps} \hspace{0.15cm}
\includegraphics[angle=-90,scale=.118]{fB1bd.eps} \hspace{0.15cm}
\includegraphics[angle=-90,scale=.118]{fB1be.eps} \hspace{0.15cm}
\includegraphics[angle=-90,scale=.118]{fB1bf.eps} \hspace{0.15cm}
\includegraphics[angle=-90,scale=.118]{fB1bg.eps} \\
\includegraphics[angle=-90,scale=.118]{fB1bh.eps} \hspace{0.15cm}
\includegraphics[angle=-90,scale=.118]{fB1bi.eps} \hspace{0.15cm}
\includegraphics[angle=-90,scale=.118]{fB1bj.eps} \hspace{0.15cm}
\includegraphics[angle=-90,scale=.118]{fB1bk.eps} \hspace{0.15cm}
\includegraphics[angle=-90,scale=.118]{fB1bl.eps} \hspace{0.15cm}
\includegraphics[angle=-90,scale=.118]{fB1bm.eps} \\
%\includegraphics[angle=-90,scale=.118]{ps/} \hspace{0.15cm}
%\vspace{0.6cm}
\caption{
Adopted Gaussian fits (solid curved lines) of HCN and HCO$^{+}$ 
emission lines.
The abscissa is the optical LSR velocity in km s$^{-1}$ and the ordinate 
is the flux density in mJy.
The horizontal black thin dotted straight line indicates the zero flux 
level.
Fits for the J=2--1 lines of certain sources can be found in
\citet{ima22}, but are shown here again. 
\label{fig:GaussHCNHCO}
} 
\end{figure*}

\begin{figure*}
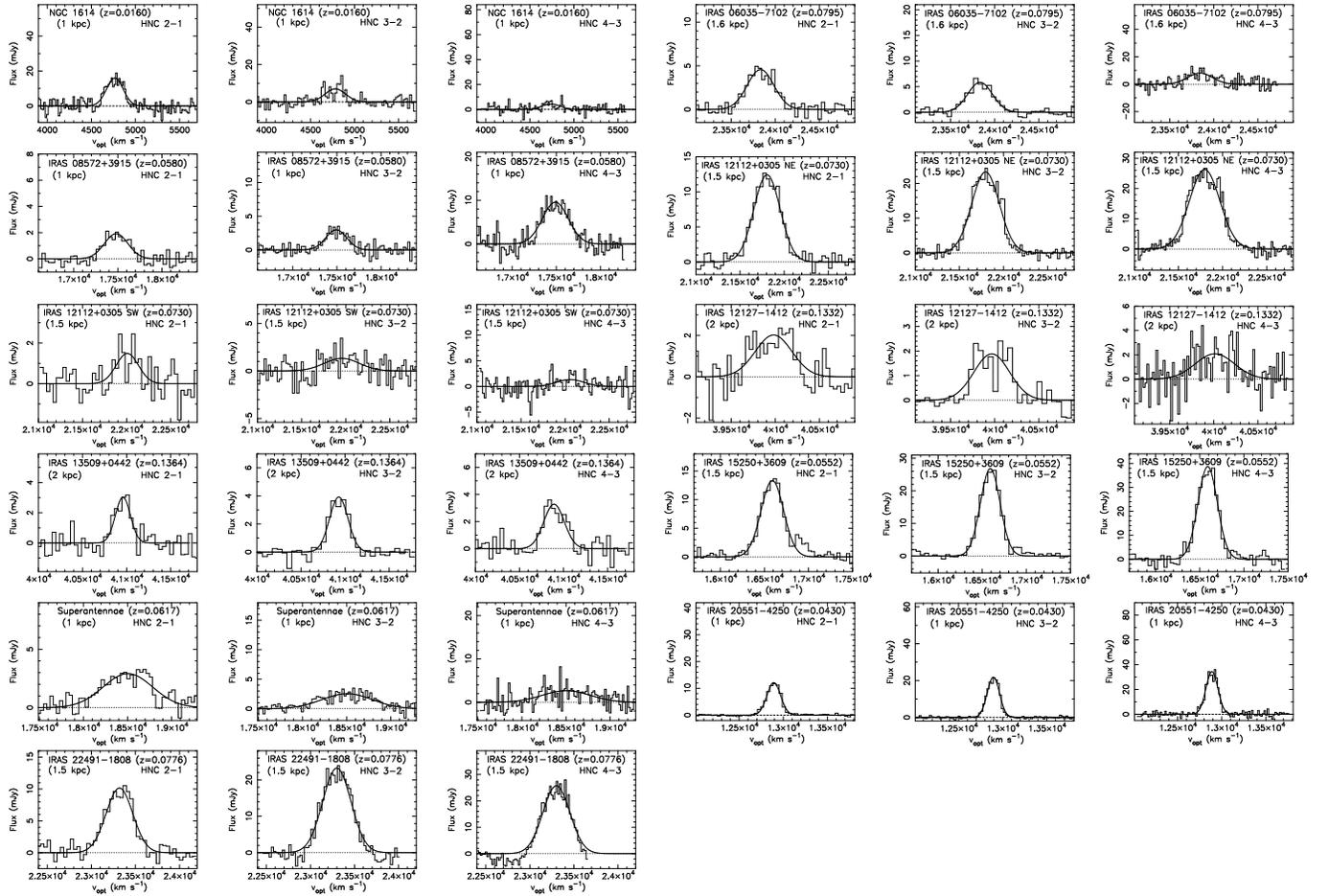

%\begin{center}
%\hspace*{-4.7cm}
\includegraphics[angle=-90,scale=.118]{fB2a.eps} \hspace{0.15cm}
\includegraphics[angle=-90,scale=.118]{fB2b.eps} \hspace{0.15cm}
\includegraphics[angle=-90,scale=.118]{fB2c.eps} \hspace{0.15cm}
\includegraphics[angle=-90,scale=.118]{fB2d.eps} \hspace{0.15cm}
\includegraphics[angle=-90,scale=.118]{fB2e.eps} \hspace{0.15cm}
\includegraphics[angle=-90,scale=.118]{fB2f.eps} \\
\includegraphics[angle=-90,scale=.118]{fB2g.eps} \hspace{0.15cm}
\includegraphics[angle=-90,scale=.118]{fB2h.eps} \hspace{0.15cm}
\includegraphics[angle=-90,scale=.118]{fB2i.eps} \hspace{0.15cm}
\includegraphics[angle=-90,scale=.118]{fB2j.eps} \hspace{0.15cm}
\includegraphics[angle=-90,scale=.118]{fB2k.eps} \hspace{0.15cm}
\includegraphics[angle=-90,scale=.118]{fB2l.eps} \\
\includegraphics[angle=-90,scale=.118]{fB2m.eps} \hspace{0.15cm}
\includegraphics[angle=-90,scale=.118]{fB2n.eps} \hspace{0.15cm}
\includegraphics[angle=-90,scale=.118]{fB2o.eps} \hspace{0.15cm}
\includegraphics[angle=-90,scale=.118]{fB2p.eps} \hspace{0.15cm}
\includegraphics[angle=-90,scale=.118]{fB2q.eps} \hspace{0.15cm}
\includegraphics[angle=-90,scale=.118]{fB2r.eps} \\
\includegraphics[angle=-90,scale=.118]{fB2s.eps} \hspace{0.15cm}
\includegraphics[angle=-90,scale=.118]{fB2t.eps} \hspace{0.15cm}
\includegraphics[angle=-90,scale=.118]{fB2u.eps} \hspace{0.15cm}
\includegraphics[angle=-90,scale=.118]{fB2v.eps} %\hspace{0.01cm}
\includegraphics[angle=-90,scale=.118]{fB2w.eps} %\hspace{0.01cm}
\includegraphics[angle=-90,scale=.118]{fB2x.eps} \\
\includegraphics[angle=-90,scale=.118]{fB2y.eps} \hspace{0.15cm}
\includegraphics[angle=-90,scale=.118]{fB2z.eps} \hspace{0.15cm}
\includegraphics[angle=-90,scale=.118]{fB2aa.eps} \hspace{0.15cm}
\includegraphics[angle=-90,scale=.118]{fB2ab.eps} \hspace{0.15cm}
\includegraphics[angle=-90,scale=.118]{fB2ac.eps} \hspace{0.15cm}
\includegraphics[angle=-90,scale=.118]{fB2ad.eps} \\
\includegraphics[angle=-90,scale=.118]{fB2ae.eps} \hspace{0.15cm}
\includegraphics[angle=-90,scale=.118]{fB2af.eps} \hspace{0.15cm}
\includegraphics[angle=-90,scale=.118]{fB2ag.eps} \\
%\includegraphics[angle=-90,scale=.118]{ps/} \hspace{0.15cm}
%\vspace{0.6cm}
\caption{
Adopted Gaussian fits (solid curved lines) of HNC emission line
displayed in the same manner as Figure \ref{fig:GaussHCNHCO}.
\label{fig:GaussHNC}
}
\end{figure*}

Our final Gaussian fitting results for molecular emission lines 
are summarized in Table \ref{tab:Gaussfit}.
In Figures \ref{fig:GaussHCNHCO} and \ref{fig:GaussHNC}, we overplot
adopted Gaussian profiles on the observed data to see the
goodness of the fits. 

%%%%%%%%%% Table B1 %%%%%%%%%

\startlongtable
\begin{deluxetable}{lcc|cccc}
%\rotate
%\tabletypesize{\small}
\tabletypesize{\scriptsize}
\tablecaption{Gaussian Fit of Dense Molecular Emission Lines \label{tab:Gaussfit}} 
\tablewidth{0pt}
\tablehead{
\colhead{Object} & \colhead{Molecule} & \colhead{Line} &  
\multicolumn{4}{c}{Gaussian fit} \\  
\colhead{} & \colhead{} & \colhead{} & \colhead{Velocity}
& \colhead{Peak} & \colhead{FWHM} & \colhead{Flux} \\ 
\colhead{} & \colhead{} & \colhead{} & \colhead{[km s$^{-1}$]} & 
\colhead{[mJy]} & \colhead{[km s$^{-1}$]} &
\colhead{[Jy km s$^{-1}$]} \\  
\colhead{(1)} & \colhead{(2)} & \colhead{(3)} & \colhead{(4)} & 
\colhead{(5)} & \colhead{(6)} & \colhead{(7)} 
}
\startdata 
NGC 1614 (1 kpc) & HCN & J=2--1 & 4765$\pm$4 & 27$\pm$1 & 243$\pm$10 & 6.8$\pm$0.4 \\
 & & J=3--2 & 4776$\pm$14 & 16$\pm$2 & 245$\pm$34 & 4.2$\pm$0.8 \\
 & & J=4--3 & 4769$\pm$10 & 16$\pm$1 & 252$\pm$24 & 4.3$\pm$0.5 \\
 & HCO$^{+}$ & J=2--1 & 4769$\pm$4 & 47$\pm$1 & 252$\pm$8 & 12$\pm$1 \\
 & & J=3--2 & 4757$\pm$8 & 42$\pm$3 & 256$\pm$13 & 11$\pm$1 \\
 & & J=4--3 & 4773$\pm$7 & 55$\pm$3 & 240$\pm$16 & 14$\pm$1 \\ 
 & HNC & J=2--1 & 4766$\pm$8 & 16$\pm$1 & 222$\pm$17 & 3.7$\pm$0.4 \\
 & & J=3--2 & 4782$\pm$24 & 7.0$\pm$1.3 & 273$\pm$60 & 2.0$\pm$0.6 
\tablenotemark{A} \\
 & & J=4--3 & 4740$\pm$32 & 4.3$\pm$1.8 & 245$\pm$94 & 1.1$\pm$0.6 
\tablenotemark{A} \\ 
\hline
IRAS 06035$-$7102 (1.6 kpc) & HCN & J=2--1 & 23857$\pm$5 & 7.8$\pm$0.2 & 371$\pm$11 & 2.9$\pm$0.1 \\
 & & J=3--2 & 23852$\pm$5 & 14$\pm$1 & 366$\pm$12 & 5.1$\pm$0.2 \\
 & & J=4--3 & 23847$\pm$26 & 23$\pm$3 & 383$\pm$62 & 8.8$\pm$1.8 \\
 & HCO$^{+}$ & J=2--1 & 23867$\pm$5 & 8.9$\pm$0.2 & 404$\pm$13 & 3.6$\pm$0.1 \\
 & & J=3--2 & 23867$\pm$4 & 18$\pm$1 & 383$\pm$9 & 6.8$\pm$0.2 \\
 & & J=4--3 & 23858$\pm$13 & 27$\pm$2 & 383$\pm$34 & 10$\pm$1 \\
 & HNC & J=2--1 & 23832$\pm$11 & 4.6$\pm$0.3 & 356$\pm$25 & 1.6$\pm$0.1 \\
 & & J=3--2 & 23833$\pm$9 & 5.9$\pm$0.3 & 330$\pm$20 & 1.9$\pm$0.2 \\
 & & J=4--3 & 23844$\pm$24 & 8.1$\pm$1.2 & 328$\pm$64 & 2.6$\pm$0.6 \\ 
\hline
IRAS 08572$+$3915 (1 kpc) & HCN & J=2--1 & 17485$\pm$11 & 3.6$\pm$0.2 & 383$\pm$27 & 1.4$\pm$0.1 \\
 & & J=3--2 & 17489$\pm$9 & 7.7$\pm$0.4 & 363$\pm$22 & 2.8$\pm$0.2 \\
 & & J=4--3 & 17476$\pm$13 & 7.6$\pm$0.6 & 367$\pm$31 & 2.8$\pm$0.3 \\
 & HCO$^{+}$ & J=2--1 & 17480$\pm$7 & 5.5$\pm$0.2 & 332$\pm$17 & 1.8$\pm$0.1 \tablenotemark{A} \\
 & & J=3--2 & 17485$\pm$8 & 11$\pm$1 & 292$\pm$17 & 3.3$\pm$0.3 \\
 & & J=4--3 & 17489$\pm$13 & 15$\pm$1 & 289$\pm$15 & 4.2$\pm$0.3 \\
 & HNC & J=2--1 & 17481$\pm$16 & 1.9$\pm$0.2 & 315$\pm$33 & 0.61$\pm$0.09 \\
 & & J=3--2 & 17493$\pm$14 & 3.0$\pm$0.3 & 278$\pm$33 & 0.84$\pm$0.13 \\
 & & J=4--3 & 17489$\pm$11 & 9.6$\pm$0.6 & 346$\pm$24 & 3.3$\pm$0.3 \\
\hline
IRAS 12112$+$0305 NE (1.5 kpc) & HCN & J=2--1 & 21811$\pm$4, 21790$\pm$5 & 17$\pm$2, $-$7.5$\pm$2.2 \tablenotemark{B} & 395$\pm$16, 180$\pm$21 & 5.5$\pm$1.0 \\
 & & J=3--2 & 21641$\pm$18, 21931$\pm$23 & 14$\pm$2, 18$\pm$1 & 253$\pm$30, 
327$\pm$52 & 9.4$\pm$1.2 \\
 & & J=4--3 & 21687$\pm$48, 21970$\pm$10 & 13$\pm$1, 18$\pm$3 & 316$\pm$101, 
164$\pm$43 & 7.2$\pm$1.7 \\
 & HCO$^{+}$ & J=2--1 & 21661$\pm$7, 21955$\pm$7 & 7.5$\pm$0.4, 8.4$\pm$0.4 & 201$\pm$18, 229$\pm$21 & 3.4$\pm$0.3 \\
 & & J=3--2 & 21671$\pm$9, 21965$\pm$7 & 12$\pm$1, 14$\pm$1 & 188$\pm$4, 
202$\pm$18 & 5.0$\pm$0.4 \\
 & & J=4--3 & 21656$\pm$21, 21989$\pm$12 & 11$\pm$2, 15$\pm$2 & 226$\pm$61, 
201$\pm$54 & 5.5$\pm$1.2 \\
 & HNC & J=2--1 & 21805$\pm$6 & 12$\pm$1 & 352$\pm$14 & 4.3$\pm$0.2 \\
 & & J=3--2 & 21796$\pm$4 & 23$\pm$1 & 370$\pm$10 & 8.6$\pm$0.3 \\
 & & J=4--3 & 21796$\pm$5 & 26$\pm$1 & 407$\pm$13 & 11$\pm$1 \\
\hline
IRAS 12112$+$0305 SW (1.5 kpc) & HCN & J=2--1 & 21984$\pm$20 &
1.8$\pm$0.2 & 278$\pm$45 & 0.49$\pm$0.10 \\ 
 & & J=3--2 & 21957$\pm$27 & 2.8$\pm$0.6 & 272$\pm$60 & 0.76$\pm$0.23 \\
 & & J=4--3 & --- & --- & --- & --- \\
 & HCO$^{+}$ & J=2--1 & 21976$\pm$16 & 2.8$\pm$0.3 & 341$\pm$47 &
0.96$\pm$0.16 \\ 
 & & J=3--2 & 21954$\pm$23 & 4.0$\pm$0.5 & 302$\pm$54 & 1.2$\pm$0.3 \\
 & & J=4--3 & 21879$\pm$34 & 6.0$\pm$1.8 & 235$\pm$84 & 1.4$\pm$0.7
 ($<$3$\sigma$) \\
 & HNC & J=2--1 & 22007$\pm$35 & 1.5$\pm$0.3 & 298$\pm$90 & 0.44$\pm$0.17 
 ($<$3$\sigma$) \\
 & & J=3--2 & 21955$\pm$71 & 1.3$\pm$0.5 & 469$\pm$138 & 0.62$\pm$0.30 
($<$3$\sigma$)  \\
 & & J=4--3 & 22057 (fix) & 1.3$\pm$0.5 & 366 (fix) & 0.49$\pm$0.19
($<$3$\sigma$) \\ 
\hline
IRAS 12127$-$1412 (2 kpc) & HCN & J=2--1 & 39967$\pm$32 & 2.1$\pm$0.3
& 463$\pm$80 & 0.94$\pm$0.21 \\ 
 & & J=3--2 & 39958$\pm$13 & 2.7$\pm$0.1 & 524$\pm$32 & 1.3$\pm$0.1
\tablenotemark{A} \\ 
 & & J=4--3 & 39907$\pm$44 & 2.0$\pm$0.3 & 602$\pm$98 & 1.2$\pm$0.2 \\
 & HCO$^{+}$ & J=2--1 & 39937$\pm$35 & 2.1$\pm$0.3 & 474$\pm$93 & 0.95$\pm$0.23 \tablenotemark{A} \\
 & & J=3--2 & 39979$\pm$15 & 2.3$\pm$0.2 & 454$\pm$31 & 0.97$\pm$0.09 \tablenotemark{A} \\
 & & J=4--3 & 39953$\pm$65 & 1.4$\pm$0.3 & 614$\pm$186 & 0.81$\pm$0.30
($<$3$\sigma$) \\
 & HNC & J=2--1 & 39980$\pm$37 & 2.0$\pm$0.3 & 490$\pm$81 & 0.93$\pm$0.20 
\tablenotemark{A} \\
 & & J=3--2 & 39966$\pm$29 & 1.9$\pm$0.2 & 476$\pm$66 & 0.84$\pm$0.15 \tablenotemark{A} \\
 & & J=4--3 & 40012$\pm$52 & 2.1$\pm$0.4 & 522$\pm$120 & 1.0$\pm$0.3 \\
\hline
IRAS 13509$+$0442 (2 kpc) & HCN & J=2--1 & 40941$\pm$15 & 3.9$\pm$0.4 & 284$\pm$41 & 1.0$\pm$0.2 \\
 & & J=3--2 & 40936$\pm$6 & 4.5$\pm$0.2 & 266$\pm$15 & 1.1$\pm$0.1 \\
 & & J=4--3 & 40922$\pm$12 & 3.2$\pm$0.3 & 277$\pm$27 & 0.83$\pm$0.11 \\
 & HCO$^{+}$ & J=2--1 & 40900$\pm$15 & 4.4$\pm$0.5 & 248$\pm$38 & 1.0$\pm$0.2 \\
 & & J=3--2 & 40923$\pm$6 & 5.9$\pm$0.3 & 251$\pm$13 & 1.4$\pm$0.1 \\
 & & J=4--3 & 40918$\pm$14 & 4.2$\pm$0.4 & 294$\pm$30 & 1.2$\pm$0.2 \\
 & HNC & J=2--1 & 40958$\pm$15 & 3.0$\pm$0.4 & 200$\pm$40 & 0.57$\pm$0.14 \\
 & & J=3--2 & 40928$\pm$7 & 3.9$\pm$0.2 & 248$\pm$14 & 0.91$\pm$0.07 \\
 & & J=4--3 & 40892$\pm$15 & 3.3$\pm$0.3 & 275$\pm$32 & 0.84$\pm$0.13 \\
\hline
IRAS 15250$+$3609 (1.5 kpc) & HCN & J=2--1 & 16581$\pm$4 & 13$\pm$1 & 272$\pm$10 & 3.5$\pm$0.2 \\
 & & J=3--2 & 16576$\pm$4 & 20$\pm$1 & 277$\pm$11 & 5.5$\pm$0.3 \\
 & & J=4--3 & 16562$\pm$7 & 27$\pm$1 & 271$\pm$15 & 7.5$\pm$0.5 \\
 & HCO$^{+}$ & J=2--1 & 16562$\pm$9 & 7.5$\pm$0.6 & 223$\pm$18 & 1.7$\pm$0.2 \\
 & & J=3--2 & 16568$\pm$9 & 11$\pm$1 & 182$\pm$23 & 2.0$\pm$0.3 \\
 & & J=4--3 & 16567$\pm$13 & 15$\pm$3 & 162$\pm$29 & 2.5$\pm$0.7 \\
 & HNC & J=2--1 & 16590$\pm$7 & 13$\pm$1 & 302$\pm$20 & 4.1$\pm$0.3 \\
 & & J=3--2 & 16591$\pm$4 & 27$\pm$1 & 263$\pm$10 & 7.1$\pm$0.3 \\
 & & J=4--3 & 16590$\pm$4 & 39$\pm$1 & 259$\pm$9 & 10$\pm$1 \\
\hline
Superantennae (1 kpc) & HCN & J=2--1 & 18539$\pm$19 & 5.5$\pm$0.3 & 878$\pm$54 & 4.9$\pm$0.4 \\
 & & J=3--2 & 18535$\pm$8 & 13$\pm$1 & 805$\pm$18 & 11$\pm$1 \\
 & & J=4--3 & 18570$\pm$21 & 12$\pm$1 & 1166$\pm$62 & 14$\pm$1 \\
 & HCO$^{+}$ & J=2--1 & 18505$\pm$21 & 4.3$\pm$0.3 & 665$\pm$52 & 2.9$\pm$0.3 \\
 & & J=3--2 & 18546$\pm$10 & 11$\pm$1 & 774$\pm$23 & 8.2$\pm$0.3 \\
 & & J=4--3 & 18534$\pm$23 & 8.3$\pm$0.5 & 741$\pm$57 & 6.1$\pm$0.6 \\
 & HNC & J=2--1 & 18504$\pm$22 & 2.9$\pm$0.2 & 661$\pm$57 & 1.9$\pm$0.2 \\
 & & J=3--2 & 18506$\pm$26 & 2.5$\pm$0.2 & 672$\pm$59 & 1.7$\pm$0.2 \\
 & & J=4--3 & 18503$\pm$57 & 2.7$\pm$0.5 & 640$\pm$129 & 1.7$\pm$0.5 \\
\hline
IRAS 20551$-$4250 (1 kpc) & HCN & J=2--1 & 12890$\pm$1 & 23$\pm$1 & 202$\pm$2 & 4.8$\pm$0.1 \\ 
 & & J=3--2 & 12892$\pm$1 & 35$\pm$1 & 191$\pm$3 & 6.8$\pm$0.1 \\
 & & J=4--3 & 12891$\pm$1 & 54$\pm$1 & 175$\pm$3 & 9.6$\pm$0.2 \\
 & HCO$^{+}$ & J=2--1 & 12887$\pm$1 & 34$\pm$1 & 214$\pm$3 & 7.4$\pm$0.1 \\
 & & J=3--2 & 12887$\pm$1 & 53$\pm$1 & 198$\pm$2 & 11$\pm$1 \\
 & & J=4--3 & 12882$\pm$1 & 78$\pm$1 & 184$\pm$3 & 15$\pm$1 \\
 & HNC & J=2--1 & 12892$\pm$2 & 12$\pm$1 & 182$\pm$5 & 2.3$\pm$0.1 \\
 & & J=3--2 & 12894$\pm$2 & 22$\pm$1 & 164$\pm$4 & 3.7$\pm$0.1 \\
 & & J=4--3 & 12889$\pm$2 & 34$\pm$1 & 160$\pm$5 & 5.6$\pm$0.2 \\
\hline 
IRAS 22491$-$1808 (1.5 kpc) & HCN & J=2--1 & 23320$\pm$6 & 13$\pm$1 & 435$\pm$14 & 5.5$\pm$0.2 \\
 & & J=3--2 & 23313$\pm$6 & 19$\pm$1 & 457$\pm$14 & 8.4$\pm$0.3 \\
 & & J=4--3 & 23307$\pm$7 & 30$\pm$1 & 481$\pm$18 & 14$\pm$1
\tablenotemark{C}\\ 
 & HCO$^{+}$ & J=2--1 & 23286$\pm$8 & 8.6$\pm$0.3 & 456$\pm$19 & 3.8$\pm$0.2 \\
 & & J=3--2 & 23293$\pm$15 & 9.3$\pm$0.6 & 463$\pm$40 & 4.2$\pm$0.4 \\
 & & J=4--3 & 23309$\pm$10 & 17$\pm$1 & 576$\pm$23 & 9.4$\pm$0.5
\tablenotemark{C} \\ 
 & HNC & J=2--1 & 23318$\pm$10 & 10$\pm$1 & 356$\pm$24 & 3.6$\pm$0.3 \\
 & & J=3--2 & 23299$\pm$5 & 23$\pm$1 & 376$\pm$14 & 8.6$\pm$0.4 \\
 & & J=4--3 & 23301$\pm$5 & 26$\pm$1 & 361$\pm$15 & 9.2$\pm$0.5
\tablenotemark{C} \\ 
\hline
\enddata

\tablenotetext{A}{
Certain double-peaked emission signature is observed.
We also attempted a double Gaussian fit, and found that its flux was
0--10\% smaller than that of the single Gaussian fit. 
We adopt single Gaussian fit flux.}

\tablenotetext{B}{One emission and one absorption (dip near the emission 
center) components.}

\tablenotetext{C}{Revised value from that shown by \citet{ima14}, 
after redefining the continuum flux level ($\S$3).}

\tablecomments{
Col.(1): Object name.
The adopted beam size in kpc is shown in parentheses. 
Col.(2): Molecule.
Col.(2): Rotational J-transition line.
Cols.(4)--(7): Gaussian fit of emission line in a 1--2 kpc beam  
spectrum at the continuum peak position.
Col.(4): Optical local standard of rest (LSR) velocity (v$_{\rm opt}$) 
of the emission line peak in km $^{-1}$. 
Col.(5): Peak flux in mJy. 
Col.(6): Observed full width at half maximum (FWHM) in km s$^{-1}$.
Col.(7): Gaussian-fit velocity-integrated flux in Jy km s$^{-1}$. 
Only the Gaussian fitting error (statistical uncertainty) is
considered.
}

\end{deluxetable}
%%%%%%%%%%%%%%%%%%%%%%%%%%%%%%%%%%%%

%%%%%%%%%% Figure B1 %%%%%%%%%
%%%%%%%%%%%%%%%%%%%%%%%%%%%%%%%%%%%

%%%%%%%%%% Figure B2 %%%%%%%%%
%%%%%%%%%%%%%%%%%%%%%%%%%%%%%%%%%%%

\clearpage

%%%%% Appendix C %%%%%

\section{Continuum Flux Density Measured with 1--2 kpc Beam}

\begin{figure}
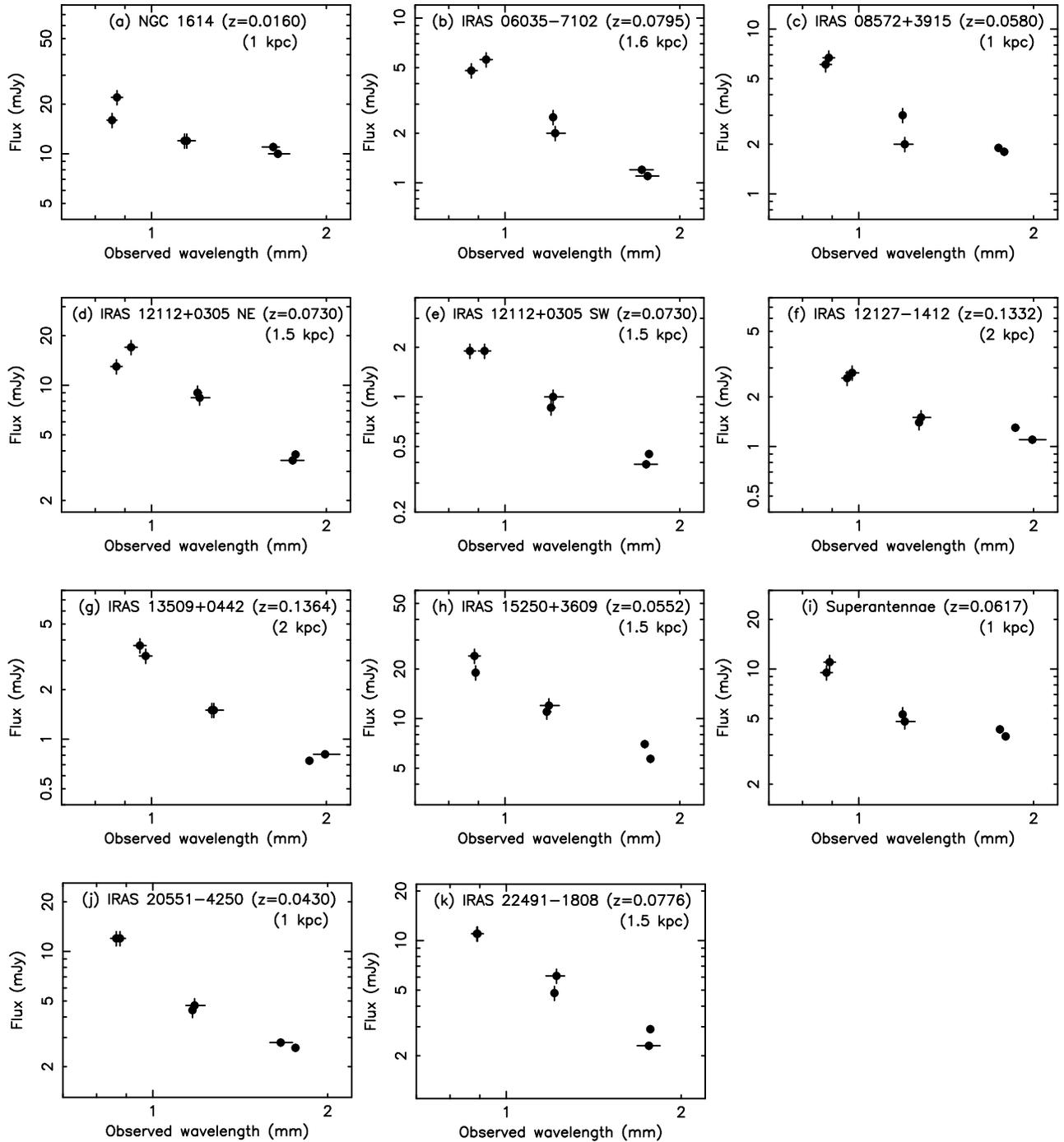

\begin{center}
\includegraphics[angle=-90,scale=.25]{fC1a.eps} 
\includegraphics[angle=-90,scale=.25]{fC1b.eps} 
\includegraphics[angle=-90,scale=.25]{fC1c.eps} \\
\vspace{0.5cm}
\includegraphics[angle=-90,scale=.25]{fC1d.eps} 
\includegraphics[angle=-90,scale=.25]{fC1e.eps} 
\includegraphics[angle=-90,scale=.25]{fC1f.eps} \\
\vspace{0.5cm}
\includegraphics[angle=-90,scale=.25]{fC1g.eps} 
\includegraphics[angle=-90,scale=.25]{fC1h.eps} 
\includegraphics[angle=-90,scale=.25]{fC1i.eps} \\
\vspace{0.5cm}
\hspace*{-5.8cm}
\includegraphics[angle=-90,scale=.25]{fC1j.eps} 
\includegraphics[angle=-90,scale=.25]{fC1k.eps} 
\end{center}
\caption{
Continuum flux density in mJy (ordinate) as a function of observed
wavelength in $\mu$m (abscissa) for the observed 11 (U)LIRGs' nuclei.
5--10\% error is added in the ordinate, by considering the possible 
absolute flux calibration uncertainty of individual ALMA observations
($\S$5.3). 
The object name, redshift, and adopted beam size in kpc are shown.
\label{fig:cont}
}
\end{figure}

Table \ref{tab:cont} and Figure \ref{fig:cont} show the continuum flux
density and spectral energy distribution measured using the adopted
1--2 kpc beam, respectively. 
Using the continuum data, we attempted to make some correction
of the possible absolute flux calibration uncertainty for individual
ALMA observations (maximum $\sim$10\%; $\S$5.3) by assuming power
law for an intrinsic continuum emission shape. 
However, in many (U)LIRGs, the observed continuum
shape was concaved in such a way that 1.1--1.4 mm (band 6) flux was
significantly smaller than the power law interpolation between 
0.8--1.1 mm (band 7) and 1.4--1.8 mm (band 5) (e.g., clearly
seen for IRAS 12127$-$1412, the Superantennae, and IRAS 20551$-$4250
in Figures \ref{fig:cont}f, \ref{fig:cont}i, and \ref{fig:cont}j).
This is most likely because a dust thermal radiation component is
significantly present in band 7 (0.8--1.1 mm) in addition to thermal
free-free and synchrotron emission, which are dominant in bands 5
and 6 (1.1--1.8 mm). 
Thus, we decided not to apply this correction, because reliable
correction turned out to be impossible.

%\clearpage

%%%%%%%%%% Table C1 %%%%%%%%%
\startlongtable
\begin{deluxetable}{llrl}
%\tabletypesize{\small}
\tabletypesize{\scriptsize}
\tablecaption{Continuum Emission in 1--2 kpc Beam\label{tab:cont}}
\tablewidth{0pt}
\tablehead{
\colhead{Object} & \colhead{Frequency} & \colhead{Flux} & 
\colhead{Peak Coordinate} \\
\colhead{} & \colhead{[GHz]} & \colhead{[mJy]} & 
\colhead{(RA,DEC)ICRS} \\  
\colhead{(1)} & \colhead{(2)} & \colhead{(3)}  & \colhead{(4)}   
}
\startdata 
NGC 1614 (1 kpc) & J21a; 181.2 (173.2--176.9, 185.4--189.1) & 
10 (7.5$\sigma$) & (04 34 00.0, $-$08 34 45) \\
& J21b; 185.4 (177.4--181.2, 189.6--193.3) & 11 (5.1$\sigma$) \\
& J32a; 263.2 (260.8--265.5) & 12 (6.9$\sigma$) \\
& J32b; 261.2 (251.8--255.5, 266.8--270.6) & 12 (5.7$\sigma$) \\
& J43a; 344.1 (336.2--338.0, 347.9--351.9)  & 22 (16$\sigma$) \\
& J43b; 351.0 (344.1--346.0, 355.9--357.9) & 16 (17$\sigma$) \\ \hline
IRAS 06035$-$7102 (1.6 kpc) & J21a; 170.9 (163.1--166.7, 175.0--178.7) & 
1.1 (10$\sigma$) & (06 02 53.9, $-$71 03 10) \\
& J21b; 174.8 (166.9--170.7, 178.9--182.6) & 1.2 (10$\sigma$)\\
& J32a; 247.8 (245.4--250.2) & 2.5 (18$\sigma$) \\
& J32b; 245.9 (237.0--240.6, 251.0--254.7) & 2.0 (24$\sigma$) \\
& J43a; 323.4 (315.4--319.0, 327.4--331.3) & 5.6 (5.2$\sigma$) \\
& J43b; 342.9 (335.0--338.9, 347.1--350.7) & 4.8 (4.4$\sigma$) \\ \hline
IRAS 08572$+$3915 (1 kpc) & J21a; 168.1 (166.2--169.9) & 1.8 (17$\sigma$) & 
(09 00 25.4, $+$39 03 54) \\
& J21b; 172.2 (170.3--174.1) & 1.9 (17$\sigma$) \\
& J32a; 252.7 (250.4--254.9) & 3.0 (21$\sigma$) \\
& J32b; 250.8 (241.8--245.4, 256.1--259.8) & 2.0 (20$\sigma$) \\
& J43a; 342.0 (334.2--337.9, 346.2--349.8) & 6.1 (39$\sigma$) \\
& J43b; 337.8 (329.9--333.6, 341.8--345.7) & 6.7 (14$\sigma$) \\ \hline
IRAS 12112$+$0305 NE (1.5 kpc) & J21a; 171.7 (163.9--167.5, 175.7--179.4) & 
3.5 (26$\sigma$) & (12 13 46.1, $+$02 48 42) \\
& J21b; 169.7 (167.9--171.5) & 3.8 (16$\sigma$) \\
& J32a; 249.1 (246.8--251.3) & 9.0 (31$\sigma$) \\
& J32b; 247.3 (238.3--242.0, 252.5--256.2) & 8.4 (39$\sigma$) \\
& J43a; 325.3 (317.3--321.0, 329.4--333.2) & 17 (37$\sigma$) \\
& J43b; 344.8 (336.9--340.8, 349.1--352.7) & 13 (30$\sigma$) \\ \hline
IRAS 12112$+$0305 SW (1.5 kpc) & J21a; 171.7 (163.9--167.5, 175.7--179.4) & 
0.39 (3.0$\sigma$) & (12 13 45.9, $+$02 48 39) \\
& J21b; 169.7 (167.9--171.5) & 0.45 (1.9$\sigma$) \\
& J32a; 249.1 (246.8--251.3) & 0.86 (3.0$\sigma$) \\
& J32b; 247.3 (238.3--242.0, 252.5--256.2) & 1.0 (4.6$\sigma$) \\
& J43a; 325.3 (317.3--321.0, 329.4--333.2) & 1.9 (3.9$\sigma$) \\
& J43b; 344.8 (336.9--340.8, 349.1--352.7) & 1.9 (4.2$\sigma$) \\ \hline
IRAS 12127$-$1412 (2 kpc) & J21a; 150.8 (142.9--146.4, 155.1--158.7) & 
1.1 (16$\sigma$) & (12 15 19.1, $-$14 29 42) \\
& J21b; 160.9 (159.0--162.7) & 1.3 (20$\sigma$) \\
& J32a; 236.1 (233.7--238.4) & 1.4 (25$\sigma$) \\
& J32b; 234.1 (225.5--229.0, 239.1--242.7) & 1.5 (36$\sigma$) \\
& J43a; 307.9 (300.0--303.7, 312.0--315.8)  & 2.8 (21$\sigma$) \\
& J43b; 314.3 (307.6--309.5, 319.1--321.0) & 2.6 (16$\sigma$) \\ \hline
IRAS 13509$+$0442 (2 kpc) & J21a; 150.4 (142.5--146.0, 154.7--158.2) & 
0.81 (11$\sigma$) & (13 53 31.6, $+$04 28 05) \\
& J21b; 160.4 (158.5--162.2) & 0.74 (9.4$\sigma$) \\
& J32a; 235.4 (233.1--237.7) & 1.5 (20$\sigma$) \\
& J32b; 233.5 (224.9--228.4, 238.4--242.0) & 1.5 (24$\sigma$) \\
& J43a; 307.1 (299.4--302.9, 311.1--314.8) & 3.2 (23$\sigma$) \\
& J43b; 314.2 (306.4--309.9, 318.3--322.0) & 3.7 (27$\sigma$) \\ \hline
IRAS 15250$+$3609 (1.5 kpc) & J21a; 168.5 (166.7--170.3) & 5.7 (19$\sigma$) &
(15 26 59.4, $+$35 58 37) \\
& J21b; 172.7 (170.8--174.6) & 7.0 (24$\sigma$) \\
& J32a; 253.4 (251.0--255.8) & 11 (31$\sigma$) \\
& J32b; 251.5 (242.5--246.1, 256.7--260.5) & 12 (35$\sigma$) \\
& J43a; 336.9 (334.9--338.8) & 19 (22$\sigma$) \\
& J43b; 338.7 (330.8--334.4, 342.6--346.6) & 24 (28$\sigma$) \\ \hline
Superantennae (1 kpc) & J21a; 167.5 (165.6--169.3) & 3.9 (14$\sigma$) &
(19 31 21.4, $-$72 39 22) \\
& J21b; 171.6 (169.7--173.5) & 4.3 (23$\sigma$) \\
& J32a; 252.0 (249.5--254.4) & 5.3 (28$\sigma$) \\
& J32b; 250.0 (240.9--244.6, 255.3--259.0) & 4.8 (34$\sigma$) \\
& J43a; 340.9 (333.0--336.8, 345.1--348.8) & 9.5 (27$\sigma$) \\
& J43b; 336.7 (328.8--332.4, 340.7--344.5) & 11 (37$\sigma$) \\ \hline
IRAS 20551$-$4250 (1 kpc) & J21a; 170.5 (168.6--172.3) & 2.6 (45$\sigma$) &
(20 58 26.8, $-$42 39 00) \\
& J21b; 180.8 (172.8--176.5, 185.0--188.7) & 2.8 (33$\sigma$) \\
& J32a; 254.2 (244.8--249.0, 259.8--263.5) & 4.7 (29$\sigma$) \\
& J32b; 256.3 (254.0--258.5) & 4.4 (38$\sigma$) \\
& J43a; 346.8 (338.9--342.8, 350.9--354.6) & 12 (28$\sigma$) \\
& J43b; 341.6 (334.6--336.5, 346.7--348.6) & 12 (26$\sigma$) \\ \hline
IRAS 22491$-$1808 (1.5 kpc) & J21a; 170.9 (163.2--166.8, 175.0--178.6) & 2.3 (15$\sigma$) & (22 51 49.4, $-$17 52 24) \\
& J21b; 169.1 (167.2--171.0) & 2.9 (12$\sigma$) \\
& J32a; 248.4 (246.1--250.7) & 4.8 (15$\sigma$) \\
& J32b; 246.6 (237.7--241.3, 251.8--255.5) & 6.1 (30$\sigma$) \\
& J43a; 336.4 (328.5--332.3, 340.6--344.3)  & 11 (34$\sigma$) \\
& J43b; 337.1 (336.1--338.0) & 11 (31$\sigma$) \\ \hline
\enddata

\tablecomments{Col.(1): Object name.
The adopted beam size in kpc is shown in parentheses. 
Col.(2): Central frequency used for continuum extraction 
for J21a--J43b (explained in the caption of Table \ref{tab:beam}) 
in GHz is shown first, and the frequency range in GHz is listed in
parentheses. 
The frequencies of obvious emission and absorption lines are removed. 
Col.(3): Continuum flux in mJy at the emission peak with adopted
beam size. 
Value at the highest flux pixel (0$\farcs$02--0$\farcs$04 pixel$^{-1}$)
is extracted. 
Detection significance relative to root mean square (rms)
noise is shown in parentheses. 
Possible systematic uncertainty arising from the absolute flux
calibration ambiguity in individual ALMA observations and 
choice of the frequency range for continuum-level determination, 
are not included. 
Col.(4): Coordinate of continuum emission peak in ICRS.
}

\end{deluxetable}
%%%%%%%%%%%%%%%%%%%%%%%%%%%%%%%%%%%

\begin{figure}
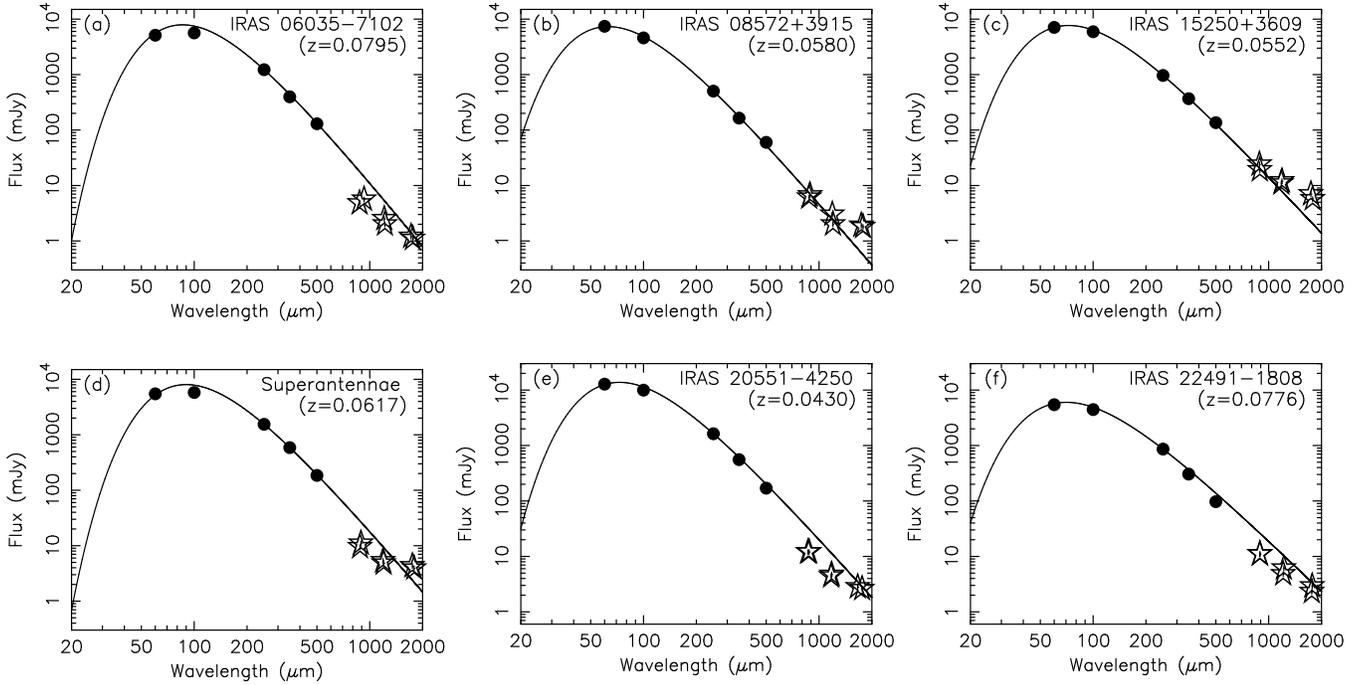

\begin{center}
\includegraphics[angle=-90,scale=.25]{fC2a.eps} 
\includegraphics[angle=-90,scale=.25]{fC2b.eps} 
\includegraphics[angle=-90,scale=.25]{fC2c.eps} \\
\vspace{0.5cm}
\includegraphics[angle=-90,scale=.25]{fC2d.eps}
\includegraphics[angle=-90,scale=.25]{fC2e.eps} 
\includegraphics[angle=-90,scale=.25]{fC2f.eps} \\
\end{center}
\caption{
Continuum flux density in mJy (ordinate) as a function of observed
wavelength in $\mu$m (abscissa) for selected ULIRGs, for which 
Herschel 250, 350, and 500 $\mu$m photometric data are available. 
Filled circle: Herschel photometric data \citep{cle18}. 
Open star: ALMA 1--2 kpc beam photometric data. 
The IRAS 60 $\mu$m and 100 $\mu$m fluxes are also plotted as filled circles.
The thick curved lines represent the best-fit graybody curves adopted by
\citet{cle18} after normalizing at the 250 $\mu$m flux.
\label{fig:contSED}
}
\end{figure}

For certain fractions of the observed ULIRGs, the photometric data 
at 250 $\mu$m, 350 $\mu$m, and 500 $\mu$m taken with the Herschel
Space Observatory are available \citep{cle18}. 
These data, together with the IRAS 60 $\mu$m and 100 $\mu$m fluxes, are
usually dominated by dust thermal radiation, and are plotted in Figure
\ref{fig:contSED}. 
We overplot our ALMA continuum data measured with a 1--2 kpc beam 
in Figure \ref{fig:contSED}.
Our ALMA continuum fluxes largely agree with, or are only slightly
smaller than, those extrapolated from shorter-wavelength 
IRAS and Herschel fluxes measured using larger beams ($\gtrsim$5$''$
or $\gtrsim$5 kpc at $z \gtrsim$ 0.05). 
This rough agreement supports the previously argued scenario that 
nearby ULIRGs are energetically dominated by compact ($\lesssim$1--2
kpc) nuclear regions \citep[e.g.,][]{soi00,dia10,ima11,per21}, which
are well covered with our ALMA data.

%%%%%%%%%% Figure C1 %%%%%%%%%
%%%%%%%%%%%%%%%%%%%%%%%%%%%%%%%%%%%

%%%%%%%%%% Figure C2 %%%%%%%%%
%%%%%%%%%%%%%%%%%%%%%%%%%%%%%%%%%%%

\clearpage

%%%%% Appendix D %%%%%

\section{Comparison of Observed HCN-to-HCO$^{+}$ Flux Ratios 
with RADEX Modeling Using Different Parameters}

\begin{figure}[h]
\begin{center}
%\vspace*{2cm}
%\hspace*{-12.3cm}
\includegraphics[angle=0,scale=.263]{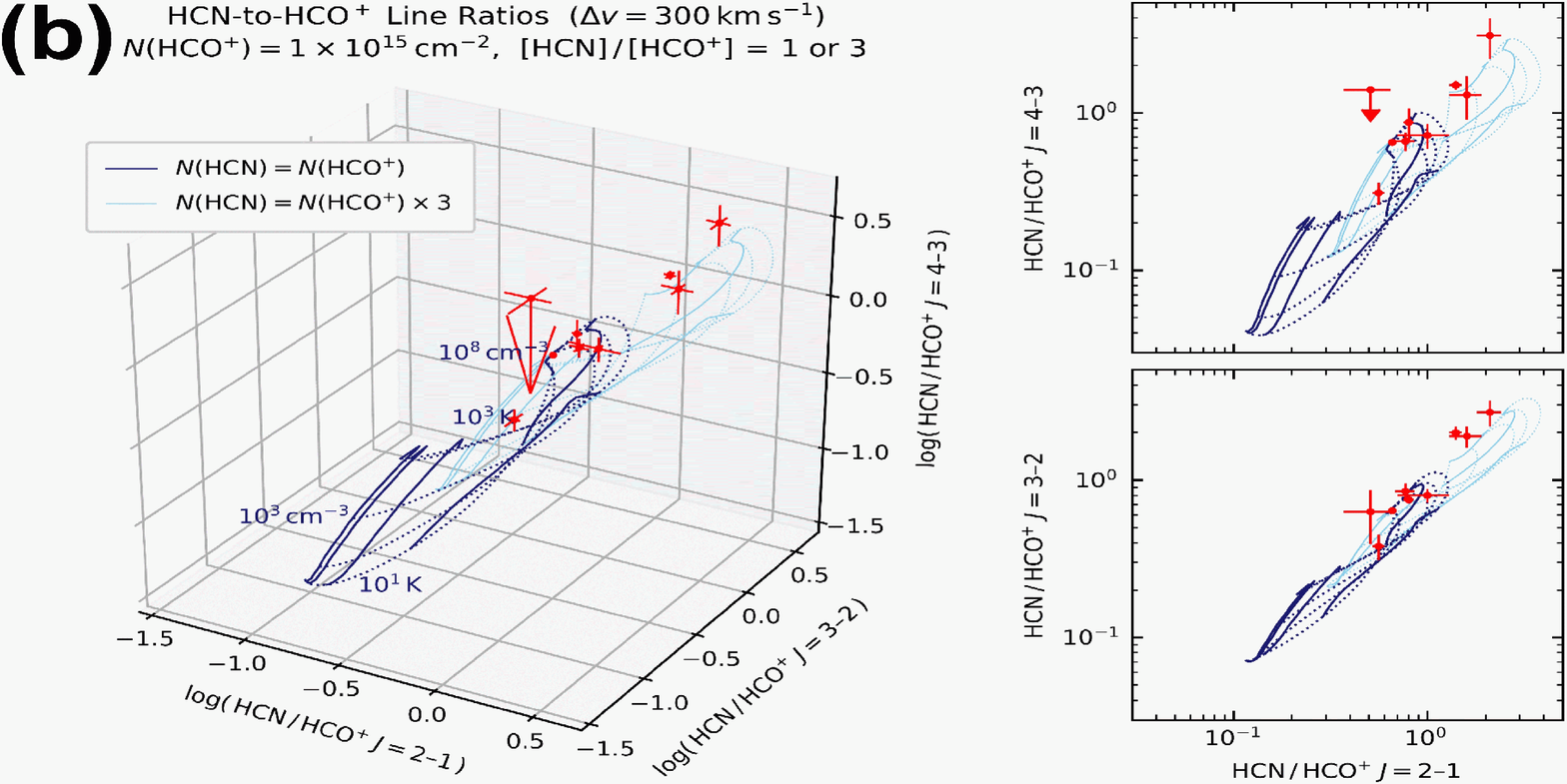} \\
\vspace*{1.5cm}
%\hspace*{-12.3cm}
\includegraphics[angle=0,scale=.263]{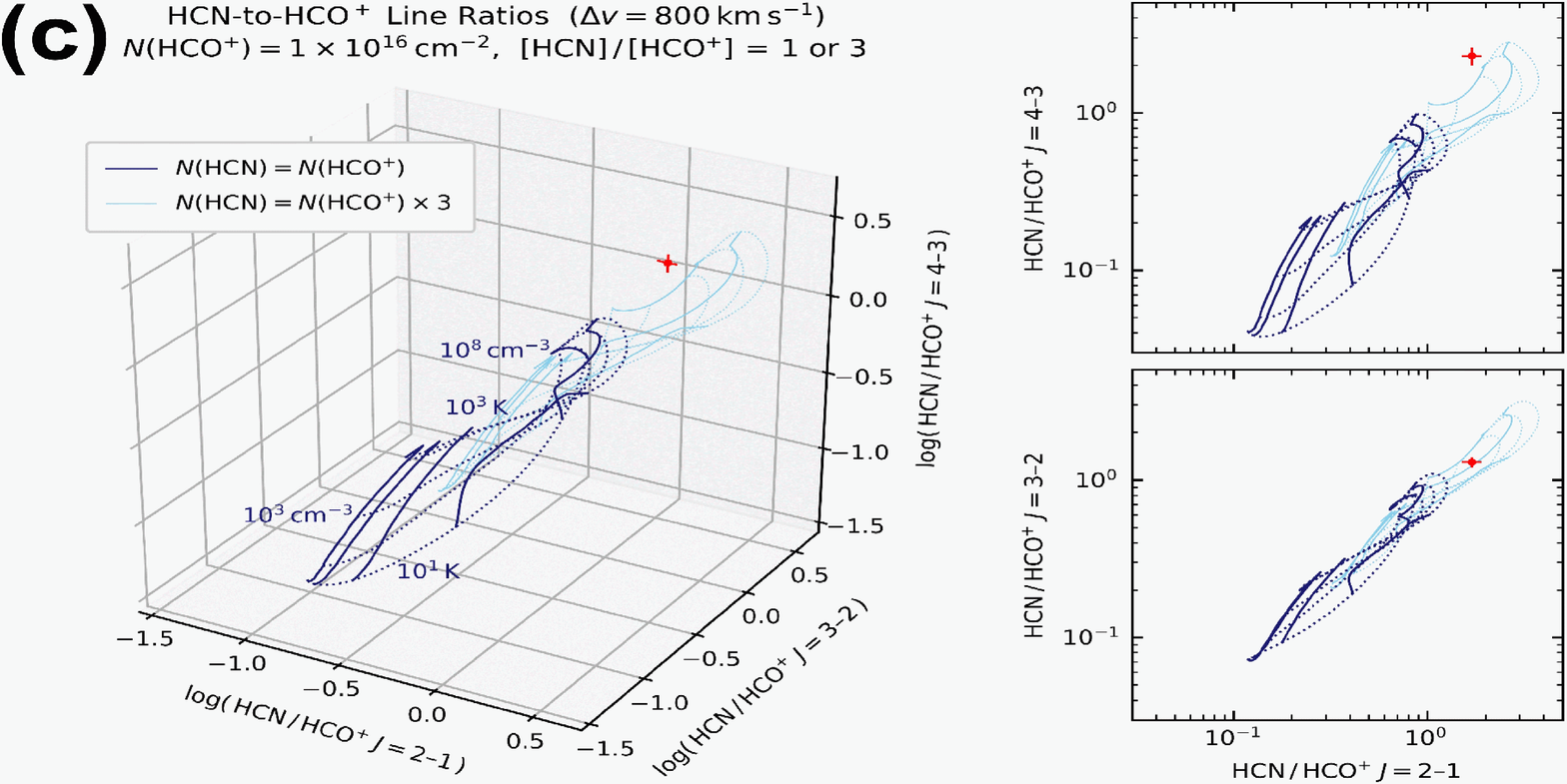} 
\end{center}
%\vspace{-1.0cm}
\caption{
Same as Figure \ref{fig:3Dplot}, but 
{\it (a)} for even higher HCN-to-HCO$^{+}$ abundance ratio of
[HCN]/[HCO$^{+}$] = 7, 
{\it (b)} for an order of magnitude smaller HCO$^{+}$ column density
of N$_{\rm HCO+}$ = 1 $\times$ 10$^{15}$ cm$^{-1}$, and {\it (c)} for
a factor of $\sim$2.3 larger line width of $\Delta$v = 800 km s$^{-1}$.
The red filled circles represent the observed HCN-to-HCO$^{+}$ flux
ratios of {\it (a,b)} the same (U)LIRGs' nuclei as plotted in Figure
\ref{fig:3Dplot} and {\it (c)} the Superantennae which displays an
exceptionally large line width compared to other (U)LIRGs' nuclei
(Table \ref{tab:flux}, column 11).  
\label{fig:3Dplotb}
}
\end{figure}

In Figure \ref{fig:3Dplotb}, the observed HCN-to-HCO$^{+}$
flux ratios of (U)LIRGs' nuclei are compared with RADEX non-LTE
modeling with significantly different values of 
HCN-to-HCO$^{+}$ abundance ratio, HCO$^{+}$ column
density, and line width, from those adopted in $\S$5.2. 
As evident, the overall distribution of the observed
HCN-to-HCO$^{+}$ flux ratios in the (U)LIRGs' nuclei are still better 
explained with an enhanced HCN-to-HCO$^{+}$ abundance ratio, as argued
in $\S$5.2.  

%%%%%%%%%% Figure D1 %%%%%%%%%
\begin{figure}[h]
\begin{center}
\vspace*{5cm}
%\hspace*{-12.3cm}
\includegraphics[angle=0,scale=.263]{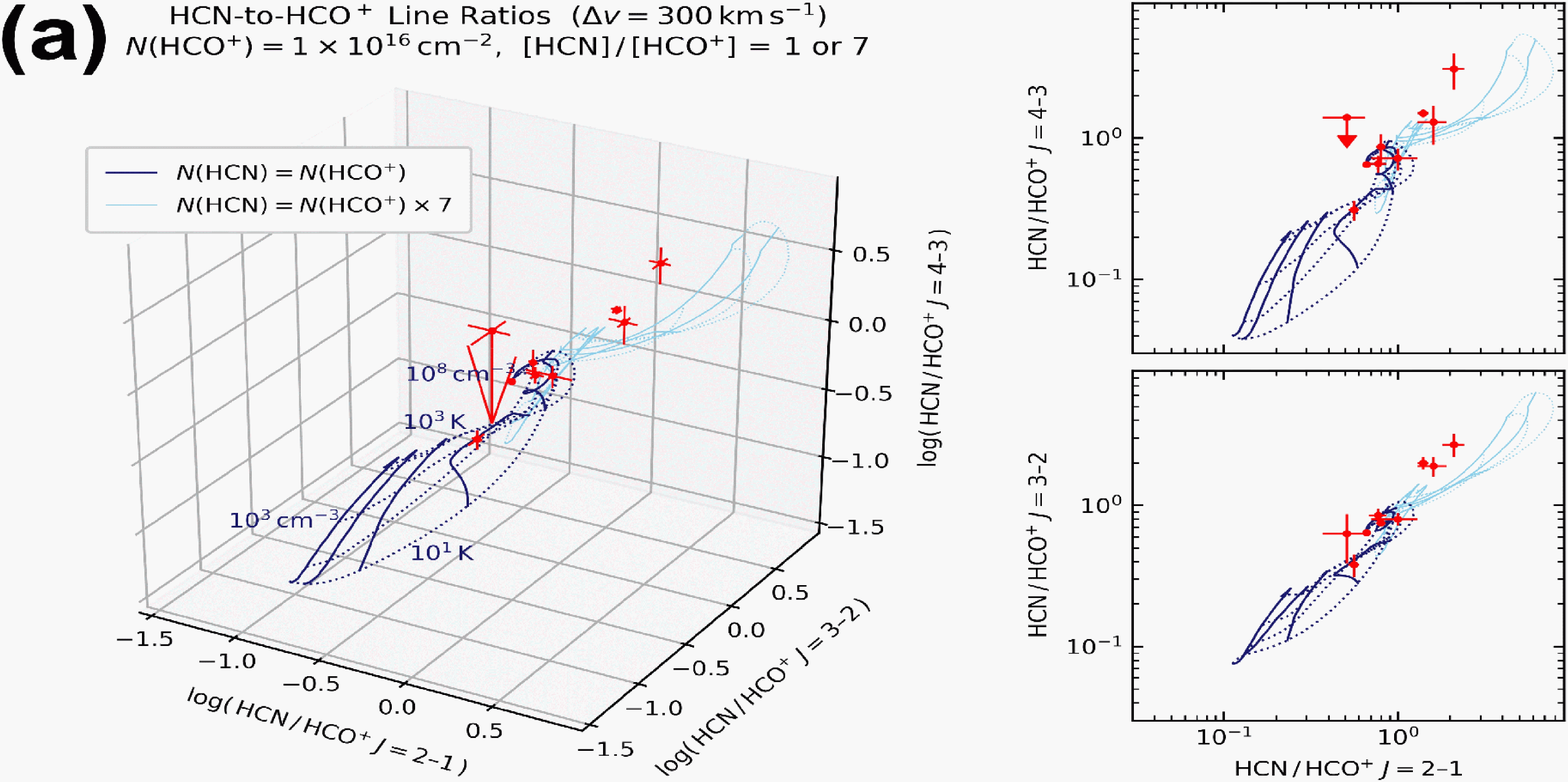} \\
\end{center}
\end{figure}

\clearpage

%%%%%%%%%%%%%%%%%%%%%%%%%%%%%%%%%%%




\end{document}